\documentclass{template}

\usepackage{amsmath}
\usepackage{amssymb}
\usepackage{bbm}
\usepackage[normalem]{ulem}

\usepackage[
backend=biber,
sorting=none
]{biblatex}

\begin{document}

\articletype{Paper} 

\title{Multivariate conformal uncertainty propagation in multitask atomistic simulation: Successes and pitfalls}

\author{Katharine Fisher$^{1,*}$\orcid{0000-0002-9655-984X}, Michael Herbst$^{2,3}$\orcid{0000-0003-0378-7921}, James Kermode$^4$\orcid{0000-0001-6755-6271} and Youssef Marzouk$^{1}$\orcid{0000-0001-8242-3290}}

\affil{$^1$Department of Aeronautics and Astronautics, Massachusetts Institute of Technology, Cambridge, MA 02139, USA}

\affil{$^2$Mathematics for Materials Modelling, Institute of Mathematics \& Institute of Materials,  École Polytechnique Fédérale de Lausanne, 1015 Lausanne, Switzerland}

\affil{$^3$National Centre for Computational Design and Discovery of Novel Materials (MARVEL), École Polytechnique Fédérale de Lausanne, 1015 Lausanne, Switzerland}

\affil{$^4$Warwick Centre for Predictive Modelling, School of Engineering, University of Warwick, Coventry CV4 7AL, United Kingdom}

\affil{$^*$Author to whom any correspondence should be addressed.}

\email{kefisher@mit.edu}


\begin{abstract}
Machine learning has become the standard tool for the design of interatomic potentials which balance efficiency and accuracy, but uncertainty quantification remains an open problem. Multiscale simulations introduce an additional challenge: robust uncertainty quantification across scales. Even within one scale, computations are often multistage, producing a sequence of target quantities, each dependent on the previous, and each with some uncertainty. Conformal methods have emerged as a model agnostic framework for recalibrating surrogate predictions to produce sets which contain the truth at a user-specified rate. For multistage workflows, we require uncertainty calibration for multiple chemical properties and atomistic configurations, and we want to propagate uncertainty sets to downstream quantities of interest. Such propagation should capture the error cancellations which occur in many downstream targets in materials science; for instance, an approximate energy difference is often more accurate than individual energy predictions. We present the first exploration of multivariate conformal methods for chemical properties, including Bonferroni-corrected hyperrectangles, hyperellipsoidal sets based on the Mahalanobis distance, and custom loss functions within conformal risk control. Calibration is applied directly to predicted energies, atomic forces, and virial stresses, then propagated to elastic constants and vacancy formation energies employing a variety of commonly considered approximate protocols in materials modeling. We highlight the benefits of building correlation predictions into the conformal procedure, making it possible to build sets which capture near symmetries and error cancellation. We conclude with a discussion of the interplay of the employed approximate computational protocol and conformal guarantees.
\end{abstract}

\section{\label{sec:intro}Introduction}

First principles methods can accurately simulate materials, but their computational cost quickly becomes astronomical as system size increases. In recent decades, machine learning surrogates have become standard in atomistic simulation because of their comparative computational efficiency~\cite{Deringer2019,Mueller2020,Kaser2023}. As a result, modern surrogate workflows can capture multiscale processes, incorporating unprecedented time and length scales. Being able to push traditional boundaries raises novel challenges: gauging the accuracy of surrogate predictions.

Both deep learning~\cite{Kaser2023} and kernel based~\cite{Bartok2010,Bartok2015,Bartok2018silicon,bartok2020} surrogates have seen remarkable success in materials simulations, but the two approaches pose different challenges for uncertainty quantification. The process by which deep learning architectures encode information from data remains mysterious~\cite{simon2026scientifictheorydeeplearning}, and the interpretation of uncertainty estimates for model parameters is not obvious~\cite{Fisher2025}.  In contrast, kernel methods, such as Gaussian process regression, offer principled predictions, but in realistic settings, data generally does not comply with the probabilistic assumptions of these surrogates,  leading to overconfident and misleading estimates. There is a need for reliable uncertainty quantification strategies which extract useful information from available data and can be easily and efficiently incorporated into multistage materials workflows.

\begin{figure}[ht!]
    \includegraphics[width=0.95\linewidth]{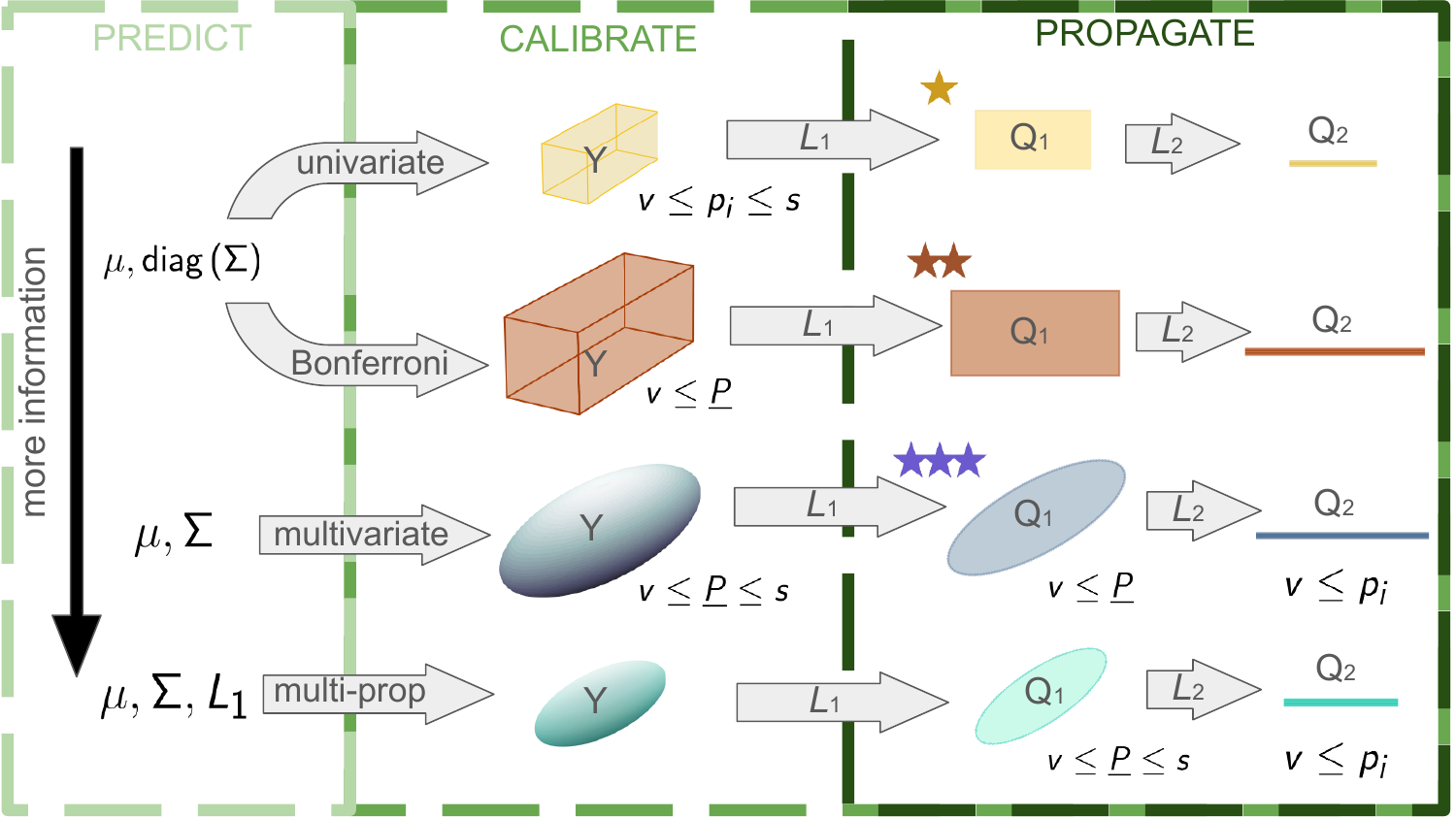}
    \caption{\label{fig:workflows} \textbf{Workflows to calibrate and propagate uncertainty in surrogate predictions}: We illustrate several conformal methods which leverage different degrees of information to construct multivariate uncertainty sets for a target $Y$. These uncertainty sets may then be propagated to downstream quantities of interest $Q_1$ and $Q_2$ via linear operators $L_1$ and $L_2$, respectively.  Where applicable, prediction sets are labeled with an abbreviation of the corresponding conformal guarantee. Let $p_i$ be the probability that an individual component of the true quantity is contained in the corresponding prediction set and $\underline{P}$ be the probability that the entire truth vector is contained in the multivariate set. The lower bound $v\leq\dots$ indicates that a \emph{coverage} (also called a validity) constraint holds; the upper bound $\dots\leq s$ indicates a \emph{sharpness} constraint, which guarantees that predictive sets will not be overly conservative. When using more information (e.g., a covariance $\Sigma$ or knowledge of the operator $L$), tighter bounds can be derived for the propagated sets. The stars accompanying sets on $Q_1$ indicate correspondence with Figure~\ref{fig:visual_sets}. }
\end{figure}
\begin{figure}[h!]
    \includegraphics[width=0.95\linewidth]{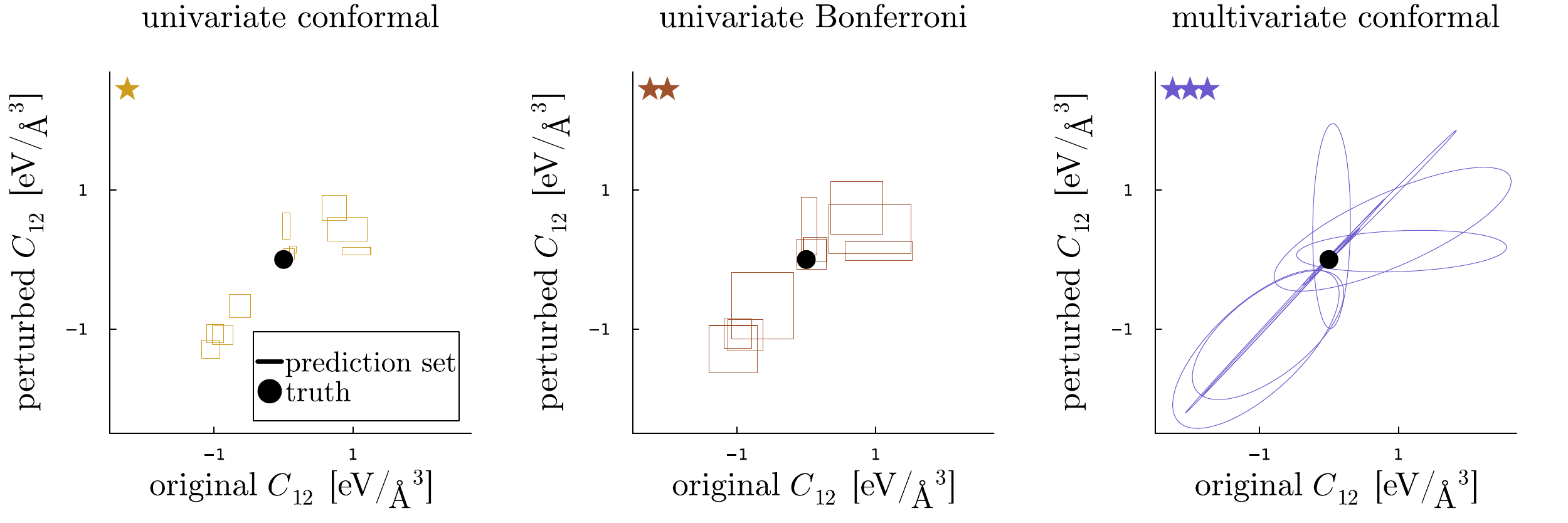}
    \caption{\label{fig:visual_sets} \textbf{Finding near symmetries}: multivariate conformal prediction can leverage the correlation between properties. As an example, the multivariate approach (purple) discovers the near symmetry of the $C_{12}$ elastic constant components for two perturbed silicon configurations. In contrast, univariate conformal methods (gold and orange) produce  uncorrelated sets which either undercover or include excessive volume. The conformal sets are calibrated on energy predictions for perturbed and relaxed supercells of Si$_{16}$ diamond crystals with tolerance $\alpha=0.25$ and uncertainty sets are projected into elastic constant space. The stars in the upper left corner show correspondence with Figure~\ref{fig:workflows}.}
\end{figure}

Conformal methods can use surrogate outputs to construct prediction sets with finite sample calibration guarantees~\cite{vovk2005,shafer2007tutorialconformalprediction,angelopoulos2022gentleintroductionconformalprediction}. The procedure can be applied post hoc to any model and essentially learns that model's prediction and uncertainty quantification capabilities. Initial explorations of these methods for atomistic simulation have demonstrated their broad compatibility and their ability to utilize model specific outputs, such as standard deviations from Gaussian process predictive distributions~\cite{Hu_2022,Best_2024,Bestwarwick,Zaverkin2024}. One limitation of existing work is the focus on scalar quantities of interest, such as global energy for a single system. Often, a multiscale workflow will require uncertainty estimates for multivariate quantities of interest, such as virial stresses. Furthermore, within multistage simulations, we will be interested in the propagation of uncertainty from multiple systems into downstream and derived quantities. For example, we may leverage the global energy of each member of an ensemble of perturbed systems to predict downstream properties such as the elastic constant tensor. Several strategies have recently been proposed to build prediction sets with multivariate guarantees, typically tested on toy data and benchmarking problems~\cite{Dheur2025,zhou2025}. To our knowledge, this paper is the first to extend these methods to material properties prediction.

In this work, we test computationally efficient methods for obtaining \emph{multivariate} prediction sets on chemical properties. We consider both standard \emph{conformal prediction} and its generalization to arbitrary, set-valued loss functions, \emph{conformal risk control} \cite{angelopoulos2025conformalriskcontrol}. Specifically, we use the correlation between chemical properties, estimated by the Gaussian process predicted covariance, to construct hyperelliptical sets \cite{johnstone2021,braun2026}. Further, we consider the propagation of these uncertainty sets to downstream quantities of interest. We compare several multistage simulation workflows, shown schematically in Figure~\ref{fig:workflows}. The choice of conformal method determines set shape and the probabilistic guarantee that the set contains the true property. In particular, we explore the behavior of hyperelliptical sets which automatically provide coverage guarantees for downstream quantities of interest. The consequences are apparent in Figure~\ref{fig:visual_sets}. Here, conformal calibration was applied to the energy of several geometrically similar Silicon crystals, then propagated to produce two-dimensional sets predicting the same elastic constant component for two different systems\footnote{More detail on this experiment will be provided in Section~\ref{ss:elasticconstant}}. Not only do the multivariate sets show superior coverage of the true values, but they reflect the near symmetry in related quantities, often producing nearly diagonal sets. We achieve this result by training the conformal procedure to recognize the correlation between related quantities. This adaptation is crucial to capturing the error cancellations which occur in the computation of many key quantities in materials science, such as trends and differences in properties.  By capturing these relationships, elliptical sets often produce sets at smaller volume than competing methods, at a given coverage level. Thus, these sets are simultaneously robust and less conservative than alternatives because they are less likely to incorporate unlikely predictions.
 
We enumerate our contributions in the following subsection, and Section~\ref{sec:litreview} puts them in the context of existing literature. In Section~\ref{sec:methods}, we provide an overview of the methods combined in our work. Our numerical experiments are detailed in Section~\ref{sec:experiments}, and we conclude in Sections~\ref{sec:discussion} and~\ref{sec:conclusion} with a discussion of the strengths and potential pitfalls of our approach. 

\subsection{\label{sec:contributions}Contributions}

\begin{itemize}
    \item We identify algorithms for conformal prediction and risk control which efficiently construct multivariate prediction sets. By combining these algorithms with interatomic potentials, we create calibrated prediction workflows.
    \item Our workflow melds Bayesian and frequentist procedures. The covariance provided by Gaussian process regression is incorporated into the conformal scores so that predictive sets capture the relationships between quantities of interest, achieving sets with frequentist validity which are not overly conservative.
    \item We combine multiple information sources using a multitask Gaussian process interatomic potential which relates the energies, forces, and stresses for bulk and vacancy atomistic systems.
    \item The test cases considered include conformal sets with $O(100)$ dimensions, one to two orders of magnitude larger than typically considered in conformal literature, with few exceptions~\cite{principato2025}.
    \item Two new loss functions for conformal risk control are formulated to promote high component-wise coverage rates for multivariate quantities of interest by leveraging covariance predictions. These strategies can avoid producing the overly conservative sets which can arise when multivariate conformal methods are applied to sufficiently high dimensional targets.
    \item We examine nontrivial examples of propagation of multivariate uncertainty sets. Specifically, we consider the prediction of elastic constants and vacancy formation energies, i.e.~quantities which involve multiple energy and force evaluations as part of a prediction workflow involving structural perturbations and relaxations.
    \item We discuss potential pitfalls which may arise when propagating conformal sets through approximate transformations.
\end{itemize}

\section{Related work\label{sec:litreview}}

\subsection{Machine learning interatomic potentials}
Due to the computational expense of simulation from first principles~\cite{Lin_Lu_Ying_2019}, machine learning has become standard in the design of potential energy surfaces and force fields. In recent years, a favorable balance of accuracy and efficiency has been achieved with data intensive, deep neural network based frameworks~\cite{Kaser2023}. Graph based architectures capture the structure of atomistic systems~\cite{Chen2022gnn,Park2024gnn} and encode equivariant transformations~\cite{Batzner2022,Yang2025}. Foundation models~\cite{Batatia2025foundation} and chemistry focused large language models~\cite{kristiadi2024llmsmaterial} demonstrate the promise of large-scale learning for general purpose prediction. While deep learning achieves state of the art performance, reliable UQ for these methods remains an open problem, as discussed in Subsection~\ref{ss:UQ_review}.

Kernel methods form another class of machine learning surrogates where the similarity between atomistic systems is explicitly modeled. The Bayesian framework was used to build Gaussian approximation potentials~\cite{Bartok2010,Bartok2015,Bartok2018silicon,bartok2020}, providing interpretable, probabilistic uncertainty quantification. In certain regimes, the accuracy of kernel methods can rival deep neural networks~\cite{arora2019harnessingpowerinfinitelywide,Lee2020,Radhakrishnan2022}, and a suite of strategies have been developed for scaling these approaches to large datasets~\cite{Ma2017,Meanti2020,liu2020gaussian}. In this work, we focus on kernel methods for their ease of implementation and principled UQ, but all of our experiments are straightforward to port to the neural network setting. Similarly, surrogates which learn from multiple information sources can be designed as neural networks or kernel models, but in the latter case similarity can be explicitly modeled.  Examples include multitask inference~\cite{Fisher2024,Khatamsaz2023}, multifidelity information fusion~\cite{Goodlett2023,Pilania2017,Batra2019,Patra2019,Fare2022}, and optimized combinations of submodels~\cite{Vinod2023,vinod2023optimized,Vinod2025,vinod2026improviseadaptovercomeonthefly}. A variant of learning from multiple data sources is $\Delta$-learning~\cite{Ramakrishnan2015,Dral2020,dral2023learning,Goodlett2023}, which directly models the difference between two datasets. This procedure can also be extended to the multitask setting~\cite{Fisher2024}. Such multitask methods inherit the UQ challenges of the underlying choice of surrogate model. 

\subsection{UQ in atomistic simulation\label{ss:UQ_review}}
A wide range of uncertainty quantification strategies devised for machine learning have been tested in the materials science setting~\cite{Frombgen2026,Wang2020UQMM,grasselli2025}. A few UQ methods have been developed in response to specific challenges in atomistic simulation, i.e.~misspecified, deterministic models~\cite{Swinburne_2025}. In this work, we focus primarily on uncertainty predicted by Gaussian process models~\cite{Bartok2010,Bartok2015,Bartok2018silicon,bartok2020} and calibrated through conformal procedures~\cite{Hu_2022,Best_2024,Bestwarwick,Zaverkin2024}. We note that both approaches offer principled theoretical guarantees, provided modeling assumptions are met, which offer interpretability. Such guarantees are challenging to access for UQ methods tailored to neural network architectures~\cite{Gawlikowski2023,He2026}, even in ostensibly Bayesian settings~\cite{Fisher2025}. Sampling the high-dimensional, multimodal posteriors of Bayesian neural networks is generally intractable. Deep ensemble members can be interpreted as posterior samples~\cite{izmailov2021} and show strong empirical performance but are computationally expensive~\cite{Tan2023,Valdenegro-Toro2023}. Last layer ensembles~\cite{lee2015mheads,Valdenegro-Toro2023} offer an efficient solution for atomistic simulations~\cite{Kellner_2024,kellner2026,Kellner2026b} and can be interpreted as a model of a single posterior mode (Laplace approximation). None of these methods guarantee calibrated UQ, and unimodal approximations are particularly vulnerable to overconfidence~\cite{Kahle2022,lu2023uncertainty,Fisher2025}.

Existing approaches to propagation of atomistic scale uncertainty generally involve pushing samples $\{y^{(i)}\}_{i=1}^n$ through a forward model  $\{f(y^{(i)})\}_{i=1}^n$. If samples are difficult to obtain or $f$ is nonlinear, this approach to propagation can incur a high computational cost. We can mitigate this expense by linearizing $f$, i.e.~via a Taylor expansion~\cite{Kellner_2024,kellner2026,Kellner2026b}. Furthermore, if the samples $\{y^{(i)}\}_{i=1}^n$ are distributed according to a Gaussian density, that distribution can be pushed through a linear model, eliminating the need for sampling~\cite{schmitz2025}. Thus, the output of Gaussian process regression models are particularly amenable to propagation, provided that we can ensure that predicted distributions are trustworthy. For this purpose, we turn to conformal methods. 

\subsection{Multivariate conformal methods}
Within the past few years, a wealth of new strategies have been proposed for conformalizing multivariate targets~\cite{Dheur2025,zhou2025}. One class of methods applies univariate algorithms to each dimension; the intersection of their predictive intervals forms a predictive hypperrectangle~\cite{neeven2018}. Tools from classical statistics, such as the Bonferroni correction~\cite{Stankeviciute2021} or multiple hypothesis testing~\cite{timans2024,timans2025}, can be used to adjust the probability that the hypperrectangle covers the true value. We use such hyperrectangular sets as a baselines in our work.

Conformal algorithms rely on a score function which rates the accuracy of surrogate predictions. Several methods function by transforming a multivariate prediction into a univariate score which is compatible with the original conformal algorithm~\cite{wang2023pcp,Dheur2025}. The score may be, for example, a probability density~\cite{Sadinle2018,Izbicki2022,sampson2025,wang2023pcp,Lei2015,wang2023pcp,Dheur2025}. Closely related is the Mahalanobis distance score, a transformation of the Gaussian density~\cite{johnstone2021,braun2026,principato2025,Xu2024ts}. We adopt this score because it is straightforward to implement and fast at prediction time. The resulting hyperelliptical sets are conducive to uncertainty propagation and capture correlation. Other explorations of hyperelliptical conformal sets learn correlation from the sample covariance~\cite{johnstone2021,principato2025,Xu2024ts} or a parameterized model~\cite{braun2026}, but we use the covariance predicted by Bayesian regression. This workflow bears some similarity to methods which build scores from the Bayesian covariance, computed using a kernel which relates prediction residuals~\cite{meyer2026}. However, the latter approach does not necessarily produce convex sets which are amenable to uncertainty propagation. The same shortcoming holds for generalizations of the Mahalanobis score to an $\ell_p$ distance where $p$ and distribution statistics are learned hyperparameters~\cite{braun2026minvol}. Further approaches also present computational challenges: leveraging optimal transport to generalize the notion of a quantile to multiple dimensions~\cite{klein2025,thurin2025,ndiaye2025}, modeling correlation structure using copulas~\cite{messoudi2021,park2025,sun2024}, performing quantile regression~\cite{feldman2022}, and optimizing transformations of hypercubes~\cite{gray2025,lutzow2025}.

\subsection{Conformal uncertainty propagation}
We focus on hyperrectangular and hyperellipsoidal sets because both are amenable to transformation. Prior exploration of conformal uncertainty propagation has been limited. The closest work~\cite{braun2026} to our own considers the propagation of ellipsoidal uncertainty sets and notes the coverage properties of downstream sets. However, the results presented consist of only toy examples or randomly sampled propagation operators applied to sets of at most $16$ dimensions, whereas we consider a practical workflow for atomistic modeling and higher dimensional calibration spaces. A few works have propagated univariate conformal sets for specific applications: for instance, pushing uncertainty in crop yield forward to a prediction interval for profit~\cite{Khan2026} or adjusting the standard deviations propagated by a Kalman filter in multiple object tracking~\cite{su2024}. Beyond these works, the most relevant field to conformal uncertainty propagation appears to be conformal decision making~\cite{Vovk2018}, where uncertainty sets inform an action. Within this category are, to the authors' knowledge, the only previous applications of \emph{multivariate} conformal prediction~\cite{bai2025mcs} and of conformal risk control~\cite{bai2026score} to materials science workflows. These applications differ from ours in that their goal is to design a diverse set of atomistic systems with low manufacturing cost, rather than to build prediction sets for chemical properties. Other works in decision making consider the settings of multiple object detection where a conformal set on labels is the basis for class conditional conformal regression on a bounding box~\cite{timans2024}, trajectory optimization where the coverage constraints are reallocated across time as the state evolves~\cite{wang2025fbcp}, hierarchical medical diagnosis based on a downstream loss function~\cite{cortesgomez2025}, and a financial model where the first stage of conformal prediction can lead to early stopping of the automated procedure and escalation to a human reviewer~\cite{Haas2025}.

\section{Methods\label{sec:methods}}

\subsection{Gaussian approximation potential (GAP)}

For each of $n$ systems, consider a $4$-tuple of properties: $\{ (R^{(i)}, E^{(i)}, F^{(i)}, \varsigma^{(i)}) \}_{i=1}^n$. $R^{(i)} \in \mathbb{R}^{M^{(i)}\times 3}$ holds the Cartesian coordinates in $3$-dimensional space for the $M^{(i)}$ atoms which compose the $i^{th}$ system. $E^{(i)}\in\mathbb{R}$, $F^{(i)} \in \mathbb{R}^{M^{(i)}\times 3}$, and $\varsigma^{(i)} \in \mathbb{R}^{3\times 3}$ are, respectively, the global energy, atomic forces, and virial stresses of the system. Note also that for our experiments, we focus on periodic systems, which also require a $3\times3$ cell matrix, conjugate to the stress. The quantities are related via
\begin{eqnarray}
    \label{eq:force} F^{(i)} &=& - \nabla_{R^{(i)}} E^{(i)}\,, \\ 
    \label{eq:rdiff} r^{(i)}_{abc} &=& R^{(i)}_{ ac}- R^{(i)}_{ bc} \,, \\
    \label{eq:stress} \varsigma^{(i)}_{\alpha\beta} &=& \frac{1}{ V^{(i)}} \sum_{a=1}^{M^{(i)}} \sum_{b>a}^{M^{(i)}} r^{(i)}_{ab\alpha}    \frac{\partial E^{(i)}}{\partial r^{(i)}_{ab\beta} } \frac{\partial r^{(i)}_{ab\beta}}{ \partial R^{(i)}_{ a\beta}} 
     = \frac{1}{ V^{(i)}} \frac{\partial E^{(i)}}{\partial \epsilon_{\alpha\beta}} \,,
\end{eqnarray}
where $V^{(i)}$ and $\epsilon$ are respectively the volume and the strain tensor of the configuration cell.

The first component of our prediction pipeline is a surrogate model which maps representations of atomistic geometry to some subset of $( E^{(i)}, F^{(i)}, \varsigma^{(i)})$. These predictions will then be calibrated with conformal methods which, conveniently, are agnostic to the form of the surrogate $f$. We choose to use Gaussian approximation potentials (GAP) which are established in chemical modeling~\cite{Bartok2010,Bartok2015,Bartok2018silicon,bartok2020}, straightforward to implement, and interpretable. In particular, we will leverage covariance predictions which are grounded in probability theory, in contrast to the comparatively ad hoc UQ methods available for neural networks~\cite{Kahle2022,Tan2023,lu2023uncertainty} as discussed in Subsection~\ref{ss:UQ_review}.

To build a GAP model, we require features $X$ which capture the similarity between configurations. Using the Smooth Overlap of Atomic Positions (SOAP)~\cite{Bartok2013,Bartok2015,Musil2021}, we construct $X^{(i)}\in\mathbb{R}^{d_X\times M^{(i)}}$ where each column $X^{(i)}$ represents atom $a$ and its interaction with neighboring atoms. Implicitly, our surrogate model will assume that the global energy is additively decomposed into local contributions
\begin{eqnarray}
    \label{eq:energy_model}
    E^{(i)} &=&  f(X^{(i)}) + \eta^{(i)} = \sum_{a=1}^{M^{(i)}} \varepsilon(X^{(i)}_{a}) + \eta^{(i)}\,, \\
    X^{(i)}_{a} &=& \xi(R^{(i)}_{a} ; \; \ell_{max}, n_{max}, r_{cut}) \,, 
\end{eqnarray}
where $\xi$ is the SOAP map which introduces hyperparameters $\ell_{max}$, $n_{max}$, and $r_{cut}$.  $\eta^{(i)}$ is a centered, Gaussian noise term, realized independently for different atomistic systems, with variance $\gamma$. Given a kernel function $k: \mathbb{R}^{d_X}\times\mathbb{R}^{d_X} \to \mathbb{R}$, we define a Gaussian process prior
\begin{eqnarray}
    f(X^*) \sim N\left( \mathbf{0}, \; \sum_{a=1}^{M^*} \sum_{b=1}^{M'} k(X^*_a, X'_b ) \right)\,.
\end{eqnarray}
We pair SOAP descriptors with a polynomial kernel because the inner product effectively computes an integral over all rotations of the descriptor power spectrum, producing a rotationally invariant covariance. Note that this prior model only describes system energy. We also want the ability to learn from and predict forces and stresses. Following Equations~\eqref{eq:force}-\eqref{eq:stress}, we derive the covariances between each pair of properties by differentiating through $k$ and $\xi$~\cite{Bartok2015}.   

Once a training set $t$ and testing set $*$ of atomistic systems and properties have been identified, let $Y_t$ be the column vector of all training data and $K$ represent the covariance given by $k$ for the sets indicated in the subscript. Gaussian process regression~\cite{gpml} provides the predictive distribution for $f$ conditioned on $Y_t$. The predictive mean and covariance are
\begin{eqnarray}
    \mu &=&  K_{*t} (K_{tt} +\gamma \mathbf{I})^{-1} Y_t \,,  \\ 
    \Sigma &=& K_{**} - K_{*t} (K_{tt} +\gamma \mathbf{I})^{-1} K_{t*} \,.
\end{eqnarray} 
These expressions require the inversion of a potentially large matrix, but the computational cost can be greatly ameliorated using methods from the well developed field of scalable Gaussian processes~\cite{Ma2017,Meanti2020,liu2020gaussian}. The final covariance $\Sigma$ is the initial test covariance $K_{**}$ minus the squared Mahalanobis distance between the training and testing data. Assuming that the multivariate Gaussian model for $f$ is justified, this adjustment is a principled update to our knowledge about the testing set.

\subsection{Multitask Gaussian process models}

There is a clear physical relationship between energy, force, and stress which guides the incorporation of these heterogeneous data types into one model. However, we may also have access to heterogeneous data where the mathematical relationship between different sets is not precisely known. For instance, for a single geometry $R^{(i)}$, we may have experimental observations of energy $E^{(i,\,\text{exp})}$ along with predictions at multiple levels of quantum chemical theory~\cite{Fisher2024}, i.e.~$E^{(i,\,\text{DFT})}$ and $E^{(i,\,\text{CCSD})}$. In our numerical experiments, we will consider an example where we learn from the energy of bulk and the vacancy configurations. Na{\"{i}}vely, each heterogeneous dataset can be used to build an independent model. We will instead consider a multitask Gaussian process approach~\cite{Kennedy2000,Bonilla2008,Leen2012,Fisher2024} which models correlations in all training data and can predict correlations in heterogeneous testing sets.

Each heterogeneous set of data will be labeled a task. We will identify one task as the primary task $p$, and the remaining tasks will be secondary tasks $s_1,\dots,s_J$. Following Equation~\eqref{eq:energy_model}, we model the energy predictions of the tasks as
\begin{eqnarray}
    E^{(i,\,p)} &=&  f^{(p)}(X^{(i,\,p)}) + \eta^{(i,\,p)} \,, \\
    E^{(i,\,s_j)} &=&  f^{(s_j)}(X^{(i,\,s_j)}) + \eta^{(i,\,s_j)}\,, \qquad j=1,\dots, J\,.
\end{eqnarray}
To allow for joint learning, we must choose an ansatz to capture the relationship between tasks. A straightforward and effective~\cite{Fisher2024} approach models the linear correlation between tasks:
\begin{eqnarray}
    \label{eq:correlation}
    E^{(i,\,s_j)}(X^{(i,\,s_j)}) =  \varrho^{(s_j)} E^{(i,\,p)}(X^{(i,\,p)}) + \delta^{(s_j)}(X^{(i,\,s_j)})\,.
\end{eqnarray}
Above, we have introduced the task specific correlation hyperparameter $\varrho^{(s_j)}$ and residual $\delta^{(s_j)}(X^{(i,\,s_j)})$. The latter prevents ``negative transfer'' by modeling behavior of task $s_j$ which is not linearly correlated to the primary task. In contrast to multifidelity fusion models~\cite{Kennedy2000}, all secondary tasks have an equivalent relationship to the primary task, so no hierarchy is encoded in $s_1,\dots,s_J$. 

To train our model, we assume Gaussian process priors for the residuals $\delta^{(s_j)}(X^{(i,\,s_j)})$ for $j=1,\dots J$. Using Equations Equations~\eqref{eq:force}-\eqref{eq:stress} and~\eqref{eq:correlation}, we can find the covariance between any pair of data points from the union of our training and test set. Crucially, we do not require full knowledge of our training examples. More concretely, if we only have data from example $(i)$ for one task, we can still incorporate that data into our model. Similarly, we do not need to have the complete set $(E^{(i)},F^{(i)},\varsigma^{(i)})$ for each $(i)$ to learn from a subset of these properties. 

\subsection{Conformal prediction}

Gaussian process regression produces a predictive distribution with uncertainty encoded in the posterior covariance: $\Sigma$. Conditioning on new data points leads to coherent updates of this uncertainty representation provided that priors, kernel models, and noise assumptions are well specified. In practical chemistry models, some level of misspecification is nearly certain, i.e.~due to interaction truncations, descriptor approximations, etc. The exact degree of misspecification is challenging to identify. Conformal methods offer a post-hoc procedure that learns to correct to raw predictions and uncertainty estimates of a surrogate model~\cite{vovk2005,shafer2007tutorialconformalprediction,angelopoulos2022gentleintroductionconformalprediction}.

Suppose we have a calibration dataset $\{(X^{(n+i)},Y^{(n+i)}\}_{i=1}^{m}$ of $m$ data points which have been set aside during training of the surrogate model. For the moment, $Y^{(i)}$ may take on the value of system energy or an element of the force or stress matrices. Given a new test point $(X^{\text{(test)}},Y^{\text{(test)}})$, conformal prediction constructs a set $\Gamma_m(X^{\text{(test)}})$ with the finite sample guarantee
\begin{eqnarray}
    \label{eq:conform}
    1-\alpha \leq \mathbb{P}\left( Y^{\text{(test)}} \in \Gamma_m(X^{\text{(test)}}) \right) \leq 1- \alpha + \frac{1}{1+m}\,,
\end{eqnarray}
provided that the calibration and testing data are exchangeable (a relaxation of independent and identically distributed). The bounded quantity is the coverage of the prediction set of the true property. We refer to the lower bound as a coverage or validity guarantee, controlled by the user's chosen tolerance $\alpha$. The upper bound is a sharpness guarantee which indicates that valid sets will not be overly conservative. 

Split, or inductive, conformal prediction is a computationally efficient strategy for building $\Gamma_m(X^{\text{(test)}})$. We define a score function $s(X,Y)$ which evaluates the ability of the surrogate to predict $Y$ given $X$. Larger scores indicate worse predictive performance. Optionally, the score may give surrogates credit for estimating high uncertainty for poor predictions. For instance, a common score applied to regression problems is
\begin{eqnarray}
    \label{eq:uni_score}
    s^{\text{uni}}(X,Y) = \frac{|\mu(X) - Y|}{\sigma(X)} \,,
\end{eqnarray}
where $\mu(X)$ and $\sigma(X)$ are the surrogate's mean and standard deviation predictions, respectively. $s(X,Y)$ is used to score each calibration data pair, and we let $\hat{q}$ be the $\lceil m^{-1} (m+1)(\alpha-1)  \rceil$ quantile of the resulting scores. Then, 
\begin{eqnarray}
    \Gamma_m(X^{\text{(test)}})   = \left\{ Y: s(X^{\text{(test)}}, Y) \leq \hat{q}  \right\}\,,
\end{eqnarray}
achieves the conformal guarantees~\eqref{eq:conform}. Choosing the score~\eqref{eq:uni_score} leads to a prediction interval: $Y\in \left[ \mu(X) - \hat{q}\sigma(X),\; \mu(X) + \hat{q}\sigma(X) \right]$. Note that for any $X$, this approach produces a $\hat{q}$ scaling of the original one standard deviation interval about the mean. Without further assumptions, conformal methods cannot  guarantee coverage conditional on $X$~\cite{angelopoulos2022gentleintroductionconformalprediction}. Rather, scores may be chosen which empirically provide good conditional coverage, procedural modifications may be made to achieve class conditional coverage, or additional assumptions may be imposed.

\subsection{Multivariate conformal prediction}

Conformal prediction was originally formulated for scalar predictions $Y$. In our materials science setting, however, we are interested in the atomic forces and virial stresses of a configuration, in addition to global energy. Applying the univariate approach to each component of a multivariate $Y\in\mathbb{R}^{d_Y}$ results in the conformal guarantee~\eqref{eq:conform} applying independently to each coordinate. The coverage probability for the entire vector is bounded below by approximately $1-d_Y^{-1}\alpha$. Based on this observation, a Bonferroni-style correction divides the original choice of $\alpha$ by $d_Y$~\cite{Stankeviciute2021}. As reviewed in Section~\ref{sec:litreview}, diverse additional strategies have been proposed for extending conformal methods to multivariate settings~\cite{Dheur2025,zhou2025}.

Some quantities of interest, such as atomic forces, scale with the number of atoms in a given system, and we are often interested in the correlation of properties across systems. Consequently, we consider higher dimensional vectors than are typically seen in literature. Save one exception~\cite{principato2025} previously reported examples only considered applications between $2$ and $16$ dimensions.
To handle the high dimensional properties in this work, we require a computationally efficient approach. The Mahalanobis distance score~\cite{johnstone2021,braun2026}
\begin{eqnarray}
    \label{eq:multi_score}
    s^{\text{multi}}(X,Y) = d_Y^{-1/2} \| \Sigma(X)^{-1/2}  (Y-\mu(X)) \|_2\,,
\end{eqnarray}
can be incorporated seamlessly into the existing conformal workflow because it maps multivariate predictions to scalar scores. Furthermore, this approach allows us to build calibration and test sets where the size of $Y$ is nonuniform, as occurs if we consider systems with different numbers of atoms. For univariate $Y$, the Mahalanobis distance reduces to $s^{\text{uni}}(X,Y)$. Unlike the univariate approach, $s^{\text{multi}}$ uses the full covariance prediction $\Sigma$. Thus, when our Gaussian process model identifies correlated predictions, multivariate conformal calibration can leverage the relationship to produce smaller sets. Specifically, for a test point, the predictive set
\begin{align*}
    s^{\text{multi}}(X^{\text{(test)}},Y) \leq \hat{q}^2\,,
\end{align*}
is a (hyper)ellipsoid. As discussed further in Subsection~\ref{ss:prop}, this shape facilitates sample-free uncertainty propagation through a multi-step workflow.

\subsection{Conformal risk control}

The choice between univariate and full vector coverage may seem restrictive, particularly as the dimension of $Y$ grows. One response may be to compensate by adjusting $\alpha$. This strategy has limitations: in the univariate setting, all correlations are discarded, risking unnecessarily large sets; in the multivariate setting in the multivariate setting, a $Y\in\mathbb{R}^{1000}$, which is not covered could imply everything between two extremes: namely that (1) all components of $Y$ are outside the component-wise intervals induced by $\Gamma_m(X^{\text{(test)}})$ or that (2) $999$ components of $Y\in\mathbb{R}^{1000}$ are contained in a projection of $\Gamma_m(X^{\text{(test)}})$ (onto the corresponding $999$ dimensional space), but only the remaining $1000^{\text{th}}$ component is not.

More bespoke guarantees are possible through conformal risk control~\cite{angelopoulos2025conformalriskcontrol}. A set $\Gamma_{m,\,\lambda}(X^{\text{(test)}})$ is constructed so that
\begin{eqnarray}
    \label{eq:control}
    \alpha - \frac{2B}{m+1} \leq \mathbb{E}\left[ \ell\left( \Gamma_{m,\,\widehat{\lambda}}(X^{\text{(test)}}) , \, Y^{\text{(test)}} \right) \right] \leq{\alpha}\,,
\end{eqnarray}
where $\ell\left( \Gamma_{m,\,\widehat{\lambda}}(\cdot)  ,\cdot\;  \right)$ is some loss function which is non-increasing in $\widehat{\lambda}$ and bounded above by $B$. Choosing $\ell$ to be $\mathbbm{1}\left\{ Y^{\text{(test)}}\in \Gamma_{m,\,\widehat{\lambda}}(X^{\text{(test)}}) \right\}$ recovers traditional conformal prediction. As previous work~\cite{angelopoulos2025conformalriskcontrol} describes in detail, $\widehat{\lambda}$ is determined from the average loss of the calibration set, analogously to $\hat{q}$ in conformal prediction.

For ease of implementation and propagation, we focus on (hyper)ellipsoidal sets
\begin{eqnarray}
    \label{eq:ineq}
     \| \Delta \|_2 = d_Y^{-1/2} \| \Sigma(X)^{-1/2}  (Y-\mu(X)) \|_2 \leq \lambda\,,
\end{eqnarray}
where for convenience, we introduce  $\Delta = (d_Y\lambda\Sigma(X))^{-1/2}  (Y-\mu(X))$. A natural choice of loss may be
\begin{eqnarray}
    \ell^2_\lambda(X,Y) = \min\left\{  B, \; \| \Delta \|_2 - \lambda  \right\} \,,
\end{eqnarray}
which applies a linear penalty to violation of~\eqref{eq:ineq} and a linear reward when the left hand side of~\eqref{eq:ineq} is smaller than $\lambda$, subject to truncation by hyperparameter $B$. In our experience, this approach produces similar behavior to the loss induced by using $s^{\text{multi}}$ in conformal prediction. Instead, we introduce two loss functions applied to $\Delta$:
\begin{eqnarray}
    \ell^1_\lambda(X,Y) &=&  \| \Delta \|_1 \,, \label{eq:1_loss} \\
    \ell^q_{\lambda,\,\beta}(X,Y) &=&  \text{quantile}_\beta\left( \Delta \right)\,. \label{eq:q_loss}
\end{eqnarray}
The above losses promotes sparsity in $\Delta$, compared to results obtained by using a loss based on $\|\Delta\|_2$. We aim to construct ellipsoidal sets with high, but not complete, component-wise coverage. 

\subsection{Conformal uncertainty propagation\label{ss:prop}}

We are interested in propagating our prediction sets to downstream quantities of interest $Q\in\mathbb{R}^{d_Q}$ where $d_Q\leq d_Y$. This task is relevant to multistage prediction pipelines. First, we are interested in uncertainty of the direct predictions of our surrogate for energies, atomic forces, and virial stresses. Then, we want to push that uncertainty to downstream quantities of interest such as elastic constants and vacancy formation energies, linear functions of surrogate predictions. Both the (hyper)rectangular sets obtained through univariate methods and the (hyper)ellipsoids of multivariate methods retain their shape under linear transformation. Furthermore, given a linear operator $L\in\mathbb{R}^{d_Q\times d_Y}$, the propagated set $L(\Gamma(X))$ can be found in closed form, so there is no need to push samples from the set through the transformation. In practice, we approximate the transformations applied to our original prediction sets. As discussed in Section~\ref{sec:experiments}, we will require our surrogate to make predictions for multiple related atomic configurations, which we form and geometrically relax. Multivariate conformal uncertainty prediction and propagation will rely on the covariance $\Sigma$ of these configurations. 

For uncertainty sets to be meaningful in propagated space, we need to understand how conformal guarantees translate. Let $P_{L\Sigma^{1/2}}=\Sigma^{1/2}L^\top\left( L \Sigma L^\top \right)^{-1} L\Sigma^{1/2}$, the projection onto the row space of $L\Sigma^{1/2}$. Then,
\begin{eqnarray}
     \| \Sigma(X)^{-1/2}  (Y-\mu(X)) \|_2   &\geq& \| P_{L\Sigma^{1/2}} \Sigma(X)^{-1/2}  (Y-\mu(X)) \|_2  \,, \notag\\
    &=& \| (L \Sigma(X) L^\top )^{-1/2}  (LY-L\mu(X)) \|_2 \,. \notag
\end{eqnarray}
Equality is achieved when $d_Y=d_Q$. We conclude that in multivariate conformal with tolerance $\alpha^{\text{multi}}$ we obtain a coverage guarantee
\begin{eqnarray}
    \mathbb{P}\left( LY \in L(\Gamma(X))  \right) \geq 1 - \alpha^{\text{multi}} \,,
\end{eqnarray}
but the upper bound which guarantees sharpness only translates when $L$ is invertible.  Similarly, in conformal risk control with upper bound $\alpha^{\text{crc}}$
\begin{eqnarray}
    \mathbb{E}\left[ \ell\left( L(\Gamma(X)) ,\, Y \right) \right] \leq \alpha^{\text{crc}} \,,
\end{eqnarray}
and the lower bound is achieved for invertible $L$. Thus, we expect propagated sets to contain the true value at a controlled rate, but the conservatism of these sets depends on the particular transformation $L$. In multivariate conformal literature, the alternative score
\begin{eqnarray}
    \label{eq:L_score}
    s^{\text{multi-L}} =  \| (L \Sigma(X) L^\top )^{-1/2}  (LY-L\mu(X)) \|_2\,,
\end{eqnarray}
has been proposed~\cite{braun2026} to preemptively correct for propagation and achieve both validity and sharpness guarantees. Notably, this score depends on $L$, i.e.~the respective downstream quantity considered. It thus implies a separate recomputation of the conformalization procedure
for each considered output quantity, which can be costly. We will discuss the tradeoffs of this score in Sections~\ref{sec:experiments} and~\ref{sec:discussion}.

\section{Experiments\label{sec:experiments}}

\subsection{Synthetic experiments\label{ss:synthetic}}

We first consider a synthetic case study which illustrates the behavior of conformal sets under propagation. To imitate the workflow of our material science set up, we train a Gaussian process model on data pairs $\{(X^{(i)}),\,Y^{(i)})\}_{i=1}^{n=1000}$. The covariates $X\in\mathbb{R}^{100\times100}$ are drawn from a centered, unit variance Gaussian distribution. Rows are independent, but between each column, there is correlation $\rho$. $X^{(i)}$ is independent of $X^{(j)}$ whenever $i\neq j$. Each column of $X^{(i)}$ is mapped to a component of $Y^{(i)}\in\mathbb{R}^{100}$ by a shallow neural network with error function activation and observation noise variance $10^{-4}$. An additional $1000$ data points are generated by the same process to be used for calibration. We compare two conformal procedures in this example: univariate conformal with score~\eqref{eq:uni_score} applied to each component of each $Y^{(i)}$ and multivariate conformal with score~\eqref{eq:multi_score} applied to each full vector $Y^{(i)}$. For both methods, we let the tolerance $\alpha$ range from $0.05$ to $0.6$. To understand the impact of dimension compression during propagation, we consider linear transformations into spaces of dimension $d_Q\in\{5,10,15,\dots,100\}$. Note that in the last case, the dimension of $Y$ is preserved under transformation. We report results averaged over $10$ random propagation operators for each $d_Q$.
\begin{figure}
    \centering
    \begin{tabular}{cc}
        \includegraphics[width=0.4\linewidth]{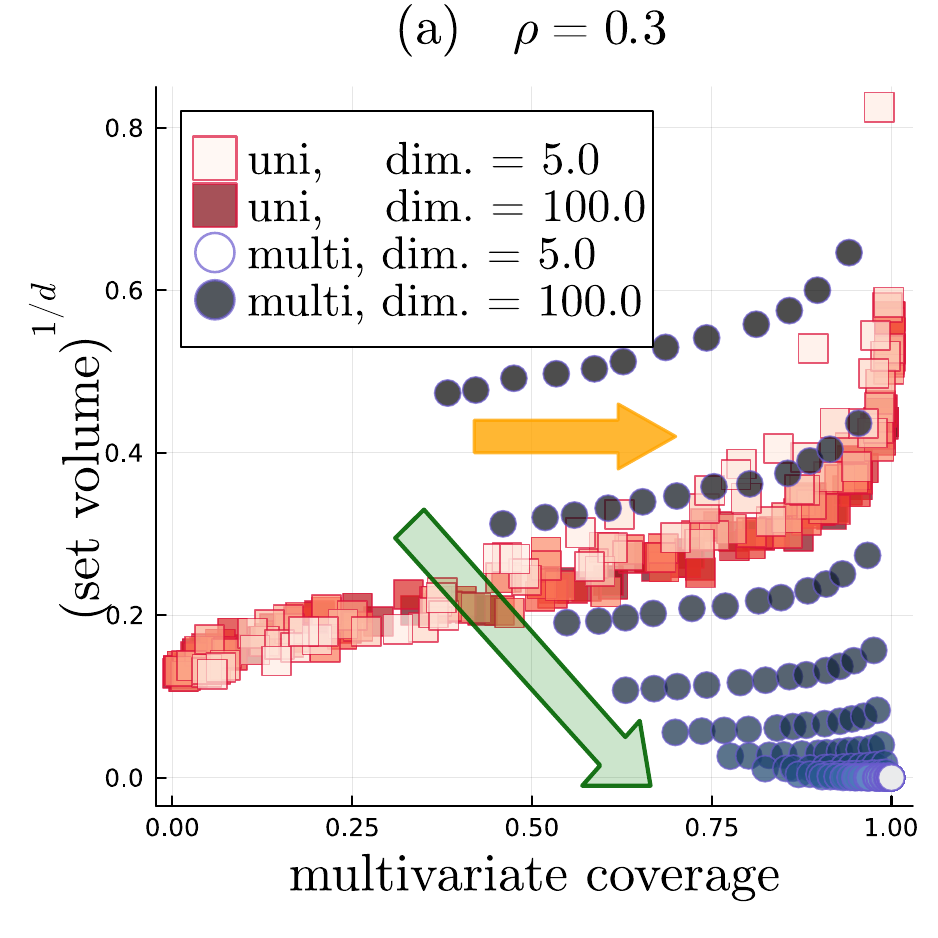} &
        \includegraphics[width=0.4\linewidth]{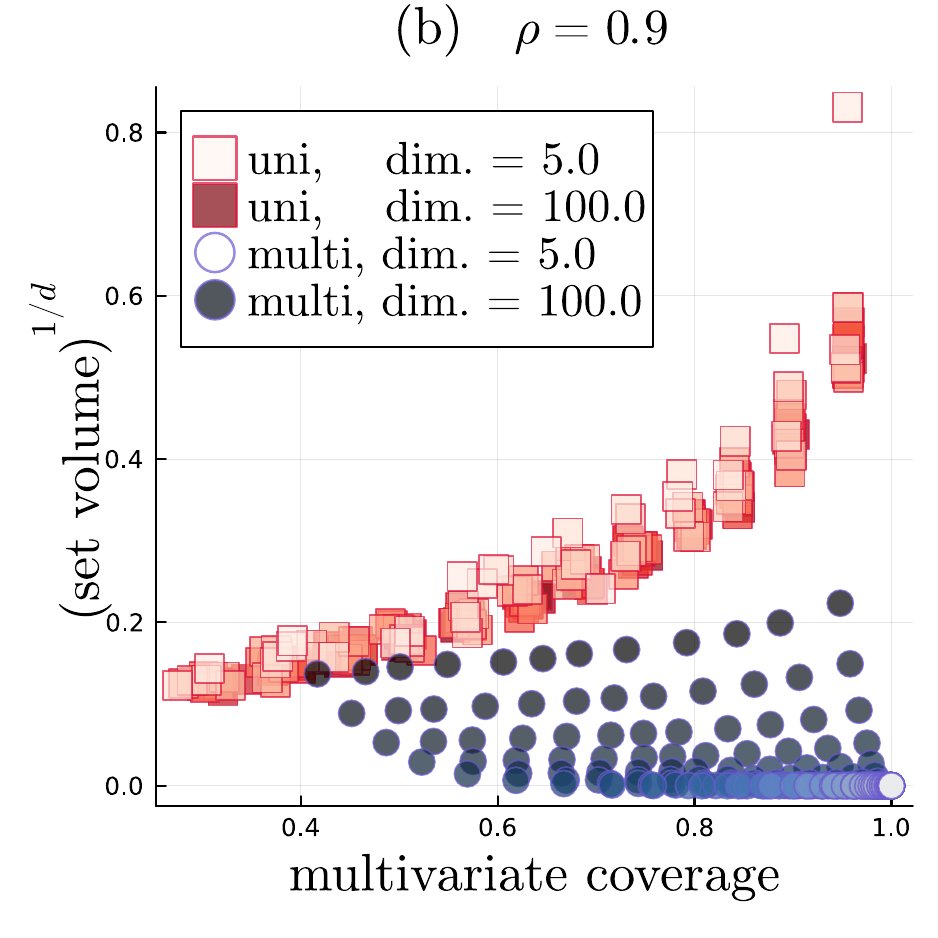} 
    \end{tabular}
\caption{\label{fig:synth_cov_vol} \textbf{Synthetic propagation study}: Comparison of full vector coverage against volume for prediction sets propagated into a dimension $d_Q$ space. We consider $d_Q=\{5,10,15,\dots,100\}$, and darker shades indicate larger $d_Q$. The dimension of the space were conformal calibration is performed (pre-propagation) is $100$. We report the average results from $10$ randomly drawn linear propagation operators. Red squares (r. blue circles) scatter points represent sets constructed via univariate (r. multivariate) conformal prediction. (a): input features $X$ of a given test point have correlation $\rho=0.3$. The solid orange arrow points toward increasing $1-\alpha$ while the transparent green arrow points toward decreasing $d_Q$. (b): input features have correlation $\rho=0.9$.}
\end{figure}

Figure~\ref{fig:synth_cov_vol} relates the full vector coverage of the propagated sets to their volume (standardized by taking the $1/d_Q^{th}$ root). Subfigure (a) shows the case when correlation $\rho$ between the covariates of each component of $Y^{(i)}$ is relatively weak while (b) shows stronger correlation $\rho=0.9$. Red squares represent univariate conformal and blue circles multivariate conformal. The shade of the markers darkens as the propagation dimension $d_Q$ increases. We can see that for univariate conformal, set volume is an increasing function of set coverage. Changing $\rho$ or the propagation dimension leads to only minor variation. In contrast, both $\rho$ and the propagation dimension have a clear impact on the behavior of multivariate conformal sets. For a given tolerance and propagation dimension, larger correlation $\rho$ produces smaller sets. We expect this result as the multivariate method can leverage correlation in $Y$. For $\rho=0.9$, nearly all of the multivariate sets are smaller than the univariate sets even though, as shown in more detail in Appendix~\ref{app:synthetic}, many of the multivariate sets produce greater coverage.  Perhaps more surprising is the observation that multivariate set volume becomes smaller as propagation dimension decreases (along transparent green arrow). While the square markers corresponding to univariate conformalized sets show no significant separation according to propagation dimension, the circular markers of the multivariate approach form clear rows for each $d_Q$, particularly evident for larger dimensions. Along each row (in the direction of the solid orange arrow), $1-\alpha$ increases, producing greater coverage. We can understand the decrease in set size with $d_Q$ by noting that in smaller dimensions, it is more likely that all components of $LY$ will be close together than in larger dimensions. However, we also noted in Subsection~\ref{ss:prop} that propagated conformal sets do not have a sharpness guarantee when the propagation reduces dimension, and we confirm in Appendix~\ref{app:synthetic} that sets for small $d_Q$ tend to overcover. We also confirm that correcting for propagation using the score~\eqref{eq:L_score} prevents overcoverage. Figure~\ref{fig:synth_cov_vol} demonstrates that while propagated multivariate conformal sets may be more conservative than necessary, this effect does not produce drastic increases in volume. 

\subsection{Dataset and representation for chemical experiments}

In our numerical experiments, we first consider a synthetic case study, then present several examples based on a silicon dataset which has previously been used to build a general-purpose Gaussian approximation potential~\cite{Bartok2018silicon}. Specifically, we use diamond-structured configurations in periodic unit cells. There are $104$ configurations containing $2$ atoms, $220$ configurations with $16$ atoms, $110$ configurations with $54$ atoms, and $55$ configurations with $128$ atoms. We also consider diamond vacancy configurations: $100$ configurations with $63$ atoms and $111$ configurations with 215 atoms. Typical modern MLIPs would be fit to thousands or, in the case of foundation models, even millions of configurations. Here, we are interested in the relatively low data setting. Note that the vacancy configurations do not correspond to the bulk configurations in the dataset--that is, they are not formed by removing an atom from one of the other configurations in the set. When constructing feature representations of the configurations, it is crucial to account for the periodic boundary conditions. We compute SOAP~\cite{Bartok2013,Bartok2015,Musil2021} features using the QUIP~\cite{Csanyi2007-py} software package and quippy~\cite{Kermode2020-wu} interface. Details on all hyperparameters can be found in Appendix~\ref{app:hyper}.

For each configuration, the silicon dataset includes density functional theory reference values for total energy, atomic forces, and virial stresses computed with the CASTEP code as described in ~\cite{Bartok2018silicon}. We use these values to train Gaussian process models, implemented in Julia, and available at \href{https://github.com/kefisher98/multivariate-conformal/}{https://github.com/kefisher98/multivariate-conformal/}. The properties included in the training set are determined on a case by case basis. Empirical evidence~\cite{Bartok2010,willow2026,Schafer2026} has shown that including gradient properties, such as forces and stresses, in machine learning models is beneficial for predicting gradients in test cases. Theoretical work~\cite{fisher2025sobolev} has suggested that gradient training data may only aid in the prediction of gradients and can detract from the prediction of function values, i.e.~energy. Furthermore, including gradients significantly increases the dimension of our quantity of interest as there are three force dimensions per atom in the system. Consequently, we train with forces only when are goal is to predict forces and train with stresses only when our goal is to predict stresses.

\subsection{Calibration of energy and atomic forces\label{ss:energy_force}}

We now consider calibrated uncertainties for high dimensional materials science quantities of interest and motivate the use of conformal risk control. Our prediction target is the global energy and all atomic forces for a given atomistic system. The training set consists of $200$ configurations randomly sampled from the silicon diamond configuration dataset. Training vectors therefore range in dimension from $7$ to $385$. For efficiency, we use the subset of regressors~\cite{smola2000} approximation to build the training covariance by selecting $30$ inducing configurations with the $k$ medoids algorithm~\cite{Kaufmann1987}. We find that minor variations in the size of the inducing set and training set produce little change in prediction performance. A further $120$ configurations are set aside as the calibration set, and $80$ configurations are used for testing. The latter two sets are randomly selected from the systems with $16$, $54$, and $162$ atoms, so the prediction vectors have dimension between $49$ and $385$.    

\begin{figure}
    \centering
    \begin{tabular}{cc}
        \includegraphics[width=0.4\linewidth]{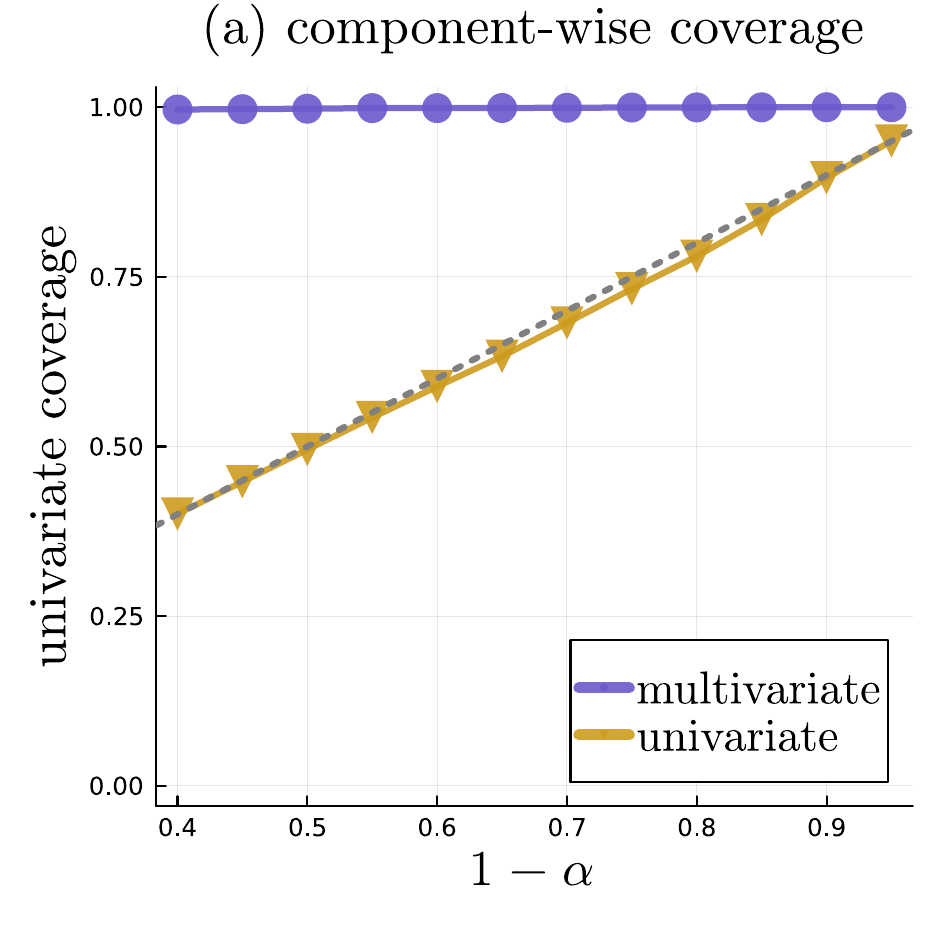} &
        \includegraphics[width=0.4\linewidth]{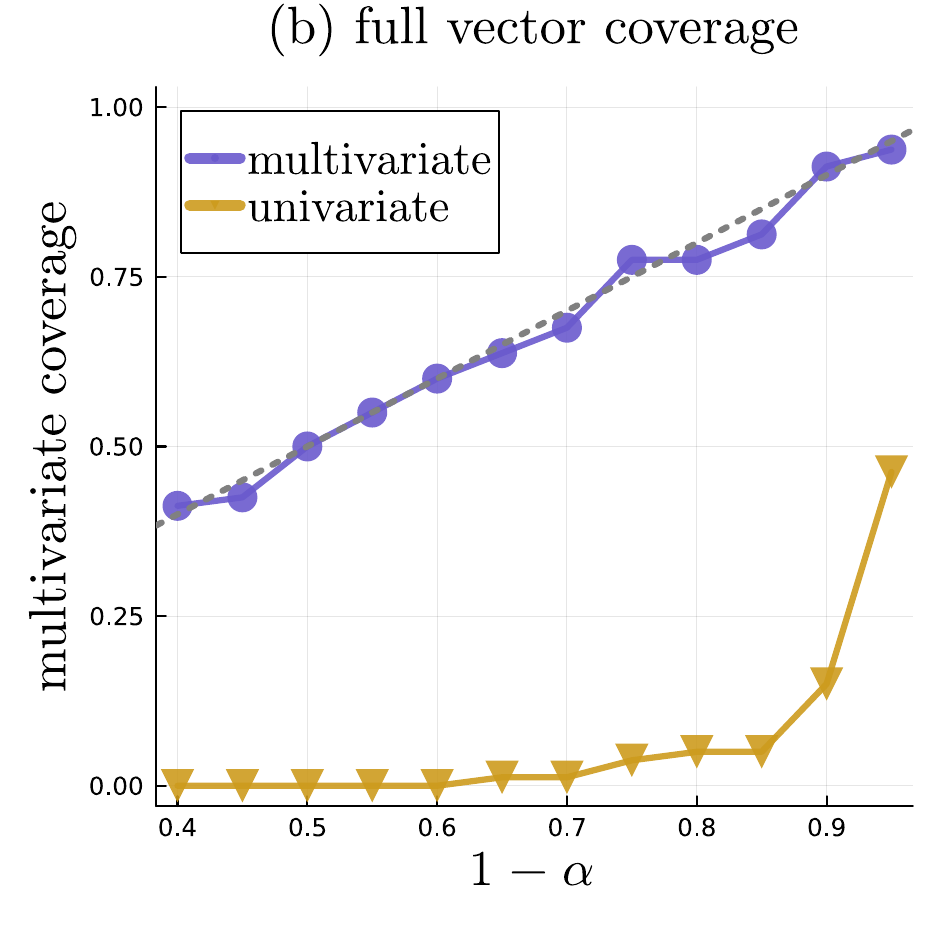} 
    \end{tabular}
\caption{\label{fig:energy_force_guarantee} \textbf{Energy and atomic force calibration}: Verification of the coverage and sharpness guarantees of conformal prediction for a given tolerance $\alpha$. Calibration is applied to a vector containing the global energy and atomic forces for a configuration. (a): fraction of vector components which fall in predictive interval obtained by univariate (gold triangles) and multivariate (purple circles) conformal prediction. (b): fraction of vectors which fall entirely within the multivariate predictive set. The $y=x$ line is dotted in gray. }
\end{figure}
\begin{figure}
    \centering
    \begin{tabular}{cc}
        \includegraphics[width=0.4\linewidth]{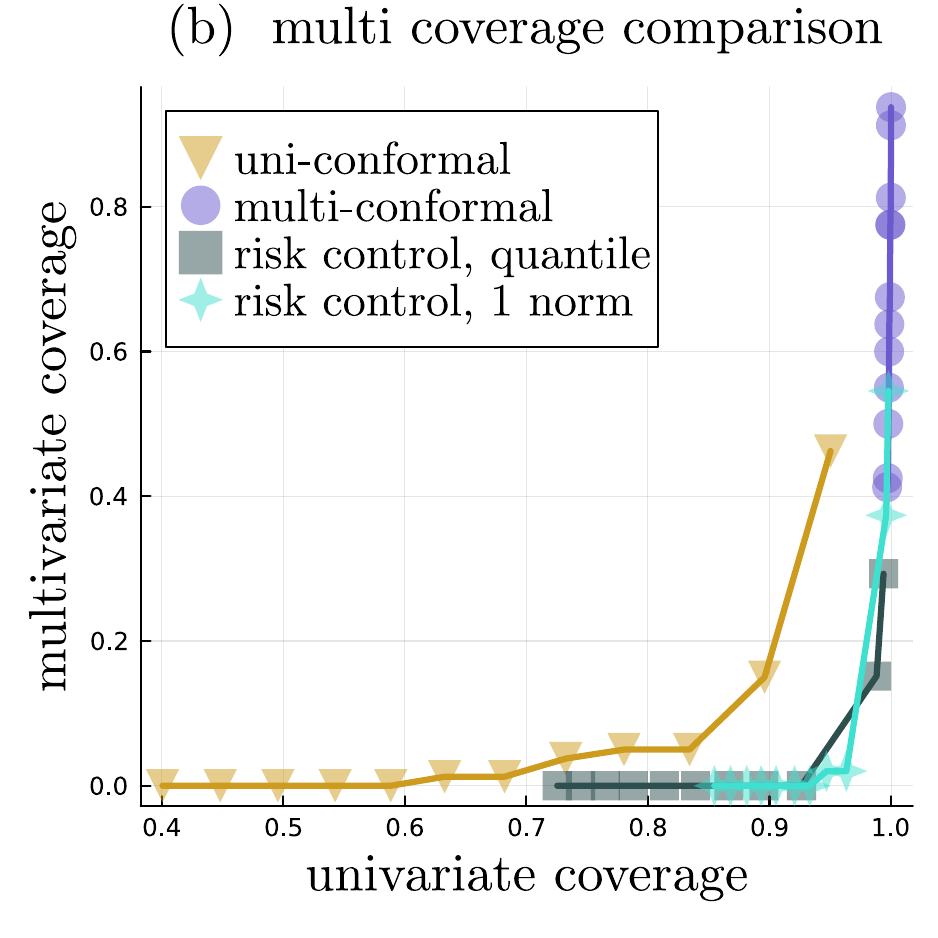} &
        \includegraphics[width=0.4\linewidth]{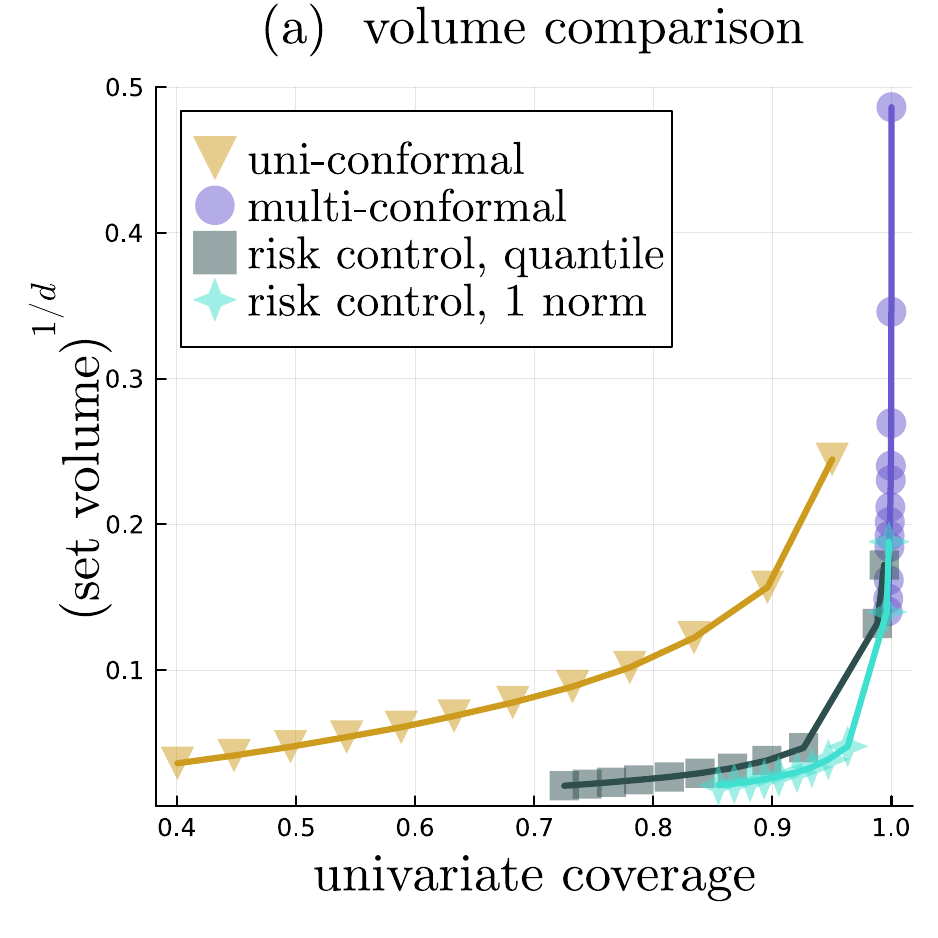}
    \end{tabular}
\caption{\label{fig:energy_force_cov_vol} \textbf{Energy and atomic force calibration}: Relationship between component-wise coverage, full vector coverage, and set volume for univariate (gold triangles) and multivariate (purple circles) conformal prediction as well as conformal risk control with $1$-norm (turquoise stars) and quantile (green squares) loss. Both calibration and test datasets contain atomistic systems of multiple sizes. Consequently, in the test set, vectors containing the global energy and all atomic forces for a system take on dimensions $d_Y\in\{49,163,385\}.$}
\end{figure}

Figure~\ref{fig:energy_force_guarantee} verifies the univariate and multivariate conformal guarantees. As expected, both univariate and multivariate conformal achieve their corresponding validity and sharpness guarantees: the gold triangles representing the performance of the univariate method lie on the $y=x$ line in plot (a), and the purple circles for the multivariate method trace this line in the plot (b). There is slightly more noise in the multivariate line in (b) compared to the univariate line in (a) because the multivariate method effectively has a smaller calibration set. The $m=120$ calibration vectors contribute one score each to the multivariate conformal algorithm. In univariate conformal, the energies and forces are treated separately though both are considered in the evaluation of coverage. Thus, there are also $m=120$ calibration points for the energy predictions, but the univariate conformal algorithm has $>10,000$ scores to calibrate the force predictions. Figure~\ref{fig:energy_force_guarantee} also provides insight into the component-wise performance of multivariate methods and the full vector performance of univariate methods. Achieving the required multivariate coverage produces ellipsoidal sets that have near perfect coverage of individual components (measured by projecting sets into lower dimensions). In contrast, univariate methods produce zero multivariate coverage for large enough tolerance $\alpha$. This behavior makes sense given that the coverage probability for a $d_Y$ dimensional vector is $1-d_Y\alpha$. Lowering $\alpha$ to enforce a high probability of univariate coverage (i.e.~applying a Bonferroni correction) eventually forces nonzero multivariate coverage, but there is not fine grained control over multivariate performance.

Conformal risk control with an appropriate loss function offers more flexibility when calibrating high dimensional data. We consider two options: the $1$-norm loss given by~\eqref{eq:1_loss} and the $0.9$-quantile loss given by~\eqref{eq:q_loss}. Appendix~\ref{app:energy_force} confirms that both losses achieve the risk control guarantees~\eqref{eq:control}. We set the truncation parameter $B$ to the $0.95$-quantile of the calibration losses and show other options in the appendix. Figure~\ref{fig:energy_force_cov_vol} shows how multivariate coverage and set volume relate to univariate coverage for the four options. We see that the risk control methods indeed have different incentives from the two conformal prediction approaches. Both risk control approaches can achieve high univariate coverage with low multivariate coverage. The threshold of $\alpha$ at which their multivariate coverage begins to increase is larger than for univariate conformal methods. As a consequence, the sets produced by risk control are consistently smaller than univariate conformal hyper-rectangles and generally smaller than the multivariate hyper-ellipsoids.  

\subsection{Uncertainty propagation to elastic constant prediction\label{ss:elasticconstant}}

The elastic constant tensor $C$ characterizes the behavior of a given material under deformation. Both energy and virial stresses can be mapped to the elastic constant via linear differential operators
\begin{eqnarray}
    \label{eq:elastic_constant}
     C  = \frac{\partial \varsigma}{\partial \eta} \bigg|_{\eta *}\,, \qquad 
            \varsigma(\eta) = \frac{1}{V(\eta)}\frac{\partial E}{\partial \eta}\,,
\end{eqnarray}
where $\eta$ is an applied strain and $V(\eta)$ is the deformed volume. Consequently, using our GAP surrogate and finite difference schemes, we have two routes to estimating elastic constant components: (1) through a second difference approximation applied to energy predictions and (2) via a first difference approximation applied to stress predictions. We consider both approaches in this subsection, applying conformal prediction to the GAP predictions (either energies or stresses) and propagating uncertainty sets into elastic constant space.

For cubic crystals like silicon, there is additional structure in the elastic constant that we use to simplify our computations. Specifically, there are several symmetries and zero components:
\begin{eqnarray}
    \begin{bmatrix}
        \varsigma_1 \\ \varsigma_2 \\ \varsigma_3  \\ \varsigma_4 \\ \varsigma_5 \\ \varsigma_6 
    \end{bmatrix} = \begin{bmatrix}
        c_{11} & c_{12} & c_{12} & 0      & 0      & 0      \\
        c_{12} & c_{11} & c_{12} & 0      & 0      & 0      \\
        c_{12} & c_{12} & c_{11} & 0      & 0      & 0      \\
        0      & 0      & 0      & c_{44} & 0      & 0      \\
        0      & 0      & 0      & 0      & c_{44} & 0      \\
        0      & 0      & 0      & 0      & 0      & c_{44} \\
    \end{bmatrix} \begin{bmatrix}
        \eta_1 \\ \eta_2 \\ \eta_3  \\ \eta_4 \\ \eta_5 \\ \eta_6 
    \end{bmatrix}\,.
\end{eqnarray}
The stress and strain vectors are presented in Voigt notation. Supposing that we have knowledge of this structure, the value of each unique component can be determined by simultaneously perturbing $\eta_1$ and $\eta_4$. (The derivative found in this scenario will be $[c_{11} \; c_{12} \; c_{12}  \; c_{44} \; 0 \; 0 ]^\top$). Thus, we can solve for all three elastic constants based on the difference in stress of two configurations, $a$ and $b$.

If we use route (1) based on energies, we will estimate the six unique stress components for $a$ and $b$ using a central difference scheme. Each central difference scheme is implemented by perturbing either $a$ or $b$ to create two further configurations. Using GAP, the geometry of these configurations is relaxed and the energy is predicted. The relaxation is necessary because of the multilattice structure of the silicon crystals: Cauchy-Born non-affine shifts between the two sublattices cannot be predicted solely from applied strain, so atomic positions must be relaxed with respect to a fixed cell. We require predictions for energy at a total of $24$ configurations, perturbed from an equilibrium configuration, to compute the elastic constant. Thus, we will apply conformal calibration to a $24$ dimensional vector, then project the prediction sets to six dimensional elastic constant space. We further project to $3$-dimensional space by selecting the unique, nonzero values $c_{11}$, $c_{12}$, and $c_{44}$.

If, instead, we use route (2), we will first relax the geometries of configurations $a$ and $b$, then predict their stresses with the GAP model. This step produces a twelve-dimensional quantity of interest which we project to the $3$-dimensional elastic constant space.

\begin{figure}
    \centering
    \begin{tabular}{ccc}
        \includegraphics[trim={0.55cm 0 0.55cm 0},clip,width=0.31\linewidth]{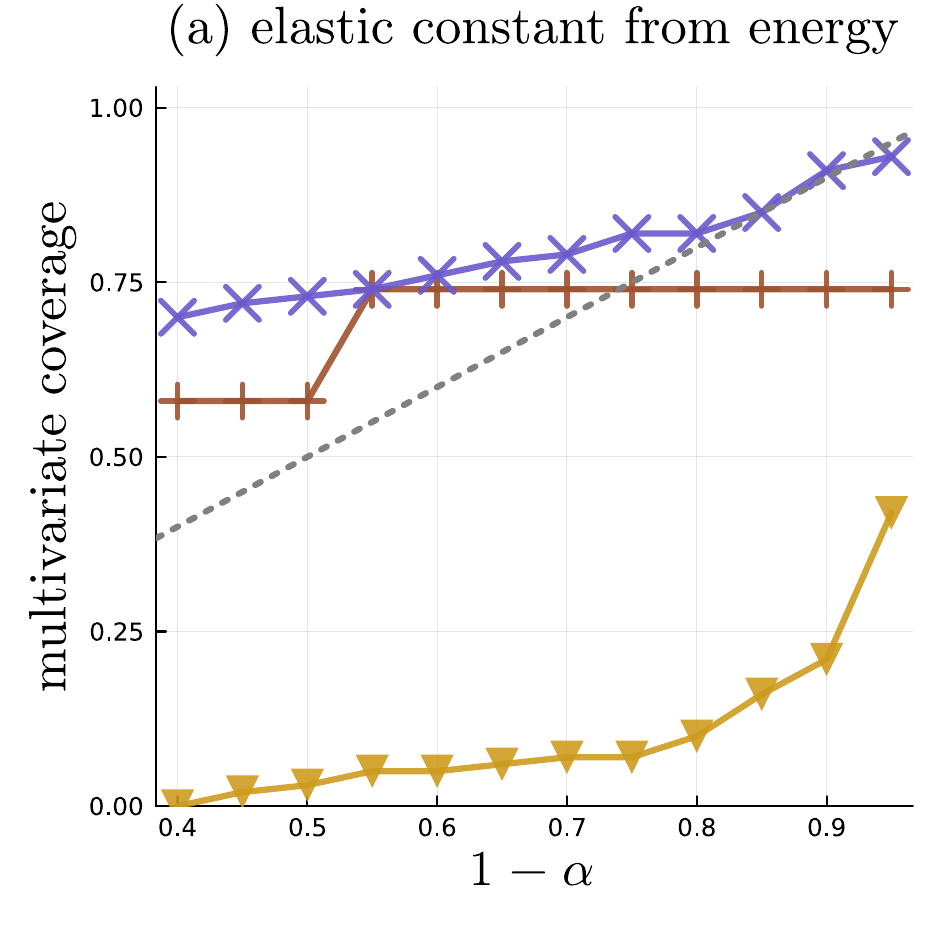} &
        \includegraphics[trim={0.55cm 0 0.55cm 0},width=0.31\linewidth]{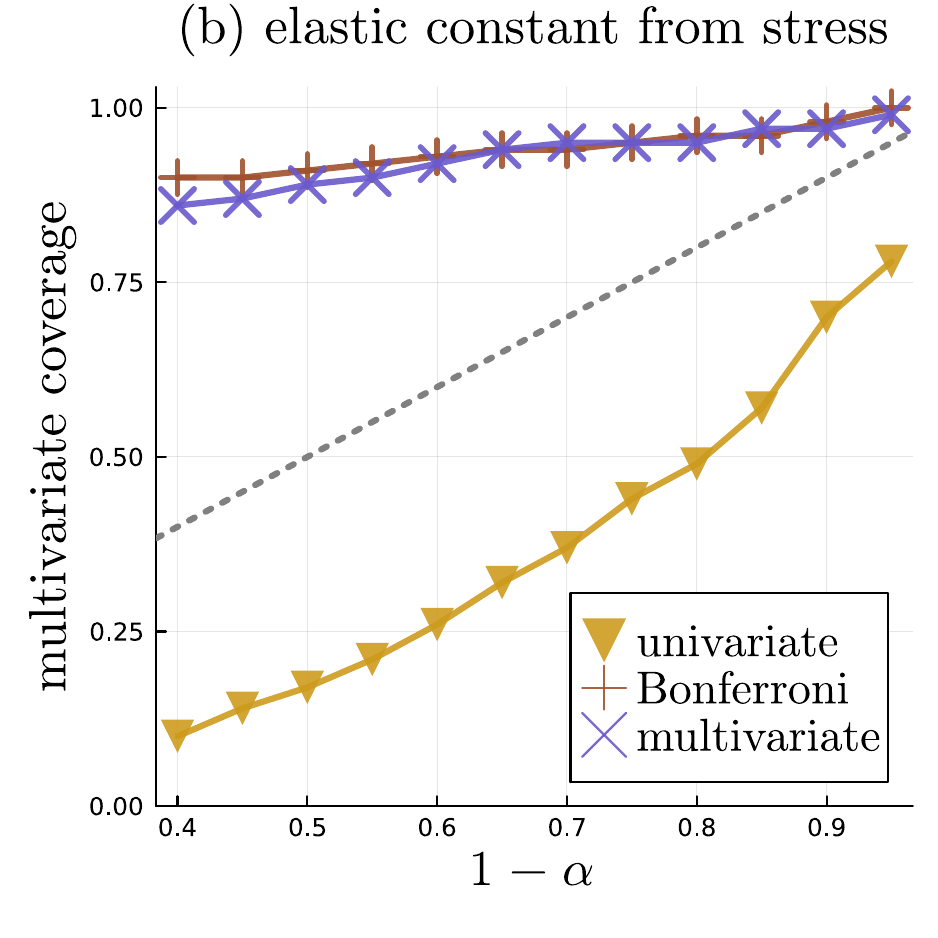} &
        \includegraphics[trim={0.55cm 0 0.55cm 0},width=0.31\linewidth]{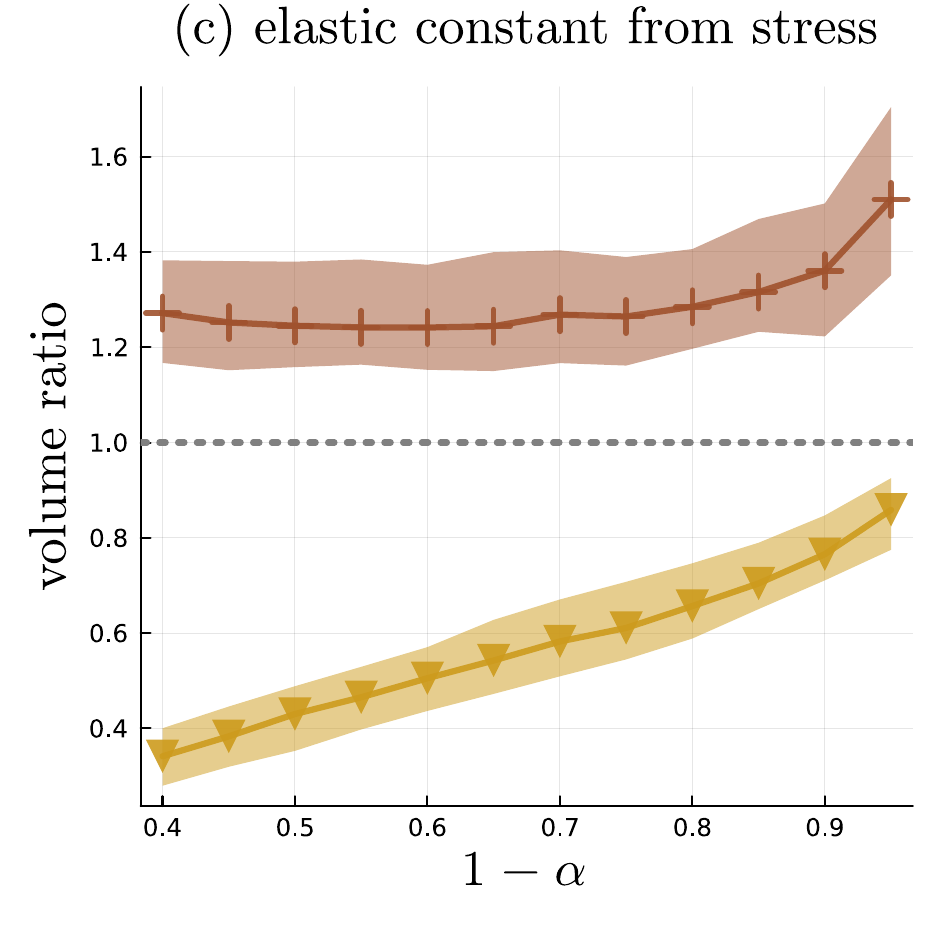} \\[6pt]
    \end{tabular}
\caption{\label{fig:elastic_constant} \textbf{Propagation to elastic constant uncertainty}: Comparison of prediction sets in the space of the three unique nonzero components of the elastic constant of silicon. $\alpha$ is the specified miscoverage tolerance. The dotted gray line marks $x=y$. (a): full vector coverage for component-wise conformal (gold triangles), Bonferroni-corrected conformal (red +s), and multivariate conformal (purple xs). Calibration was performed in the space of energies. (b): Same quantities as (a) resulting from calibration performed in the space of stresses. (c): Ratio between the cube root volume of prediction sets. Red +s (gold triangles) mark the ratio between Bonferroni-corrected (component-wise) sets and multivariate sets. Ribbons shade the region between the $0.25$ and $0.75$ quantiles. }
\end{figure}

Figure~\ref{fig:elastic_constant} summarizes the performance of propagated prediction sets for elastic constant values, obtained using vanilla univariate conformal prediction, Bonferroni-corrected conformal, and multivariate conformal. Predictions are made by perturbing the equilibrium configuration of a $16$ atom silicon diamond configuration. Results are estimated based on $100$ random draws of both the training and calibration data sets. GAP models which are used to predict energy are trained on $350$ diamond configurations. The calibration set is based on an additional $100$ configurations. To implement multivariate conformal, $200$ random bags of $24$ configurations are drawn with replacement from the calibration set. This approach is more computationally efficient than perturbing each calibration configuration $24$ times (as is done for test points) and appears to work well. Further discussion can be found in Appendix~\ref{app:ec}. For GAP models which predict stress, we can include both energy and stress data in our training set. To compensate for additional computational cost, we use a training set of a total of $300$ atomistic systems where energy data is included for each system but stress data is included for only $15$ systems. We find that this training set composition is sufficient to achieve good performance provided the stress training data is sampled from systems sufficiently close to equilibrium. To calibrate the stress configurations, $350$ random bags of size $2$ where drawn with replacement from the calibration set, and the stress predictions of the two configurations were concatenated. Conformal calibration was applied with tolerances ranging from $0.4$ to $0.95$. 
 
Plots (a) and (b) within Figure~\ref{fig:elastic_constant} show empirical coverage found for a given lower bound $1-\alpha$. Since we are evaluating the behavior of multivariate propagated sets, we do not expect any method to be exactly on the $y=x$ line (simultaneously satisfying both validity and sharpness constraints). Rather, we expect the multivariate and Bonferroni-corrections to be above $y=x$, meeting their validity guarantee. For the univariate approach, neither guarantee is expected to hold though we expect the stress example to show more coverage at each $1-\alpha$ than the energy example because its finite difference propagation operator $L$ has two nonzeros per row rather than four. The empirical results confirm these expectations, and we see that the full vector coverage of the univariate approach is considerable lower than the other two methods. For the stress example, we see that both the multivariate and Bonferroni-corrected methods meet the expected guarantee.   The energy example shows somewhat different behavior: the multivariate method meets the expected guarantee, but the Bonferroni-corrected coverage is only above the $y=x$ line for smaller $1-\alpha$. The plateau in the Bonferroni corrected curve occurs because for $\alpha\geq0.55$, the Bonferroni-correction in the energy case results in zero tolerance for miscoverage. There is no mechanism for improving coverage by decreasing $\alpha$. 

The underlying reason for the miscoverage in the Bonferroni curve is likely partially due to the limited size of our dataset and partially due to the evaluation procedure. Coverage plots such as (a) and (b) within Figure~\ref{fig:elastic_constant} confirm coverage by determining the fraction of test configurations for which the \emph{DFT reference} for elastic constant lies within the 3-dimensional prediction set. While Equation~\eqref{eq:elastic_constant} shows that the elastic constant is obtained from linear differentiation operators applied to the true energy or true virial stresses, one set of approximations is required to obtain the \emph{DFT prediction}, and another set of approximations is made by our \emph{GP based} workflow. Furthermore, approximate prediction requires geometric relaxation of perturbed configurations: a nonlinear minimization problem which determines the energy values to be conformalized. Separate relaxations are performed for the \emph{DFT reference} and the \emph{GP workflow}, each respectively employing force predictions from the corresponding method. Thus, we do not expect conformal guarantees to hold exactly when we evaluate coverage of the DFT references. For more context, we include cases in Appendix~\ref{app:ec} where both the Bonferroni and the multivariate approaches undercover while still producing better coverage than the vanilla univariate method.

We can also compare the volume of the propagated prediction sets. Plot (c) of Figure~\ref{fig:elastic_constant} plots ratios of the volume (corrected for dimension by taking the cube root) for pairs of methods resulting from case ii) prediction via stresses. Specifically, the red $+s$ indicate the ratio of Bonferroni-corrected set volume to multivariate set volume, and the gold triangles represent the ratio of univariate set volume to multivariate volumes. The dashed line at unity indicates equal volume. We can see that univariate conformal produces the smallest prediction sets. This finding is not surprising, considering that this approach systematically fails to achieve the desired coverage guarantees in contrast to the other methods. Interestingly, in Appendix~\ref{app:ec}, we show cases of prediction of the vector $[c_{11} \; c_{12} \; c_{12}  \; c_{44} \; 0 \; 0 ]^\top$ where the multivariate approach produces better coverage than the univariate approach as well as smaller sets. The latter result occurs because the multivariate method can leverage the symmetries of repeated elements. In our present setting, the comparison of the methods with similar coverage, the Bonferroni-corrected and multivariate approaches, shows that multivariate sets are consistently smaller than Bonferroni sets. Thus, leveraging covariance predictions to build ellipsoidal sets produces a size benefit compared at a given coverage level.

\subsection{Uncertainty propagation to vacancy formation energy prediction\label{ss:vacancy}}

For our final example, we estimate the energy required to remove an atom from a bulk crystal, creating a vacancy in the configuration. Recalling that $M^{(i)}$ is the number of atoms in the $(i)^{th}$ configuration, the vacancy formation energy is the difference
\begin{eqnarray}
    E_{\text{formation}}^{(i)} = E_\text{vacancy}^{(i)} - \left( \frac{M^{(i)}-1}{M^{(i)}}\right)E_\text{bulk}^{(i)}\,.
\end{eqnarray}
The energies on the right hand side are the global energies before ($E_\text{bulk}^{(i)}$) and after ($E_\text{vacancy}^{(i)}$) the removal of the atom. We will train a multitask GAP model where the primary task is learning the energy of bulk diamond configurations, and the secondary task is the energy of the diamond vacancy systems. Note that our silicon dataset contains examples for both tasks, but for a given $(i)$, we have either $E_\text{bulk}^{(i)}$ or $E_\text{vacancy}^{(i)}$, not both.

The multitask framework can be successfully implemented in this setting, but we would expect a performance improvement if we filled in the gaps in the training set~\cite{Fisher2024}. For a test structure $(j)$, we relax the vacancy geometry $R_\text{vacancy}^{(j)}$ using the GAP model, then make predictions for the mean and covariance of $[E_\text{bulk}^{(j)}\; E_\text{vacancy}^{(j)} ]^\top$. We apply conformal calibration to the two-dimensional energy predictions, then propagate the resulting sets to obtain interval predictions for the system's  vacancy formation energy.    
\begin{figure}
    \centering
    \begin{tabular}{cc}
        \includegraphics[width=0.4\linewidth]{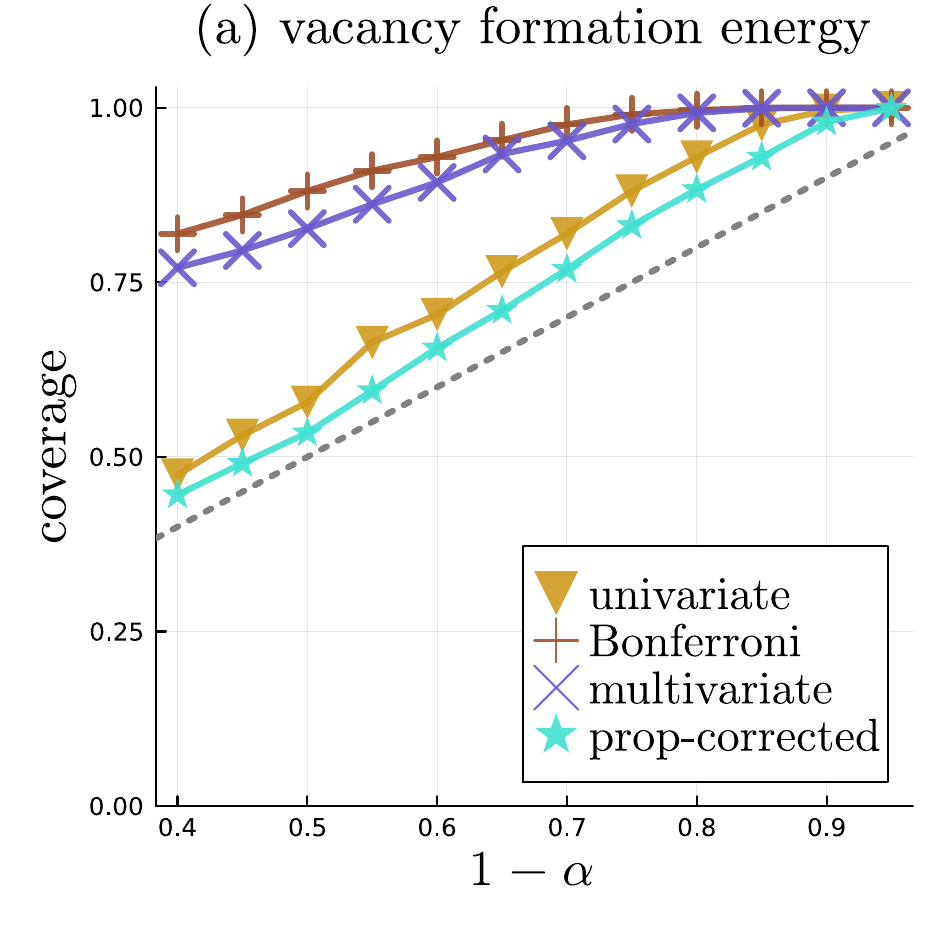} &
        \includegraphics[width=0.4\linewidth]{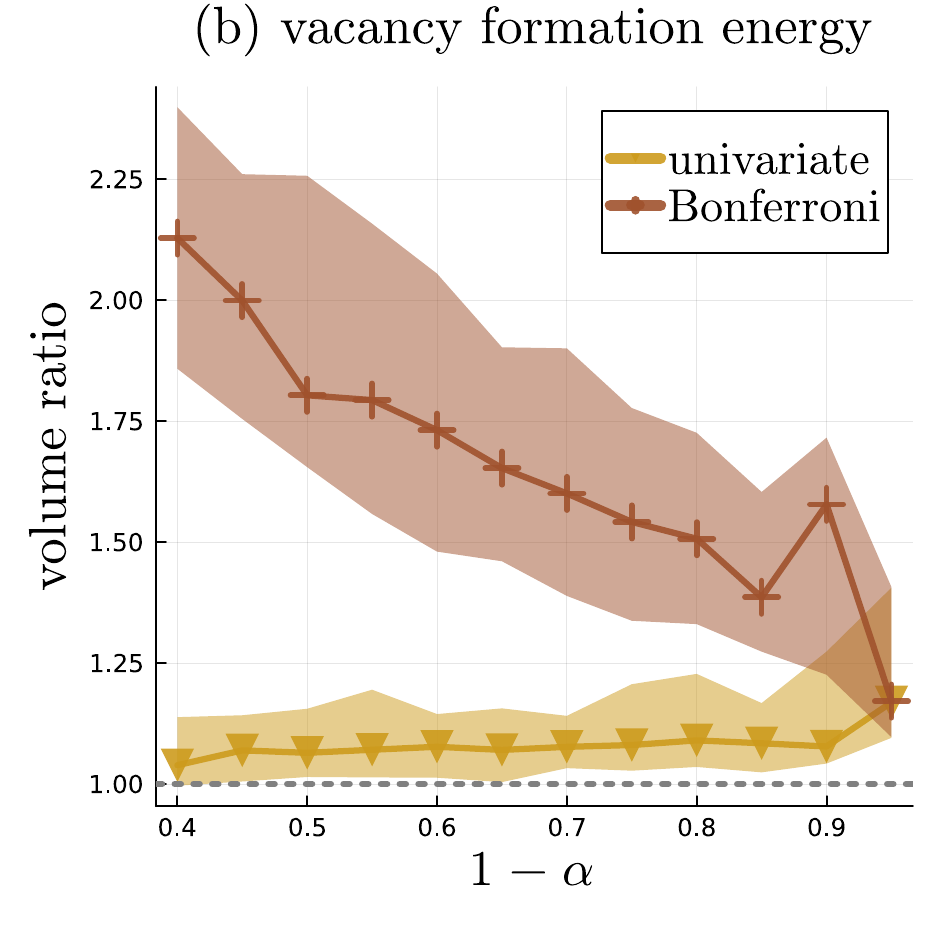} 
    \end{tabular}
\caption{\label{fig:vacancy_formation} \textbf{Propagation to vacancy formation energy uncertainty}: Coverage and relative volume of prediction sets propagated to univariate intervals over vacancy formation energy. (a): coverage of propagated conformal sets produced by univariate (gold triangles), Bonferroni-corrected univariate (red +s), multivariate (purple xs), and propagation-corrected multivariate (turquoise stars) approaches at tolerance $\alpha$. The $y=x$ line is dotted in gray. (b): Gold triangles mark the ratio between univariate and projection corrected multivariate prediction interval length. Red $+s$ compare Bonferroni-corrected univariate to the  projection corrected multivariate method. A dotted horizontal line marks a ratio of $1$. Ribbons shade the region between the $0.25$ and $0.75$ quantiles.  }
\end{figure}

In Figure~\ref{fig:vacancy_formation}, we compare the performance of four approaches to calibration: vanilla univariate conformal, Bonferroni-corrected univariate conformal, multivariate conformal, and propagation-corrected multivariate conformal. The latter uses the score~\eqref{eq:L_score}. We consider this example to be a reasonable use case for the propagation-corrected score because the transformation from global energies to vacancy formation energy only depends on system size---unlike our previous example, where the transformation to compute the elastic constant depends on the volume of perturbed configurations. Results are averaged over $25$ training and calibration sets randomly drawn for each of $55$ near-equilibrium test configurations with either $54$ or $128$ atoms. For training, we use the energies of $350$ diamond bulk configurations and $150$ diamond vacancy configurations. For calibration, we use $100$ additional diamond configurations and $40$ additional vacancy configurations, constructing $200$ random pairings between the two. We note that we have relatively few vacancy configurations in the silicon dataset, and $\approx 150$ training configurations appear necessary to achieve secondary task accuracy comparable to primary task accuracy. More details on the experiment, and results for a smaller primary training set are reported in Appendix~\ref{app:vfe}.

Plot (a) of Figure~\ref{fig:vacancy_formation} demonstrates that all methods achieve coverage greater than $1-\alpha$, even though conformal calibration does not theoretically guarantee this bound. There are two clear groupings based on coverage performance: the multivariate approach is similar to the Bonferroni-correction, and the propagation-corrected approach is similar to the vanilla univariate method. Thus, the approaches which provide similar full vector coverage at the calibration stage also provide similar coverage after propagation. Within each pair of methods, modeling seems to produce somewhat more conservative sets. For $\alpha=0.05$, all methods provide complete coverage of test vacancy formation energies. In plot (b), we compare the width of the prediction interval produced by the propagation-corrected multivariate approach to both the vanilla and Bonferroni-corrected multivariate methods. The propagation-corrected intervals are consistently smaller than their univariate counterparts, but only by a small factor. Unsurprisingly, the Bonferroni-corrected intervals are as much as $2.3$ times as large as the propagation-corrected intervals, and the difference narrows as the coverage of the two methods converges with decreasing $\alpha$. We see an outlier in the Bonferroni-corrected volume at $1-\alpha=0.9$, around where the method achieves complete coverage. Notably, when $1-\alpha=0.95$ and all methods reach complete coverage, the widths of the sets produced by the different methods do not converge. The propagation-corrected multivariate approach produces the sharpest set.

\section{Discussion\label{sec:discussion}}

We use this section to expose the approximations, assumptions, and sources of error present in our workflow. Based on these observations, we discuss the relevant considerations for choosing a conformal approach for practical, multistage problems. 

\subsection{Limited Data}
Several challenges arise from data limitations. For one, we implement our materials science examples based on fewer than one thousand configurations, not all of which are relevant to each problem. These configurations must be divided between subproblems: hyperparameter optimization, surrogate training, calibration, and evaluation. While conformal methods provide finite sample guarantees, the sharpness of the predictive sets depends on the size of the calibration dataset. 

\subsection{Exchangeability}

Perhaps a greater challenge than dataset size: for the methods we investigate, the conformal guarantees rely on the assumption that the calibration dataset is statistically \emph{exchangeable} with our test cases. This assumption is difficult to directly verify for atomistic systems. The generating probability distribution is unknown, and the practical procedure for sampling new configurations often involves simulations that make a series of approximations~\cite{Bartok2018silicon}. 

Rather, we can argue that the success of conformal calibration for sets of energies and forces reported in Figure~\ref{fig:energy_force_guarantee} provides empirical evidence for the exchangeability of the diamond configurations in our silicon dataset. However, exchangeability of this dataset is not sufficient for our uncertainty \emph{propagation} tests, where the final quantity of interest is based on predictions for \emph{perturbations} of the configurations in our original dataset. For instance, one approach to computing the elastic constant requires us to apply $24$ perturbations to one configuration, relax the resulting geometries using our GAP surrogate, and predict the mean $\mu$ and covariance $\Sigma$ for the energies of these configurations. To calibrate, we use $24$ configurations chosen randomly from the original silicon set; we do not perturb one configuration $24$ times because data generation would be computationally expensive, and our goal is to investigate conformal methods for an easy to implement workflow. Consequently, we expect \emph{more} correlation in our test cases than our calibration set. As synthetic experiments reported in Appendix~\ref{app:synthetic} demonstrate, this setting leads multivariate conformal approaches to \emph{overcover}. 

\subsection{Approximate workflow}

It is likely that our empirical results show a balance of factors which promote \emph{over-} and \emph{under-coverage}. Theoretically, we know that propagated multivariate sets meet a validity guarantee but \emph{not} a sharpness guarantee. Our synthetic experiments, shown in more detail in Appendix~\ref{app:synthetic}, demonstrate that propagation produces overly conservative sets, and this effect increases with $d_Y/d_Q$, the ratio of prediction dimension during conformalisation to the dimension of the propagated quantity of interest. This behavior likely combines with the tendency toward overcoverage produced by the non-exchangeability of our calibration and test set. 

These factors may counteract a tendency towards \emph{uncercoverage} produced by our propagation and evaluation procedure. Specifically, the propagation operator $L_{\text{true}}$ used to obtain the reference density functional theory (DFT) quantities is \emph{not} the same finite difference operator $L_{\text{approx}}$ which we use apply to our conformal sets. To get a sense of the impact of $L_{\text{approx}}$, note that we apply different approximate operators to predict the elastic constant based on GAP energy predictions and based on GAP stress predictions, but for both scenarios, we compare to the same reference DFT value. In the energy based example, which employs an additional stage of approximation, the coverage is lower. Further note that in Appendix~\ref{app:ec}, we consider a workflow which predicts the same elastic constant using a larger silicon system and an example which uses a smaller finite difference step size which performs well in hyperparameter tuning but is farther from the step size which is expected to balance approximation errors. In these examples, propagated set coverage is partially or completely below the expected validity guarantee. Relative effects are still present: we see greater coverage in the stress example than the energy example, and the vanilla univariate approach typically produces the smallest coverage for a given $\alpha$. Further, the difference in the performance of multivariate sets and Bonferroni-corrected sets demonstrates that using the full covariance $\Sigma$ is a meaningful source of information for conformal scores. 

From these results, we draw several conclusions. First, while the conformal methods are trained on the errors of GAP surrogate predictions for energy, force, and stress against DFT reference, it does not capture the full effect of configuration perturbation, surrogate relaxation, and approximate propagation. Second, the effect of these results is sufficiently small that for reasonable settings, we achieve expected coverage. Finally, even in the presence of approximation error, we find insight into the relative behavior of conformal strategies.

\subsection{Choosing a conformal method}

Each conformal approach offers performance tradeoffs, and the choice of score should be informed by the problem at hand. Basic \emph{univariate} methods may be appropriate if component-wise coverage is the goal or if covariance information is unavailable. If the covariance is insignificant, but full vector coverage is required, a \emph{Bonferroni correction} to a univariate procedure may be suitable. When covariance estimates with some signal are available, \emph{multivariate} conformal methods based on a Mahalanobis distance score are easy to implement to provide similar coverage to the Bonferroni approach with smaller set volume. The resulting sets can capture error cancellations common in target quantities within materials science. These (hyper)ellipsoidal sets are straightforward to propagate through a linear operator $L$, but resulting sets may be overly conservative or impacted by approximations in $L$. 

If $L$ is known at calibration time, it can be built into the conformal score. Such an approach essentially propagates the calibration set. Unfortunately, when $L$ depends on the test case, as is common in materials science, the conformal procedure will require frequent retraining. Furthermore, in materials science, we are often interested in uncertainty at \emph{each} stage of a multi-step workflow. In the general case, applying propagation-corrected conformal prediction produces a pre-propagation prediction ellipse without validity guarantees, and for additional transformations applied following $L$, the resulting prediction sets only have a validity guarantee with no sharpness guarantee---the same as vanilla multivariate propagation. 

Finally, we consider the flexibility offered by \emph{conformal risk control}. We find that these methods can be promising for \emph{high dimensional} calibration targets where covariance information is available. Tailoring a loss function to our goals can create sets with theoretical guarantees and high univariate coverage but comparatively low volume. There is a large design space for new loss functions, but the corresponding danger is that methods may be developed which are overly adapted to a given problem or dataset. A more complete understanding of the role of conformal risk control in materials workflows is a compelling direction for future research.

\section{Conclusion\label{sec:conclusion}}

Our experiments uncover both the utility and potential pitfalls that arise when conformal methods are built into a multivariate, multistage workflow. We aim to build a bridge between the theoretical development of conformal strategies and their practical use within materials science. For this reason, we focus on methods which can be efficiently applied to high dimensional targets $Y$ and propagated to downstream quantities of interest $Q$. As baselines, we consider two approaches which calibrate each component of $Y$ individually: one to achieve target univariate coverage of $Y$ and another to achieve target full coverage. Neither of these methods use surrogate predictions for the correlation in $Y$, so the resulting sets may include unnecessary corners which inflate their volume. We show that multivariate approaches based on the Mahalanobis distance are easy to implement and produce ellipsoidal sets conducive to uncertainty propagation. These approaches can leverage estimates of the correlation between different configurations and between properties, in some cases discovering near symmetries and thereby considerably reducing set volume. Thus, elliptical conformal sets can capture error cancellations common in predictions of relative properties in materials science. Finally, we examine the validity of propagated sets to predict elastic constant components and vacancy formation energies, discussing the impact of approximations in our workflow.

There are many viable directions for future inquiry to build on our results. While our experiments consider the calibration of Gaussian process surrogates, conformal prediction is model agnostic, so these methods can straightforwardly be incorporated in a neural network workflow. Further, much work~\cite{podkopaev2021,tibshirani2020,gendler2022,clarkson2025,Cauchois2024,aolaritei2025} has explored conformal approaches robust to distribution shift which will be useful for many workflows based on atomistic data. For interpretation of results, it will be useful to analyze class conditional or relative representations of uncertainty given by conformal sets. Finally, the generalization to conformal risk control allows for great flexibility in the development of new uncertainty guarantees. We believe that intricate materials science prediction pipelines will draw from an extensive toolbox of methods.
\ack{  

The authors thank Stephen Bates and Niklas Schmitz for productive conversations. Further, the authors acknowledge the MIT Libraries for the resources provided and the MIT Office of Research Computing and Data for providing high performance computing resources that have contributed to the research results reported within this paper.
}

\funding{

MFH acknowledges support by the Swiss National Science Foundation (SNSF, Grant No.~10002757) as well as the NCCR MARVEL, a National Centre of Competence in Research, funded by the SNSF (Grant No.~205602). KEF expresses gratitude for the support by the Don Loomis Galusha (1904) Fellowship from the MIT Office of Graduate Education as well as the National Science Foundation Graduate Research Fellowship (Grant No. 1745302). KEF and YMM acknowledge support from the Department of Energy (DOE), National Nuclear Security Administration PSAAP-III program (award number DE-NA0003965) and PSAAP-IV program (award number DE-NA0004266).

}

\data{

The data that support the findings of this study are openly available in multivariate-conformal at \href{https://github.com/kefisher98/multivariate-conformal/}{https://github.com/kefisher98/multivariate-conformal/}.

}

\printbibliography[
heading=bibintoc,
title={References}
]

\newpage

\appendix

\section{Hyperparameter settings\label{app:hyper}}

Since our workflow combines several computational tools, we have multiple hyperparameter categories: 
\begin{itemize}
    \item \emph{SOAP hyperparameters}: We set these hyperparamters based on best practices in literature~\cite{Bartok2018silicon,Musil2021} and previous experience~\cite{Fisher2024}. The cutoff radius which defines atomic neighborhoods is set to $5$ angstroms. Each atom is represented by a Gaussian function with variance $0.25$. Finally, the spherical harmonics expansion is truncated according to $n_{max}=8$ and $\ell_{max}=6$.
    \item \emph{Gaussian process kernel hyperparameters}: We choose a polynomial kernel with exponent $4$, which has been shown to be effective for Gaussian approximation potentials~\cite{Bartok2010,Bartok2015,Bartok2018silicon,bartok2020}. The kernel function introduces hyperparameters variance $v$ and normalization $c$ in addition to the data noise standard deviation $\gamma$. Note that because the model's predictive covariance $\Sigma$ is rescaled by conformal calibration, the absolute scale of $\Sigma$ does not impact our final predictions. Then, the predictive $\mu$ and $\Sigma$ depend only on the composite scalar hyperparameter $\gamma^2/(cv)$. We find that the accuracy of the Gaussian process potential is sensitive to the scale of this parameter, but optimization of the precise value produces diminishing returns and risks overfitting. Consequently, we set aside a small optimization dataset of $20$ configurations and perform a line search on powers of $\gamma^2/(cv)=10^a$ for $a\in\{-12,\dots, 0,1\}$, evaluating performance on energy prediction with leave one out cross validation. Ultimately, we find $a=-8$. 
    \item \emph{Multitask hyperparameters}: In our examples, the bulk diamond configurations will inform the primary task, and the vacancy configurations will inform the secondary task. This designation is made because the primary tasks hyperparameters can be optimized independently; thus, we are able to reuse the optimal values already found for the bulk diamond configurations, as described above. For both tasks, we will use polynomial kernels with quartic exponents. Again noting that conformal calibration makes our results insensitive to the absolute scale of the predictive covariance, we find that the remaining hyperparameters are the ratio of primary and secondary kernel variance $v_p/v_s$ and the correlation $\varrho_s$. Using the $20$ configuration primary optimization set and $10$ additional vacancy configurations set aside for secondary task optimization, we perform a $2$-dimensional line search using leave one out cross validation. We find a narrow basin of low error for $\varrho_s\geq 0.95$ and large $v_p/v_s$ and a wide basin of low error about $\varrho_s=0.5$ and $v_p/v_s\approx 1$. To avoid overfitting, we set $v_p/v_s=1$ and $\varrho_s=0.55$.
    \item \emph{Finite difference hyperparameters}: We will propagate uncertainty sets using a central difference approximation and must choose the step size. For one set of tests, we will use the differences of energies for perturbed configurations to predict stress, then use the differences of stresses to predict the elastic constant tensor for silicon. For another set of tests, we directly predict stress with a Gaussian process model, then apply a finite difference scheme to predict the elastic constant. Since each configuration has the same elastic constant and it is our target for prediction, we cannot optimize hyperparameters based on their performance predicting the elastic constant. Rather, we optimize based on error in stress prediction, using leave one out cross validation on the $20$ configuration optimization set. To balance errors in an order $m$ finite difference approximation, the step size should be chosen to be about $1/(m+1)$ root of the surrogate error~\cite{Driscoll2017}. Since central difference is a second order approximation and the Gaussian process produces errors on order of ~$0.005$ eV, we perform a line search of step sizes between $10^{-4}$ and $0.5$. For our experiments, we choose step size $0.05$, and in Appendix~\ref{app:ec}, we present additional results for step size $0.01$, another well performing value. 
    \item \emph{Conformal hyperparameters}: The character of conformal sets is controlled by the tolerance $\alpha$. Conformal risk control may introduce additional loss function parameters, including the upper bound $B$. In general, we will choose $B$ to be some quantile of the calibration losses. We perform tests for a range of conformal parameters and present representative results.
\end{itemize}
Overall, we aim to select hyperparameters using minimal data and to avoid overfitting. 

\newpage

\section{Additional results of synthetic experiment\label{app:synthetic}}

\begin{figure}[h]
    \centering
    \begin{tabular}{cc}
        \includegraphics[width=0.35\linewidth]{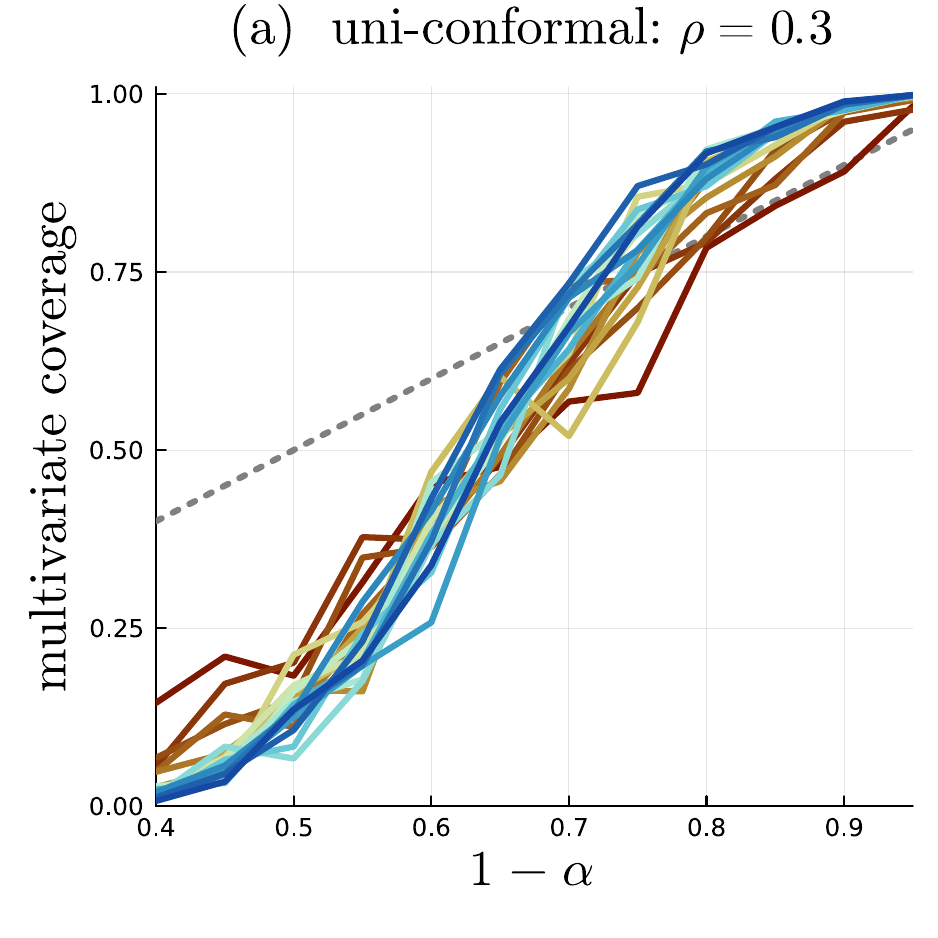} & \includegraphics[width=0.35\linewidth]{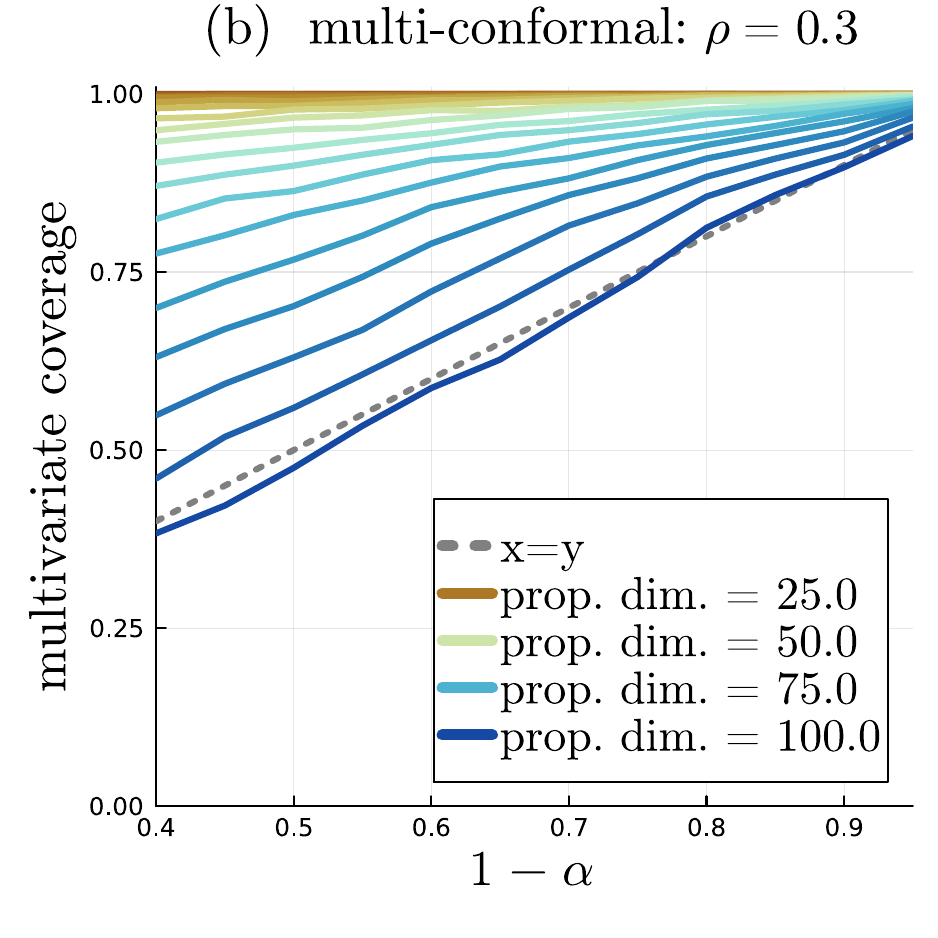}  \\
        \includegraphics[width=0.35\linewidth]{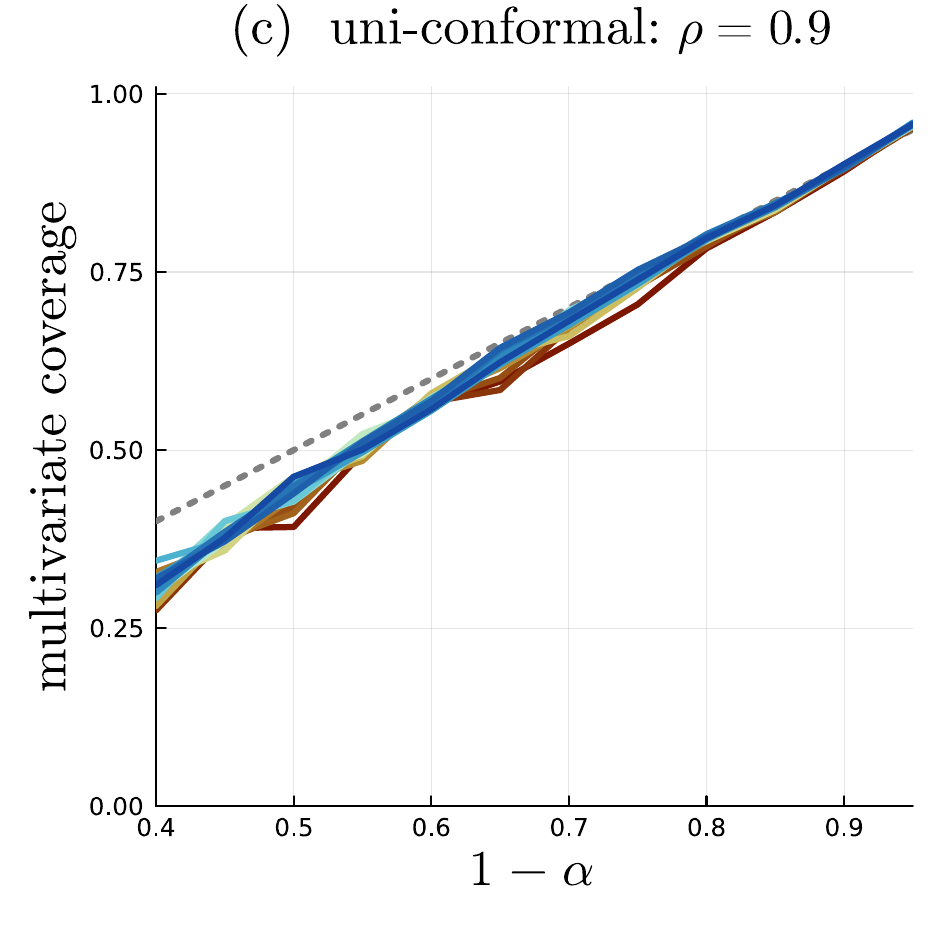} & 
        \includegraphics[width=0.35\linewidth]{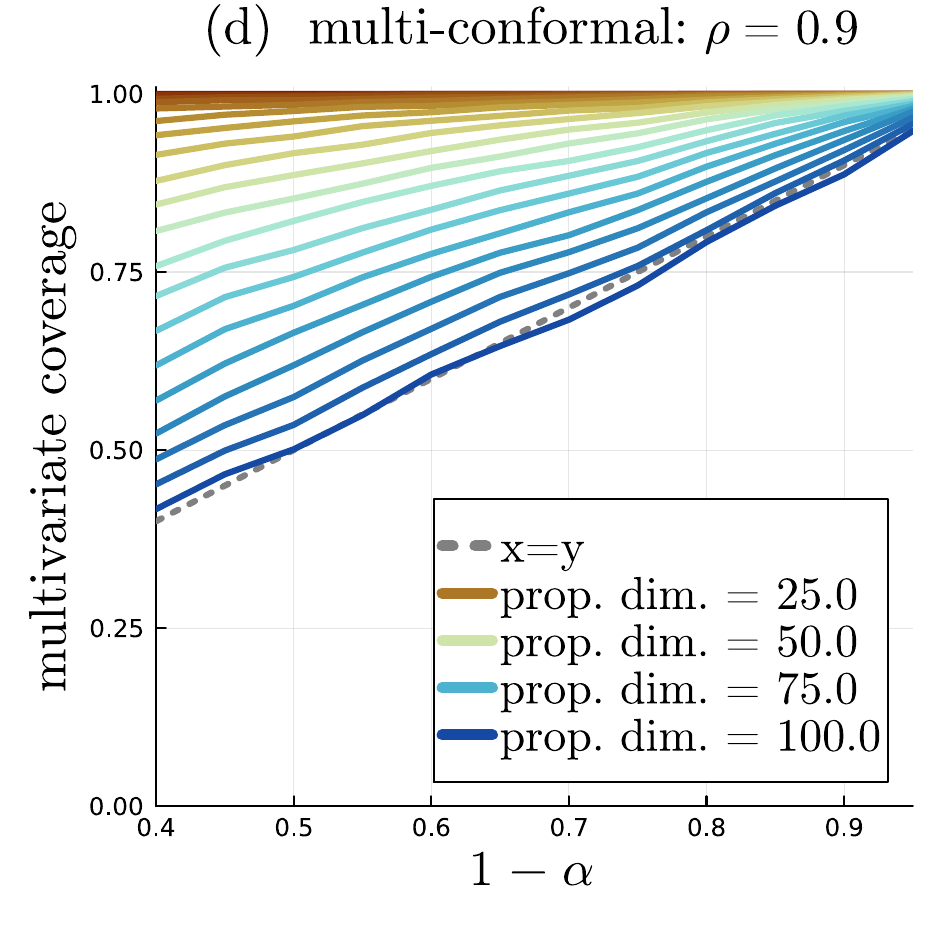} \\
    \end{tabular}
\caption{\label{fig:synth_prop} \textbf{Synthetic propagation study}: Full vector coverage of propagated conformal sets. Calibration is applied to a $100$ dimensional vector, and sets are propagated via linear transformation dimension $d\in\{5,10,\dots,95,100\}$. Color indicates $d$, more brown is smaller and more blue is larger. Results are averaged over $10$ random propagation operators. The performance of component-wise conformal calibration is reported in the left column, and multivariate conformal results are in the right. $\rho$ indicates the correlation of the covariates of the components of each vector which is calibrated.  }
\end{figure}

The results in Figure~\ref{fig:synth_prop} complement those shown in Section~\ref{ss:synthetic}. Specifically, we examine the coverage a propagated prediction set where the propagation dimension $d_Q$ varies from $5$ to $100$. The left column shows the performance of univariate conformal prediction, while the right column reports the results for multivariate methods. The propagated univariate conformal sets often fail to achieve the desired coverage level $1-\alpha$. This effect is exacerbated for smaller covariate correlation $\rho=0.3$ or larger tolerance $\alpha$. As $\alpha\to 0$, the low tolerance forces the univariate conformal prediction sets to achieve high coverage, but the price is the larger volume sets reported in Figure~\ref{fig:synth_cov_vol}. The propagated sets produced by multivariate conformal methods achieve the desired coverage, but as $d_Q$ descreases, these sets become increasingly conservative.

\begin{figure}
    \centering
    \includegraphics[width=0.5\linewidth]{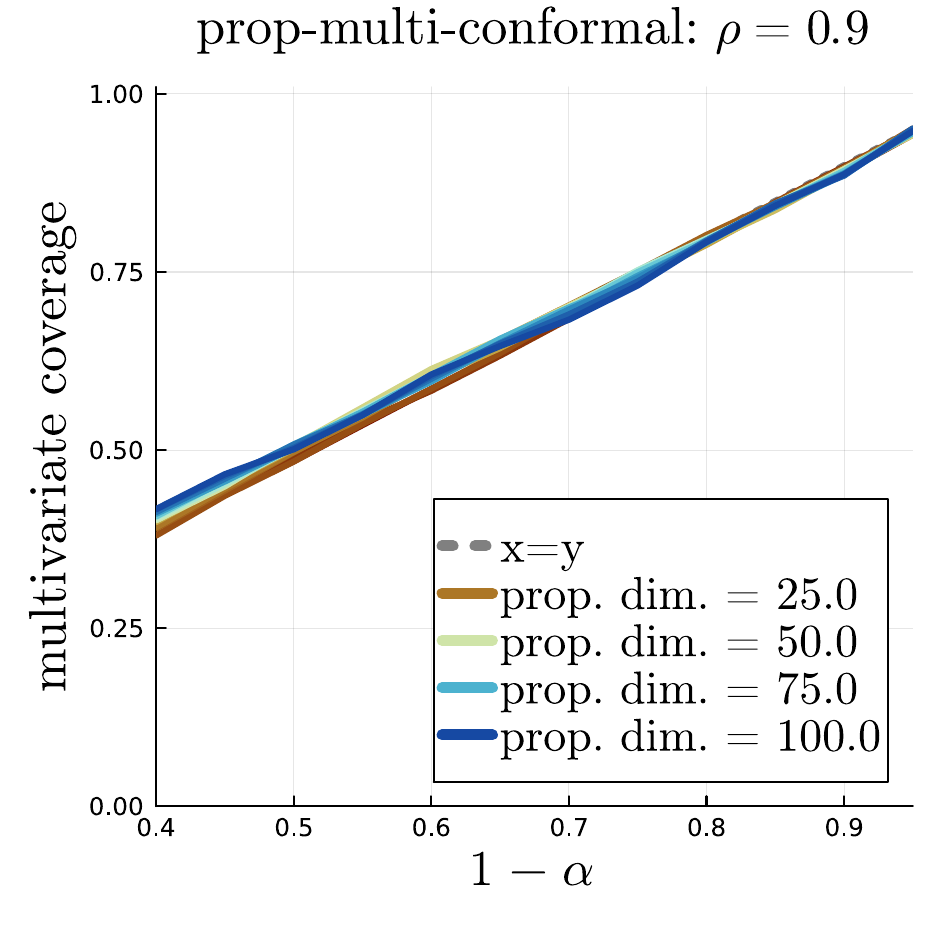}
\caption{\label{fig:M_synth_prop} \textbf{Synthetic propagation study}:  Full vector coverage of multivariate conformal calibration with propagation correction. As in Figure~\ref{fig:synth_prop}, calibration is performed on a $100$ dimensional vector which propagated into space with dimension $d\in\{5,10,\dots,95,100\}$.}
\end{figure}

Using propagation-corrected multivariate conformal prediction is expected to alleviate this conservatism. This approach provides a sharpness bound on the propagated set, but it requires knowledge of the propagation operator at the time of calibration. Figure~\ref{fig:M_synth_prop} confirms that for all $d_Q$ considered, the propagation-corrected approach produces sets that satisfy both the coverage and sharpness guarantees. 

\begin{figure}[h]
    \centering
    \begin{tabular}{cc}
        \includegraphics[width=0.4\linewidth]{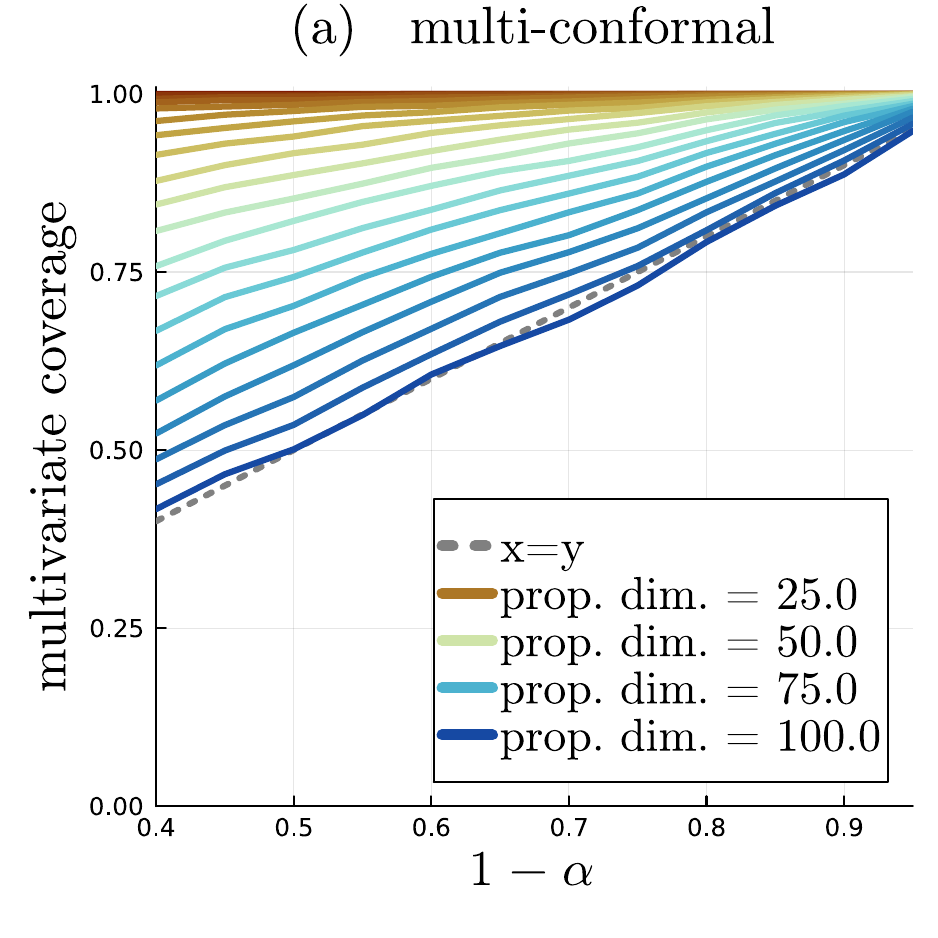} & \includegraphics[width=0.4\linewidth]{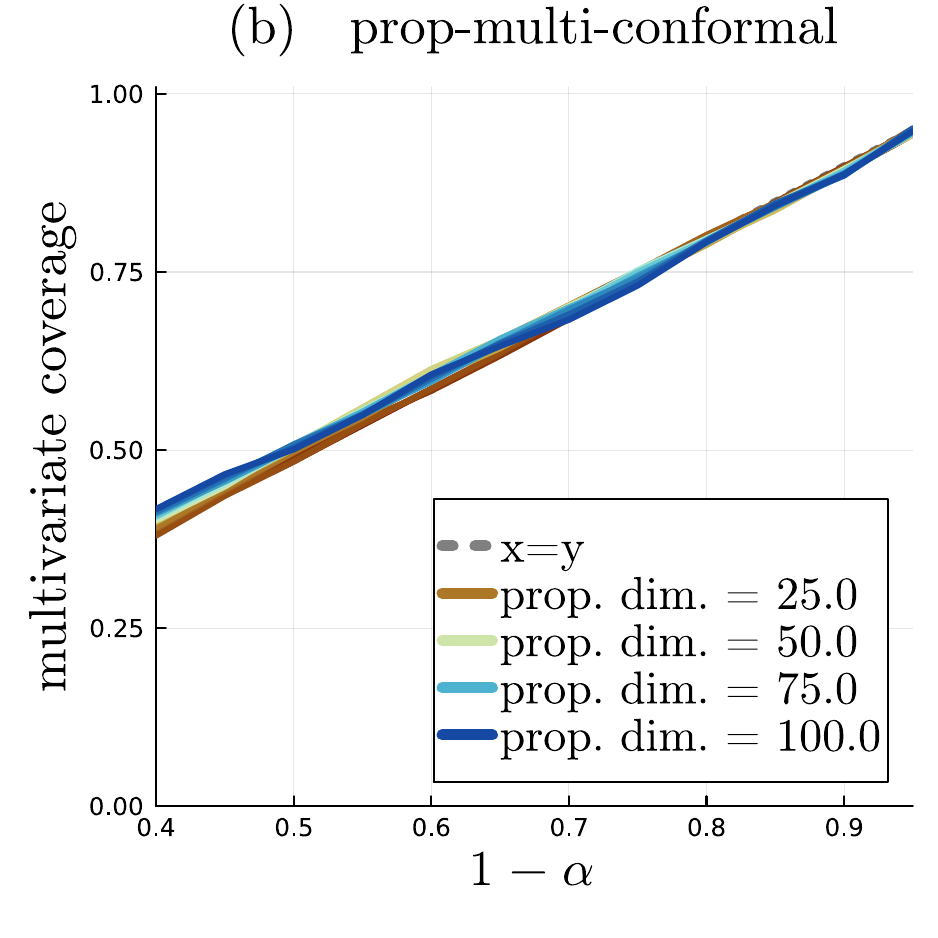}  
    \end{tabular}
\caption{\label{fig:synth_prop_uncorrelated} \textbf{Synthetic propagation study}: Coverage post-propagation when the calibration data set has uncorrelatd covariates, and the test set has highly correlated $(\rho=0.99)$ covariates.  As in Figure~\ref{fig:synth_prop}, calibration is performed on a $100$ dimensional vector which propagated into space with dimension $d\in\{5,10,\dots,95,100\}$. (a): performance of multivariate conformal prediction. (b): multivariate conformal prediction with propagation correction.}
\end{figure}

For our synthetic tests, we have assumed that for both the calibration and testing data, the elements in each multivariate covariate have the same correlation level $\rho$. We now consider the case where there is $\rho=0$ correlation in the covariate elements for the calibration set and $\rho=0.99$ correlation in the covariate elements of the test set. These settings represent an extreme case, but we do expect some disparity in the correlation of our calibration and test sets for our elastic constant experiment. This disparity arise because the test set is created by applying small perturbations to the geometry of one system, and the difference in geometry of any two calibration structures is larger than this perturbation. In Figure~\ref{fig:synth_prop_uncorrelated}, we see that the propagated sets achieve the desired coverage level but are consistently over-conservative. This effect occurs for both the basic (a) and propagation-corrected (b) forms of multivariate conformal prediction. The misspecification in calibration set covariance reduces the difference in performance between the two methods.

\section{Additional results of energy and force calibration\label{app:energy_force}}

\begin{figure}
    \centering
    \begin{tabular}{cc}
        \includegraphics[width=0.4\linewidth]{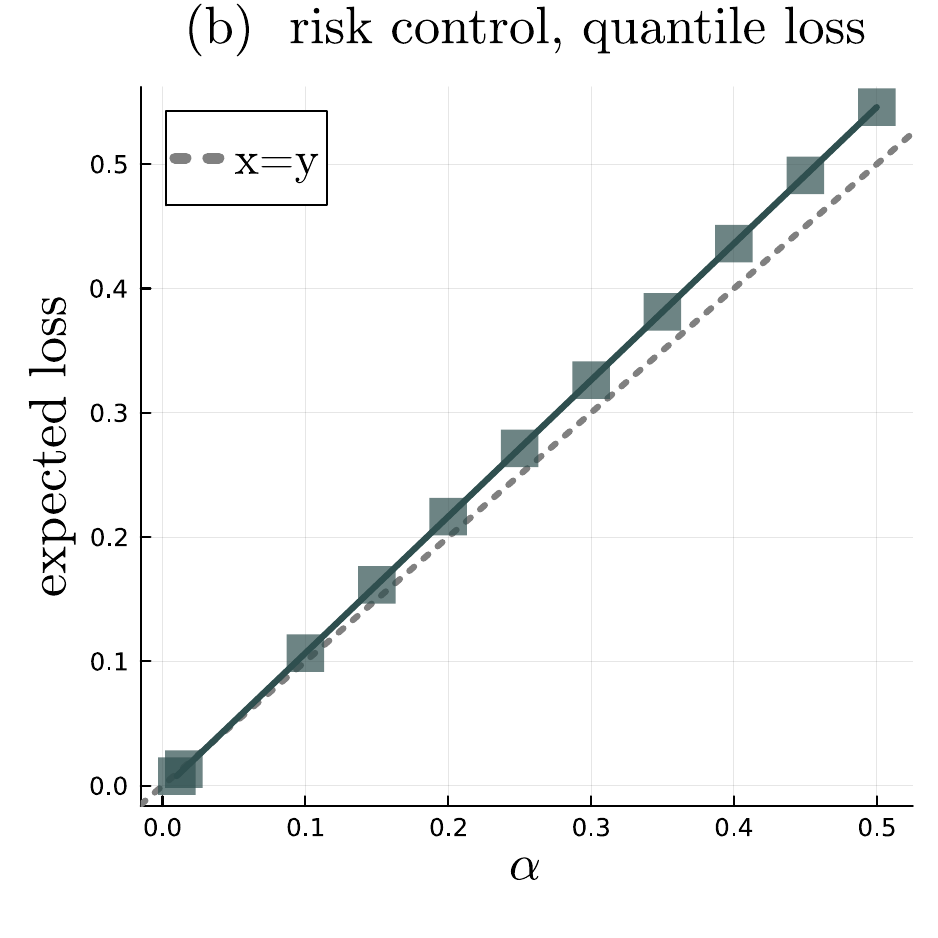} &
        \includegraphics[width=0.4\linewidth]{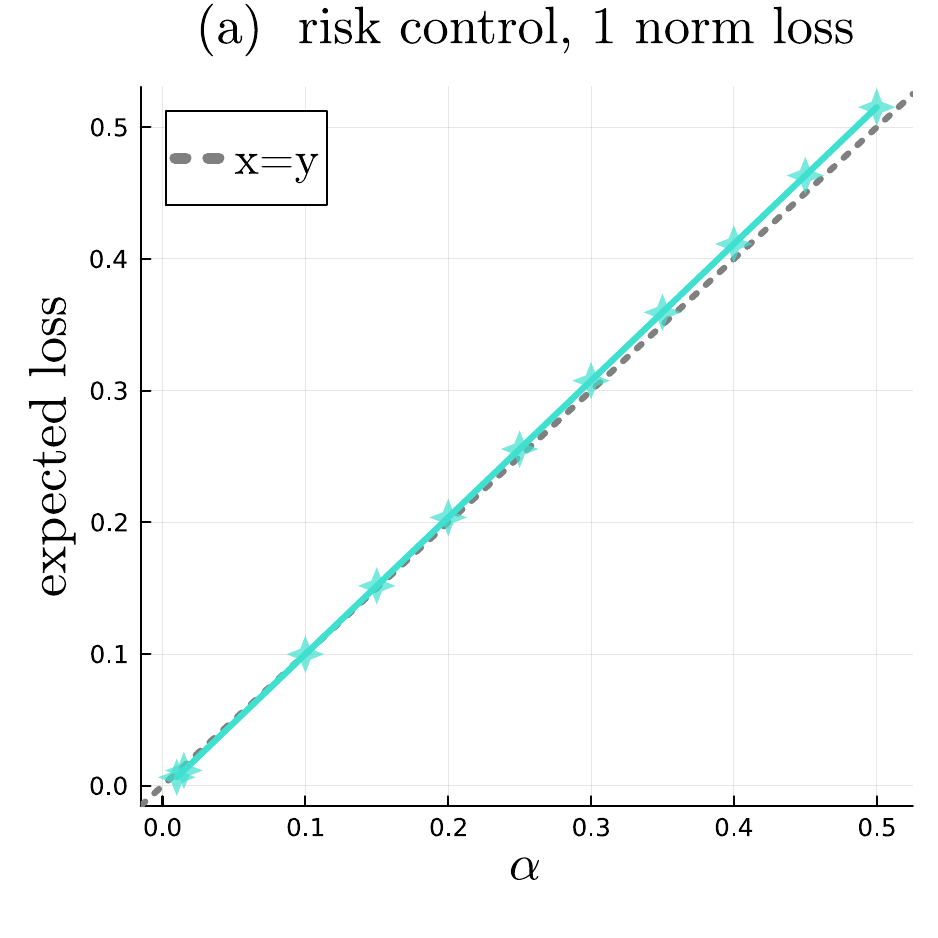} 
    \end{tabular}
\caption{\label{fig:rc_guarantee} \textbf{Energy and atomic force calibration}: verification that multivariate conformal risk control achieves guaranteed loss bounds for tolerance $\alpha$. (a): loss attained by the risk control applied with loss based on the $\ell_1$ norm of the score vector. (b): risk control based on the $0.9$ quantile of the score vector. The $x=y$ line is dotted in gray.}
\end{figure}
\begin{figure}[h]
    \centering
    \begin{tabular}{cc}
        \includegraphics[width=0.4\linewidth]{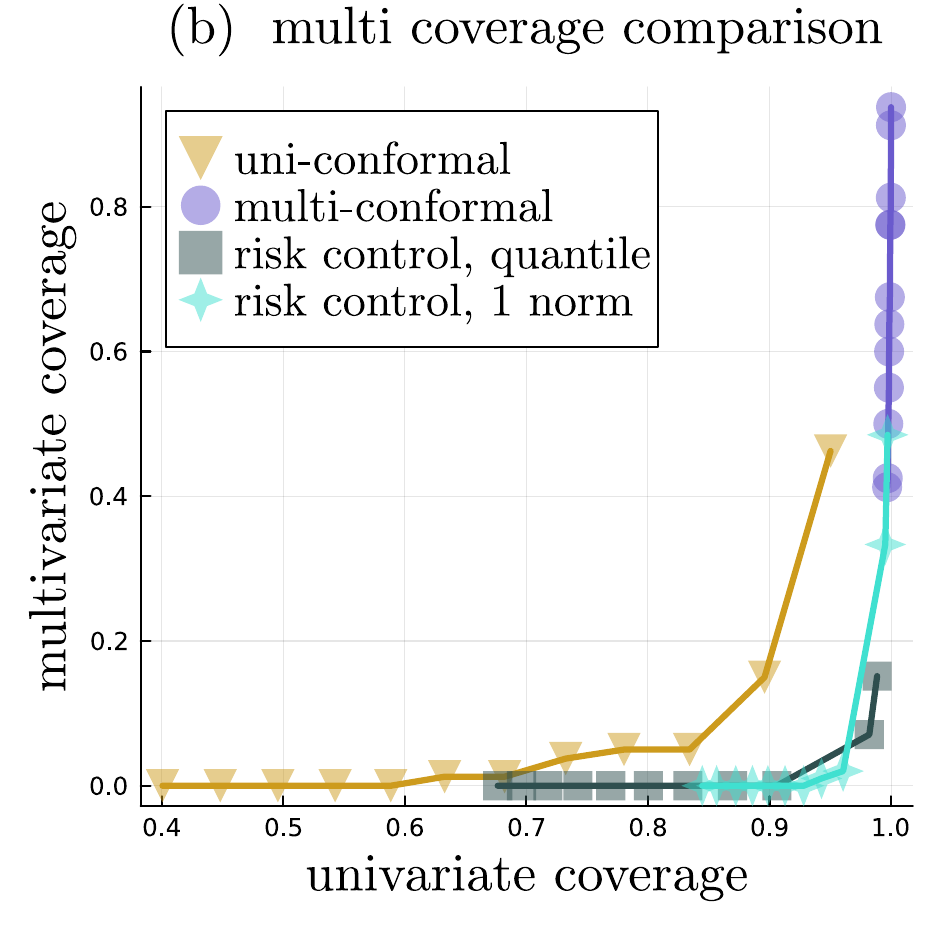} &
        \includegraphics[width=0.4\linewidth]{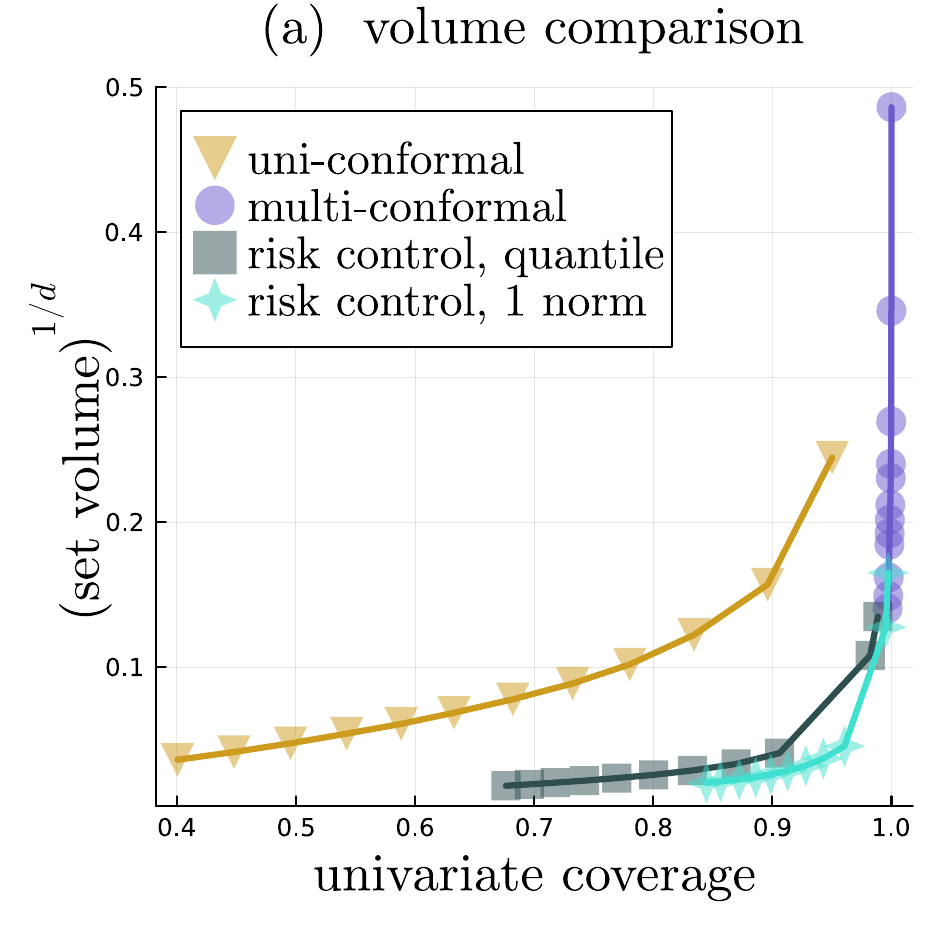} 
    \end{tabular}
\caption{\label{fig:energy_force_cov_vol_08} \textbf{Energy and atomic force calibration}: Replication of Figure~\ref{fig:energy_force_cov_vol_08} for upper bound on loss function set to the $0.8$ quantile of the calibration scores rather than the $0.95$ quantile. Relationship between component-wise coverage, full vector coverage, and set volume for univariate (gold triangles) and multivariate (purple circles) conformal prediction as well as conformal risk control with $1$-norm (turquoise stars) and quantile (green squares) loss. Both calibration and test datasets contain systems of multiple sizes. Consequently, in the test set, vectors containing the global energy and all atomic forces for a system take on dimensions $d\in\{49,163,385\}.$}
\end{figure}

Conformal risk control guarantees that the average loss of our prediction sets will be bounded, as shown in~\eqref{eq:control}. Figure~\ref{fig:rc_guarantee} verifies these guarantees for the experiments presented in Section~\ref{ss:energy_force}. The subfigures correspond to different user defined loss functions: the $1$ norm loss and quantile loss, respectively. In both cases, we see that the empirical expectation of the loss---calculated over random draws of the train, calibration, and test sets---closely tracks the tolerance $\alpha$. Deviation from the $x=y$ reference line likely occurs due to the finite size of the calibration dataset.

Conformal risk control also includes an additional hyperparameter when compared with conformal prediction: the upper bound $B$ of the loss function. In the main text, this hyperparameter is set to $0.95$ of the maximum loss value achieved on the calibration dataset. To show the insensitivity of our results to this choice, Figure~\ref{fig:energy_force_cov_vol_08} reproduces the experiment with $B$ set to $0.8$ of the maximum loss value achieved on the calibration dataset.

\section{Additional results of elastic constant prediction\label{app:ec}}

Table~\ref{tab:ec_error} records the accuracy of the elastic constant predictions made by our Gaussian process model. As described in Section~\ref{ss:elasticconstant}, we apply perturbations to the geometry of a silicon diamond structure and the resulting properties to build finite difference approximations to elastic constant components. We consider one case where the input to the finite difference approximation is the energy of each perturbed system and a second case where the inputs are virial stresses. The choice of perturbations lead us to predict the three unique components of silicon's elastic constant tensor ($C_{11}$, $C_{12}$,$C_{44}$) in addition to redundancies and zero components ($C_{21}=C_{12}$, $C_{45}=C_{54}=0$). We report the error of all size predicted values for both the energy and stress based workflows. The relative error is reported for all components with a nonzero reference ($C_{11}$, $C_{12}$, $C_{21}$, $C_{44}$). For the zero components ($C_{45}$, $C_{54}$) report absolute error in units of $eV/\AA^3$. We note that the main purpose of work is to evaluate an uncertainty quantification workflow and not to propose a state of the art approach to prediction, so reasonable accuracy suffices for our purposes. 

\begin{table}[h]
    \centering
     \caption{\textbf{Propagation to elastic constant uncertainty}: Relative error of elastic constant predictions against DFT reference. Note: the true value of $C_{45}=C_{54}$ is $0$, so absolute error is reported for these quantities. The $0.25$, $0.5$, $0.75$ quantiles are taken across $100$ random draws of the training and calibration sets from silicon diamond configurations with $\{2,16,54,128\}$ atoms. The top sets are obtained with a finite difference approximation of the second derivative of energy, and the bottom results are an approximation of the first derivative of virial stress. Geometric relaxation, energy, and stress prediction were preformed with a GP model. For these results the finite difference step sizes are set to $0.05$ angstrom. }
    \begin{tabular}{c|ccc}
    & & quantiles & \\
    & $0.25$ & $0.5$ & $0.75$ \\ 
    \hline
    via energy & & & \\
    $C_{11}$   &   $0.142154$    &  $0.32009$   & $0.810568$ \\
    $C_{12}$   &   $0.366259$    & $2.22343$    & $2.95303$ \\
    $C_{21}$   &   $0.151435$    & $0.395372$   & $2.79784$ \\
    $C_{45}$   &   $0.0007689$   & $0.00196889$ & $0.00340531$ \\
    $C_{54}$   &   $0.000624246$ & $0.00163359$ & $0.00289075$ \\
    $C_{44}$   &   $0.123629$    & $0.315627$   & $0.57253$ \\
    \hline 
    via stress & & & \\
    $C_{11}$   &     $0.238593$    & $0.3784$     & $0.569042$ \\
    $C_{12}$   &     $0.204968$    & $0.39699$    & $0.625762$ \\
    $C_{21}$   &     $0.205036$    & $0.397864$   & $0.632561$ \\
    $C_{45}$   &     $0.00134252$  & $0.00268628$ & $0.00459271$ \\
    $C_{54}$   &     $0.000856786$ & $0.00167998$ & $0.00299521$ \\
    $C_{44}$   &     $0.0455208$   & $0.100135$   & $0.19318$ \\
    \end{tabular}
    \label{tab:ec_error}
\end{table}

\begin{figure}[h]
    \centering
    \begin{tabular}{cc}
        \includegraphics[width=0.4\linewidth]{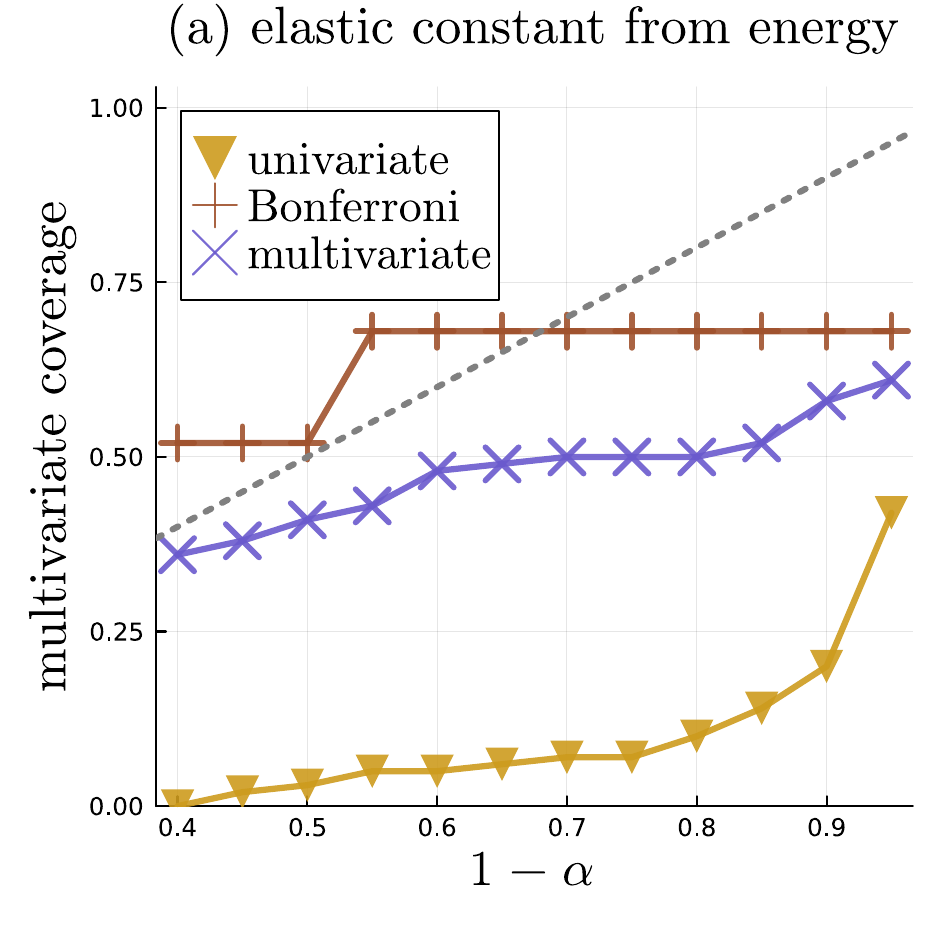} & \includegraphics[width=0.4\linewidth]{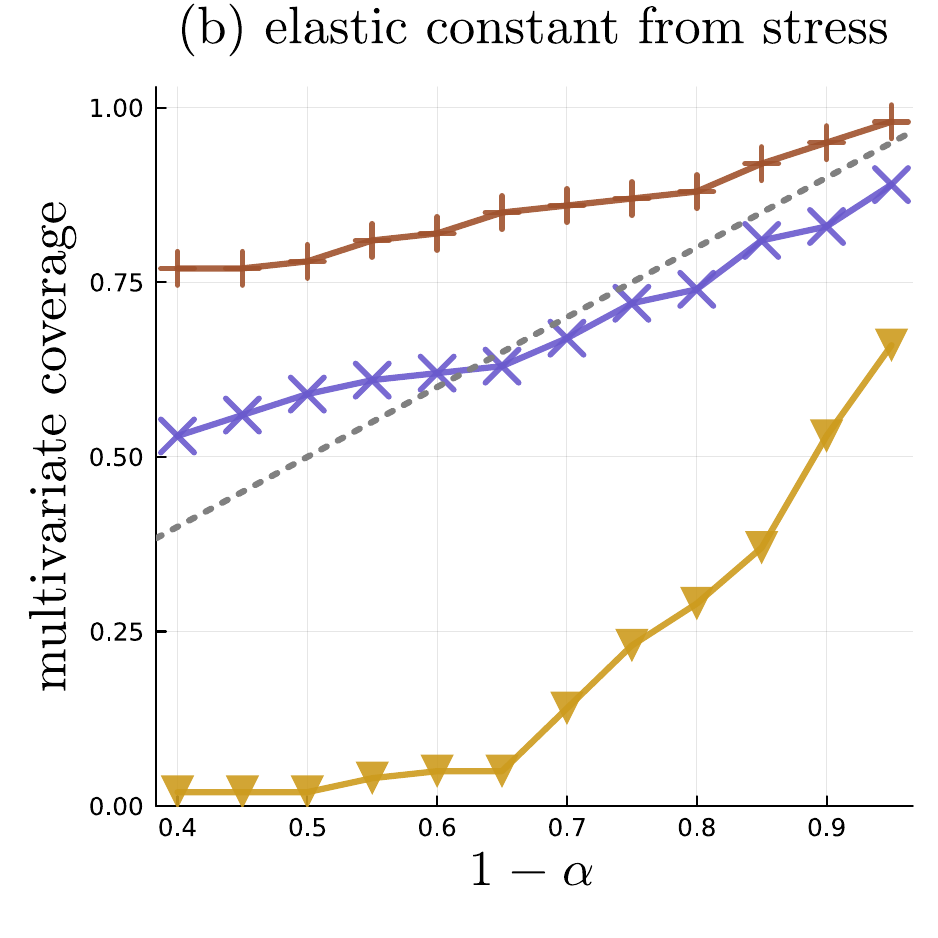}  \\
        \includegraphics[width=0.4\linewidth]{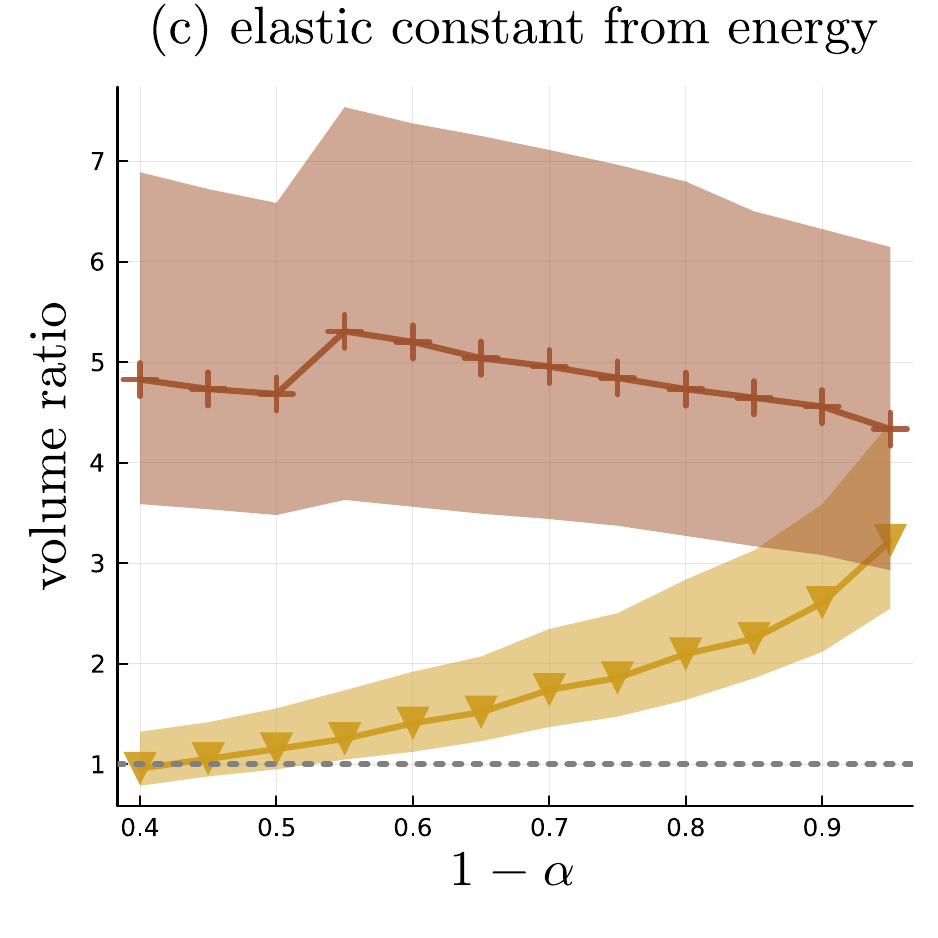} & \includegraphics[width=0.4\linewidth]{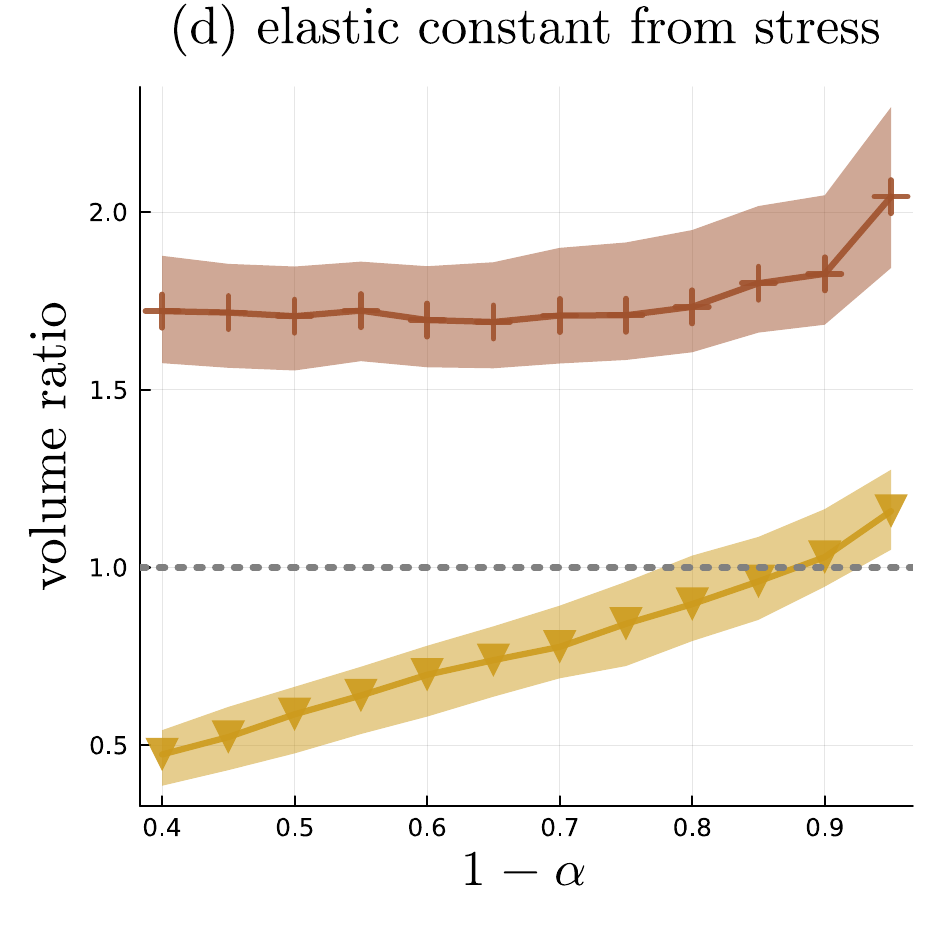} \\
    \end{tabular}
\caption{\label{fig:ec_full} \textbf{Propagation to elastic constant uncertainty}: Results from the experiment presented in Figure~\ref{fig:elastic_constant}. Here we consider only the projection into elastic constant space preformed by the finite difference approximation with no further projection step to isolate the non-redundant, nonzero components. As a result, the prediction sets described in these plots are $6$ dimensional, rather than $3$ dimensional. Left column: elastic constant computed as a second derivative of GP energy predictions. Right column: elastic constant computed as a first derivative of GP stress  predictions. Conformal methods: univariate (gold triangles), univariate with Bonferroni correction (red +s), multivariate (purple xs). Top row: full vector coverage for tolerance $\alpha$. Bottom row: ratio (corrected for $d=6$) of univariate sets size (gold triangles) as well as Bonferroni-corrected size (red +s) to multivariate set size.}
\end{figure}

In the main text, Figure~\ref{fig:elastic_constant} reports the performance of prediction sets propagated into the space of the three nontrivial elastic constant components ($C_{11}$, $C_{12}$,$C_{44}$). Figure~\ref{fig:ec_full} reports the coverage and volume of the conformal sets propagated into the space of all six predicted elastic constant components. In this case, the multivariate and Bonferroni-corrected approaches do not achieve the guaranteed coverage level for all $\alpha$. We attribute this behavior to the challenges discussed in Section~\ref{sec:discussion}. The univariate conformal approach still leads to systematically lower coverage than the competing approaches. Notably, in the energy based example, even while the multivariate approach leads to prediction sets with greater coverage than the univariate approach, it also systematically produces sets with smaller volume. We emphasize that the volume ratio reported in Figure~\ref{fig:ec_full} standardized by taking the root of the prediction dimension, so the true volume ratios reach $O(1000)$. This extreme difference in volume likely occurs because multivariate conformal prediction leverages covariance information to identify symmetries (such as $C_{12}=C_{21}$) in our quantity of interest.

\begin{figure*}[h]
    \centering
    \begin{tabular}{cc}
        \includegraphics[width=0.4\linewidth]{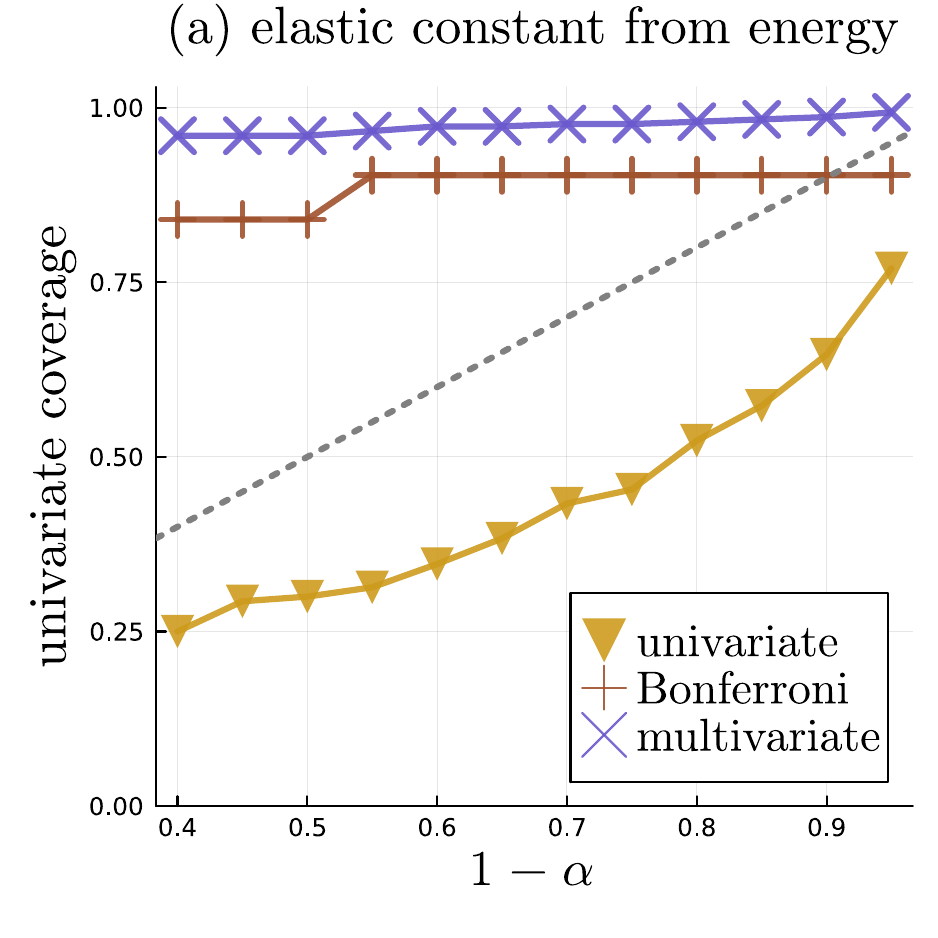} & \includegraphics[width=0.4\linewidth]{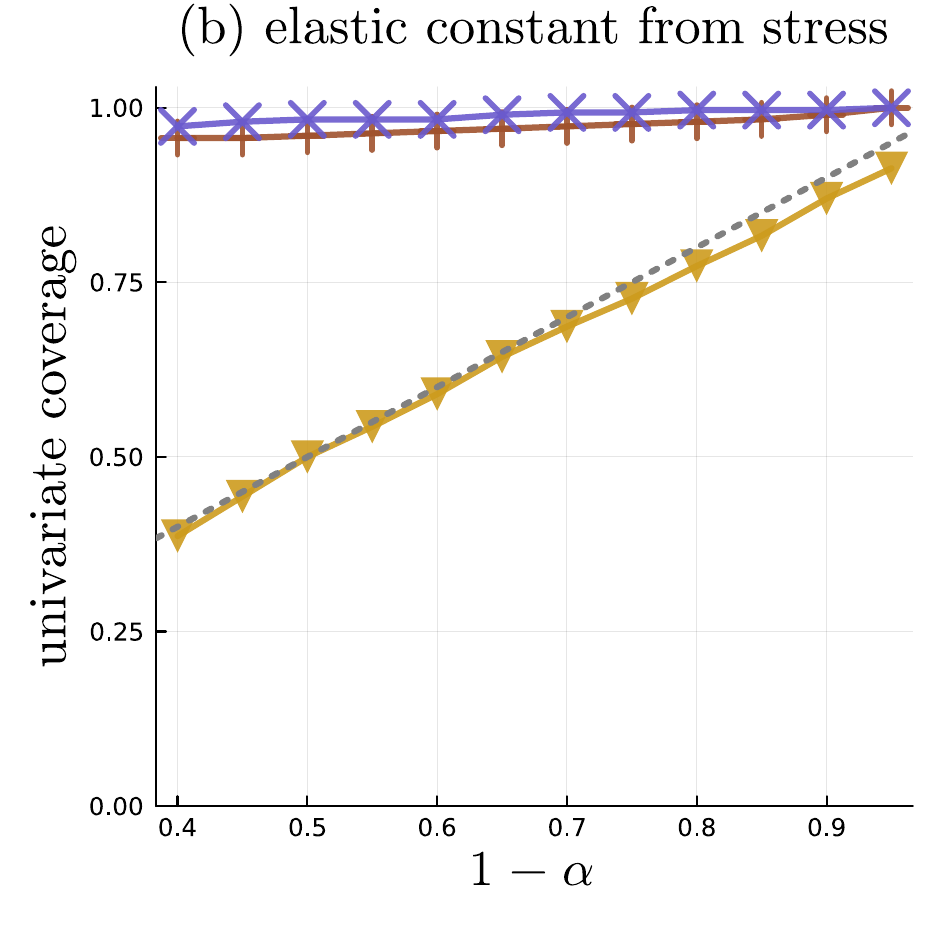}  \\[6pt]
    \end{tabular}
\caption{\label{fig:unicov} \textbf{Propagation to elastic constant uncertainty}: Results from the experiment presented in Figure~\ref{fig:elastic_constant}. Here, we present the coverage of each component of the final $3$ dimensional prediction for the elastic constant. (a): elastic constant computed as a second derivative of GP energy predictions. (b): elastic constant computed as a first derivative of GP stress predictions. Conformal methods: univariate (gold triangles), univariate with Bonferroni correction (red +s), multivariate (purple xs).}
\end{figure*}

While our primary interest is the multivariate coverage of our propagated prediction sets, we can also evaluate univariate coverage by projecting to one dimension. Figure~\ref{fig:unicov} shows the univariate coverage on the vector $[C_{11}, \; C_{12}, \; C_{44}]$ achieved by the conformal methods for both the energy and stress based workflows. As expected, the multivariate conformal approach consistently achieves high univariate coverage. This effect may be exacerbated because the components that the multivariate sets are most likely to miscover ($C_{45}=C_{54}=0$) are excluded from this test. The univariate method is expected to produce univariate coverage close to $1-\alpha$. While it achieves this guarantee in the stress based example, in falls short in the energy based example, likely due to the approximation limitations discussed in Section~\ref{sec:discussion}.

\begin{table}[h]
    \caption{\textbf{Propagation to elastic constant uncertainty}: Replication of results shown for elastic constant prediction via energy in Table~\ref{tab:ec_error} with the finite difference step sizes are set to $0.01$ angstroms.  }
    \centering
    \begin{tabular}{c|ccc}
    & & quantiles & \\
    & $0.25$ & $0.5$ & $0.75$ \\ 
    \hline
    via energy & & & \\
     $C_{11}$   &$0.0554215$  &   $0.121471$ &   $0.17805$ \\
     $C_{12}$   &$0.230013$   &  $0.358458$  &  $0.531036$ \\
     $C_{21}$   & $0.178043$  &   $0.301457$ &   $0.444404$ \\
     $C_{45}$   & $0.00716034$&   $0.016069$ &   $0.0411496$ \\ 
     $C_{54}$   & $0.00524922$&  $0.0184795$ &  $0.0337626$ \\
     $C_{44}$   & $0.18665$   &   $0.257848$ &   $0.331342$ \\ 
    \end{tabular}
    \label{tab:ec_error_01}
\end{table}

\begin{figure}
    \centering
    \begin{tabular}{cc}
        \includegraphics[width=0.4\linewidth]{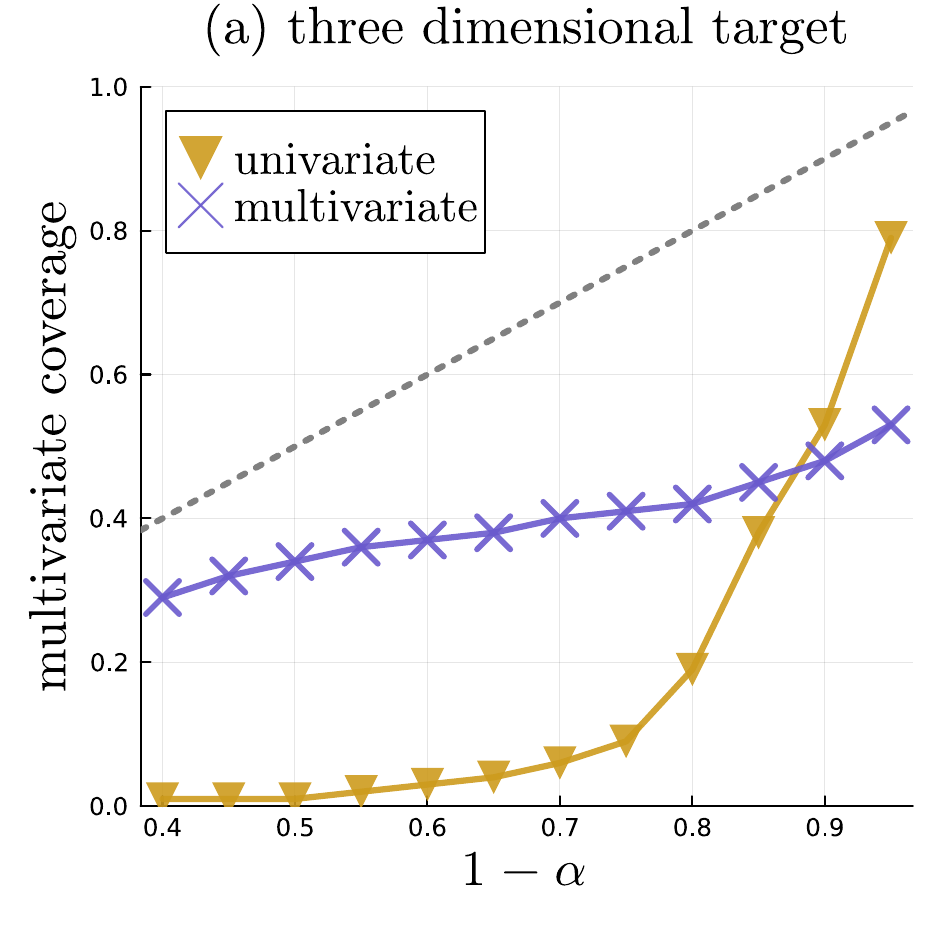} & \includegraphics[width=0.4\linewidth]{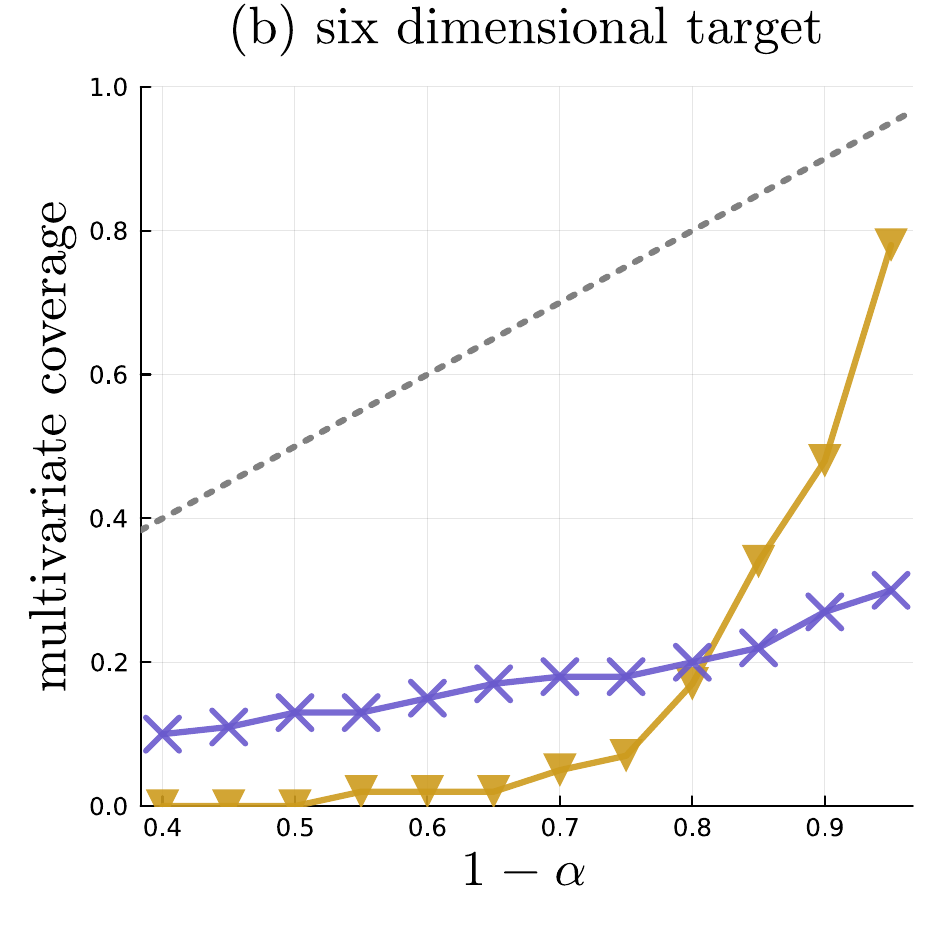}  \\[6pt]
        \includegraphics[width=0.4\linewidth]{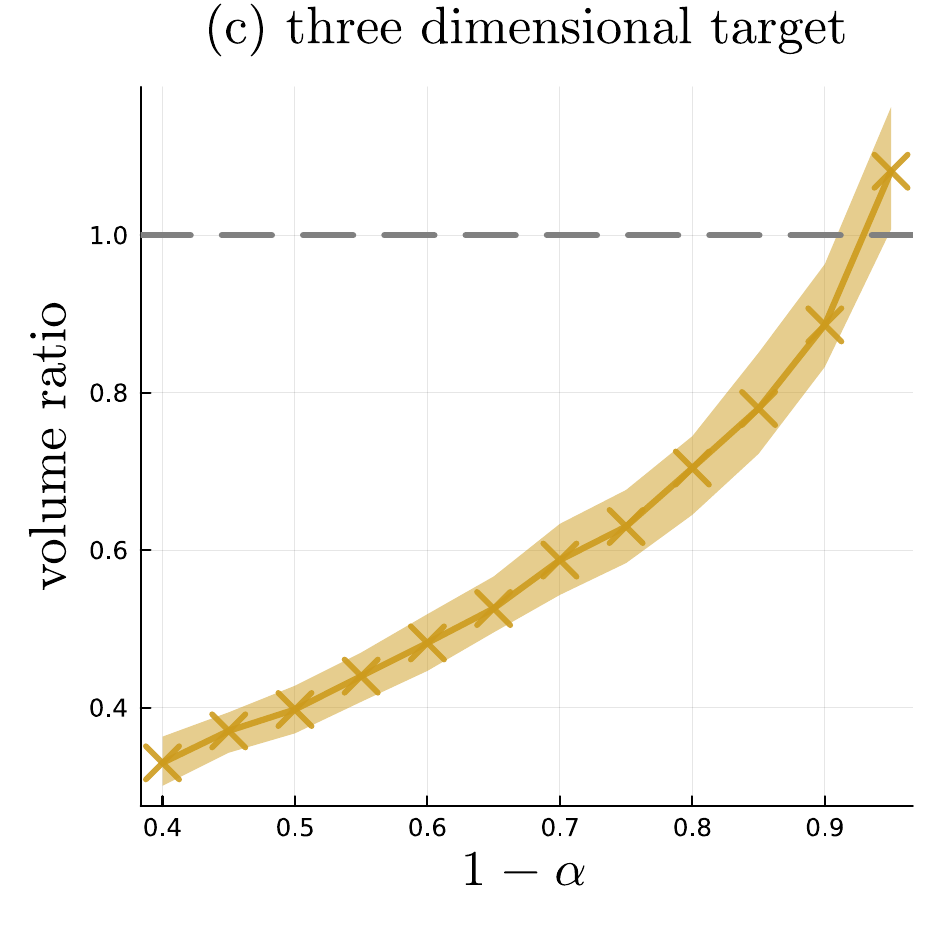} & \includegraphics[width=0.4\linewidth]{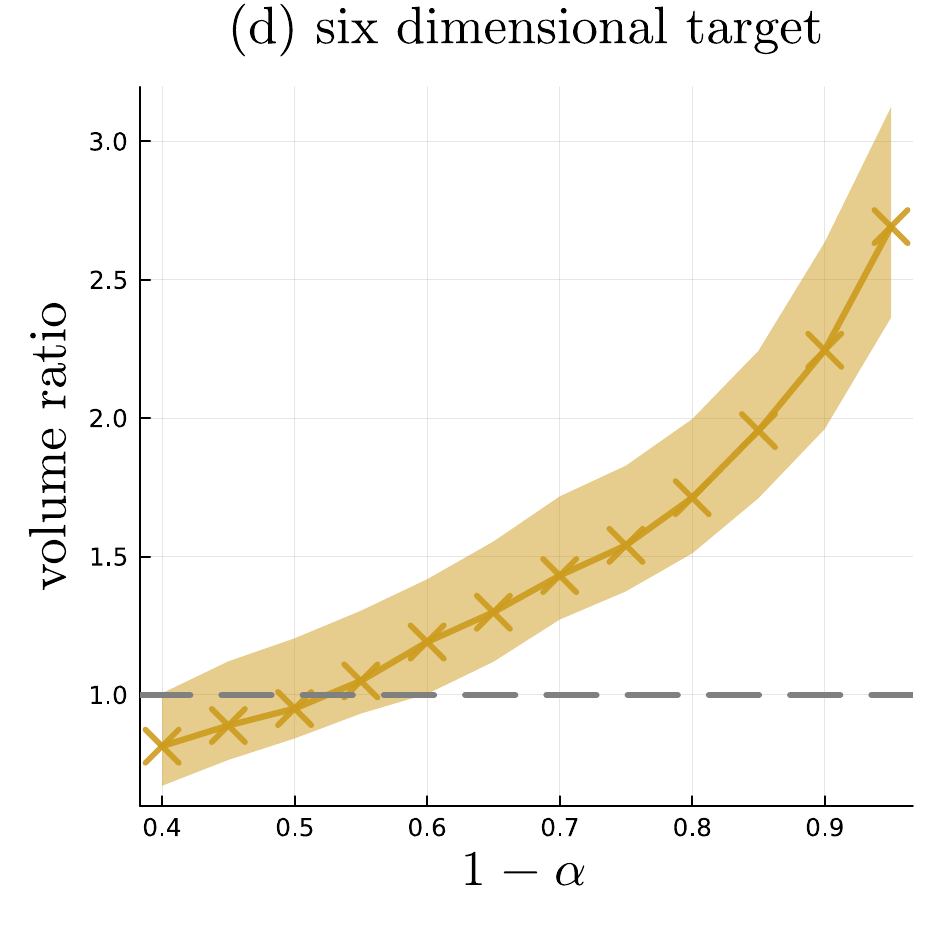} \\[6pt]
    \end{tabular}
\caption{\label{fig:energy_ec_step_01} \textbf{Propagation to elastic constant uncertainty}: Replication of energy based results presented in Figures~\ref{fig:elastic_constant} and~\ref{fig:ec_full} with finite difference step size set to $0.01$ angstrom. These results correspond to the error reported in Table~\ref{tab:ec_error_01}. Left column: coverage and volume comparison for $3$ dimensional elastic constant prediction. Right column: results for final $6$ dimensional elastic constant prediction. Top row: full vector coverage for tolerance $\alpha$. Bottom row: ratio (corrected for $d=6$) of univariate sets size to multivariate set size.}
\end{figure}

As documented in Appendix~\ref{app:hyper}, our workflow relies on many hyperparameters. In particular, our exploration of the finite difference step size suggest that a range of values may work well in practice. For most of our results, we use a step size of $0.05$ angstrom, but we also repeat our energy based experiments using a step size of $0.01$ angstroms. The accuracy is reported in Table~\ref{tab:ec_error_01}, and the coverage and volume ratios are shown in Figure~\ref{fig:energy_ec_step_01}.  The left column shows results for all six predicted components of the elastic constant, while the right column describes only the three nontrivial components.  The coverage of multivariate conformal prediction sets is reduced in this case compared with our other experiments and the expected guarantee. Section~\ref{sec:discussion} discuss the balance of errors which is likely responsible for this behavior.

\begin{table}[h]
    \caption{\textbf{Propagation to elastic constant uncertainty}: Replication of results shown for elastic constant prediction via energy in Table~\ref{tab:ec_error} for a diamond silicon configuration with $54$ atoms rather than $16$. }
    \label{tab:ec_error_config549}
    \centering
    \begin{tabular}{c|ccc}
    & & quantiles & \\
    & $0.25$ & $0.5$ & $0.75$ \\ 
    \hline
    via energy & & & \\
    $C_{11}$   &0.130652  &   0.333228  &   0.633059 \\
    $C_{12}$   &0.10643    &  0.246696  &   0.450938 \\
    $C_{21}$   &0.111804  &   0.241559   &  0.418278 \\
    $C_{45}$   & 0.000138352&  0.000262387&  0.000460765 \\
    $C_{54}$   &0.000145068&  0.000277879 & 0.000521586 \\
    $C_{44}$   &0.205293  &   0.342873    & 0.460263 \\
    \end{tabular}
\end{table}

\begin{figure}[h]
    \centering
    \begin{tabular}{cc}
        \includegraphics[width=0.4\linewidth]{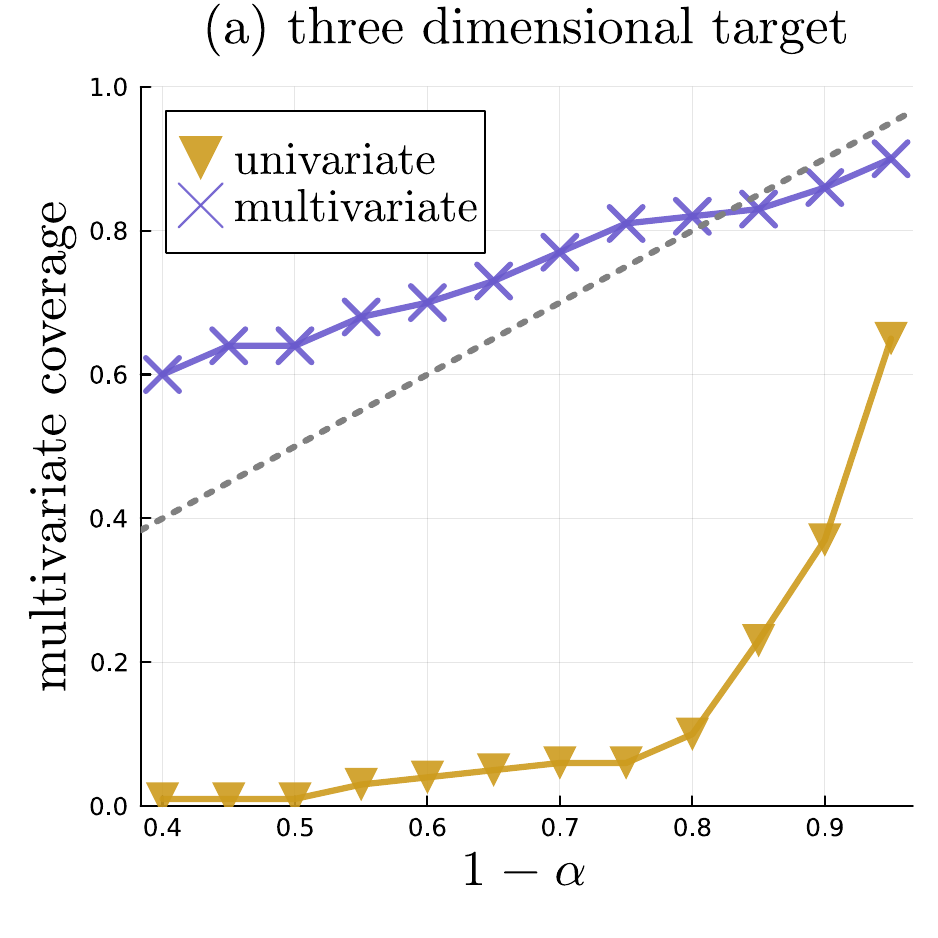} & \includegraphics[width=0.4\linewidth]{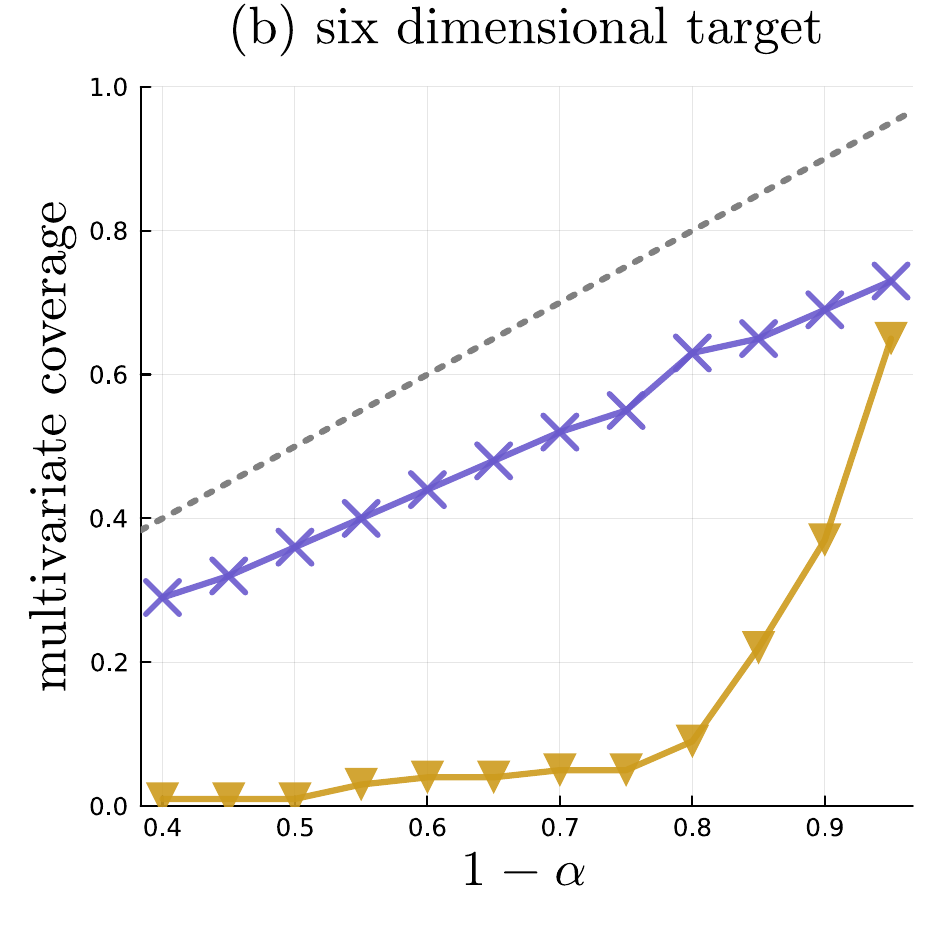}  \\
        \includegraphics[width=0.4\linewidth]{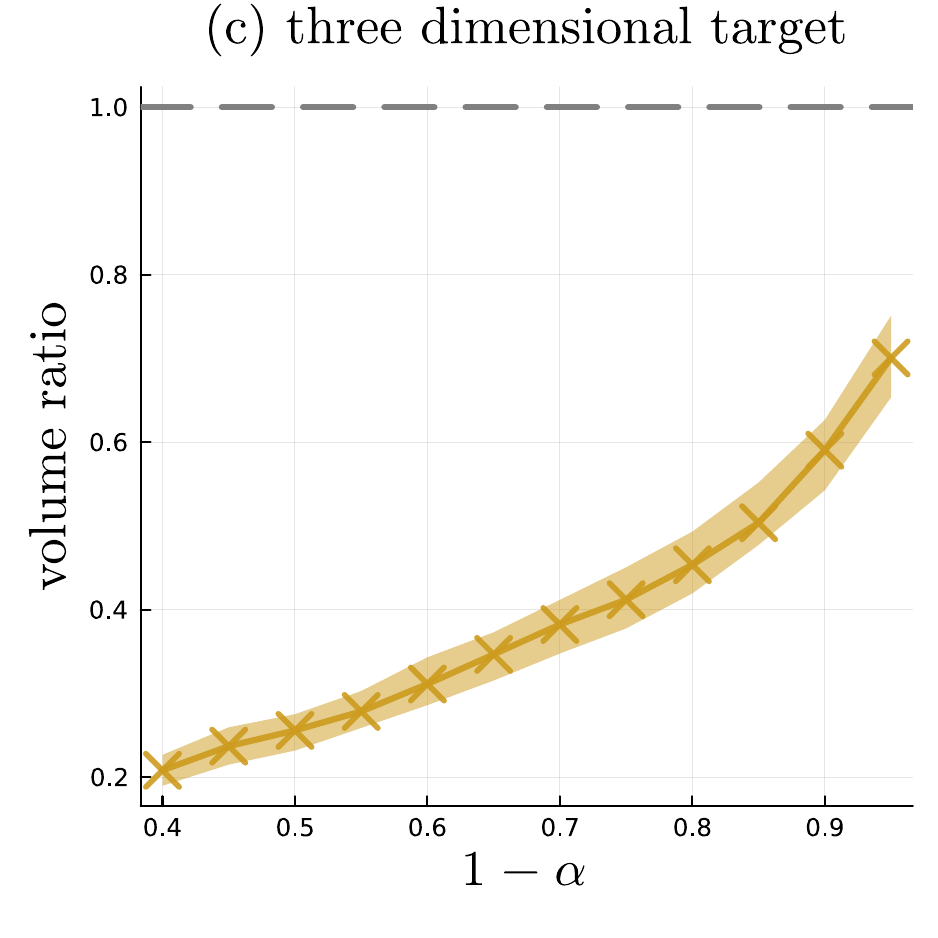} & \includegraphics[width=0.4\linewidth]{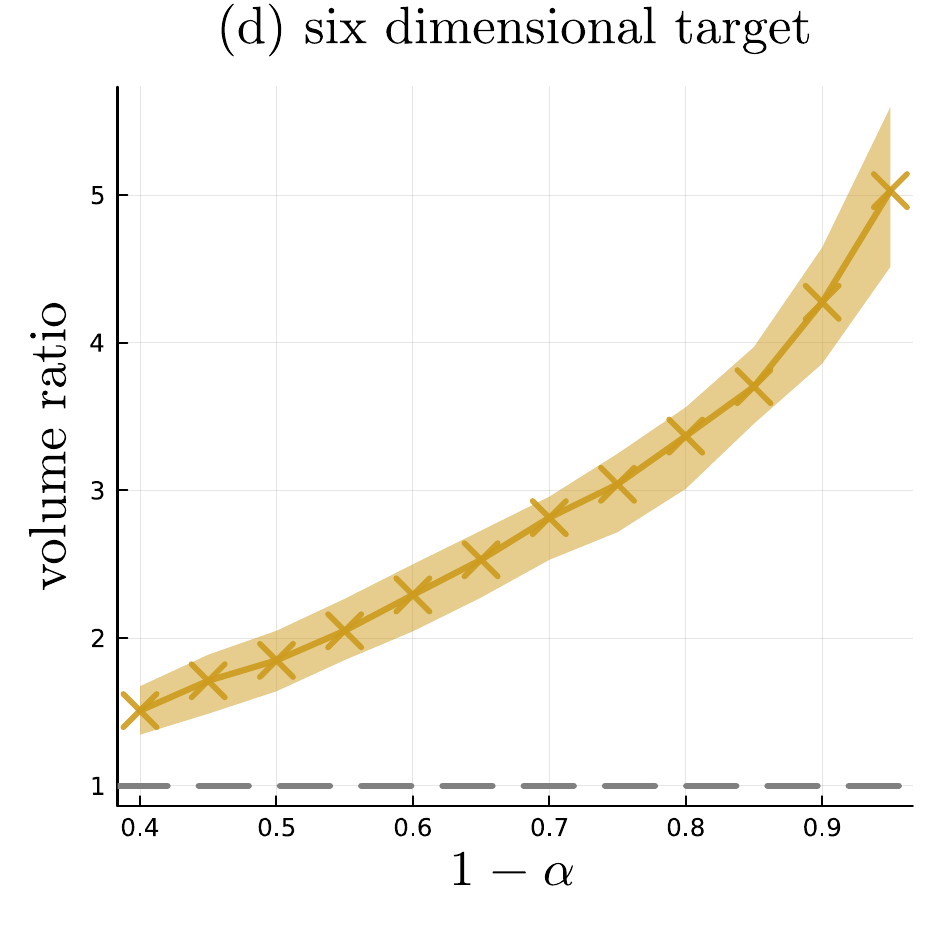} \\
    \end{tabular}
\caption{\label{fig:energy_ec_config549} \textbf{Propagation to elastic constant uncertainty}: Replication of energy based results presented in Figures~\ref{fig:elastic_constant} and~\ref{fig:ec_full} for a diamond silicon configuration with $54$ atoms rather than $16$. These results correspond to the error reported in Table~\ref{tab:ec_error_config549}. Left column: coverage and volume comparison for $6$ dimensional elastic constant prediction. Right column: results for final $3$ dimensional elastic constant prediction. Top row: full vector coverage for tolerance $\alpha$. Bottom row: ratio (corrected for $d=6$) of univariate sets size to multivariate set size.}
\end{figure}

We also repeat our energy based experiments for Si$_{54}$ rather than Si$_16$. In practice, the elastic constant should be computed using only the smaller system, as the constant is the same for all system sizes, and predictions for larger systems are more computationally expensive. We perform this test to establish that our prediction set performance is similar in both cases. We report the error of our predictions of the elastic constant in Table~\ref{tab:ec_error_config549} and show the coverage and volume in Figure~\ref{fig:energy_ec_config549}. We see qualitatively similar behavior to our previous experiments on Si$_16$. The multivariate conformal method produces prediction sets with greater coverage than the univariate method. The left column shows results for all six predicted components of the elastic constant, while the right column describes only the three nontrivial components. For six dimensional prediction, the multivariate prediction sets do not achieve the guaranteed coverage but manage to produce prediction sets considerably smaller than the univariate prediction sets. For the nontrivial components, the multivariate prediction sets achieve the coverage guarantee for almost all values of $\alpha$. We attribute this difference in behavior depending on target dimension to the difficulty of covering the zero components ($C_{45}$, $C_{54}$).

\clearpage

\section{Additional results of vacancy formation energy prediction\label{app:vfe}}

\begin{figure}[h]
\centering
    \begin{tabular}{ccc}
        \includegraphics[trim={0.55cm 0 0.55cm 0},clip,width=0.31\linewidth]{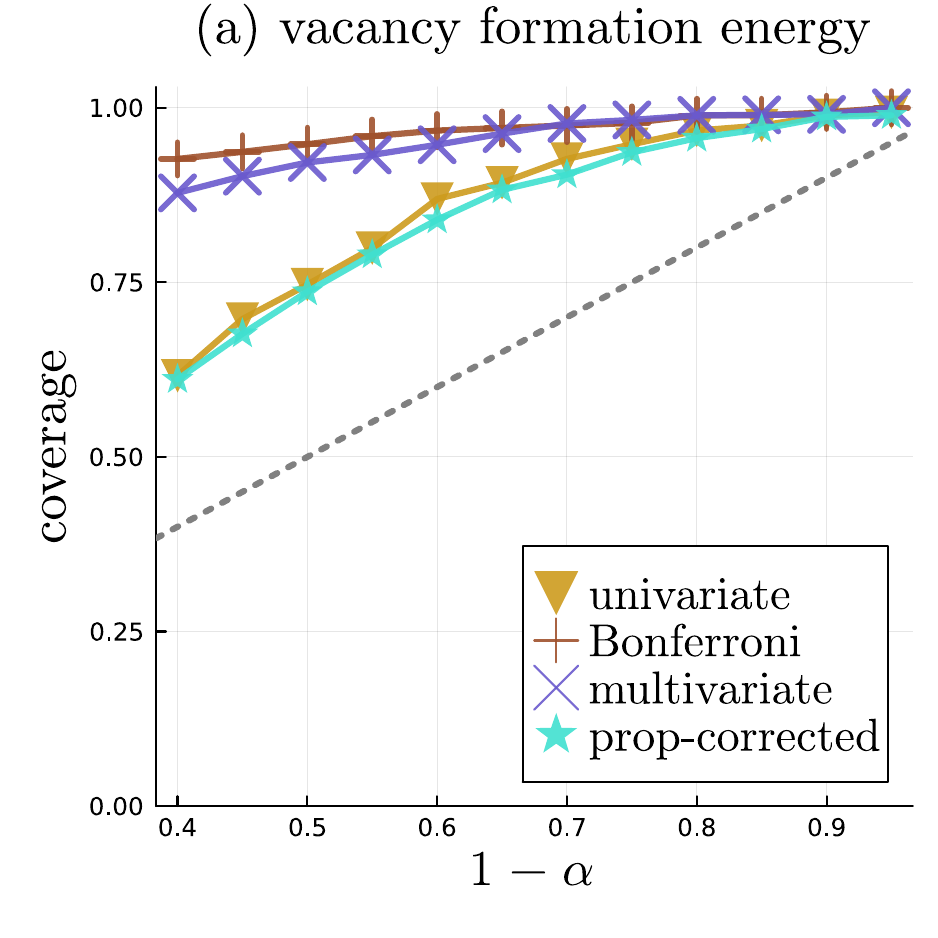} &
        \includegraphics[trim={0.55cm 0 0.55cm 0},width=0.31\linewidth]{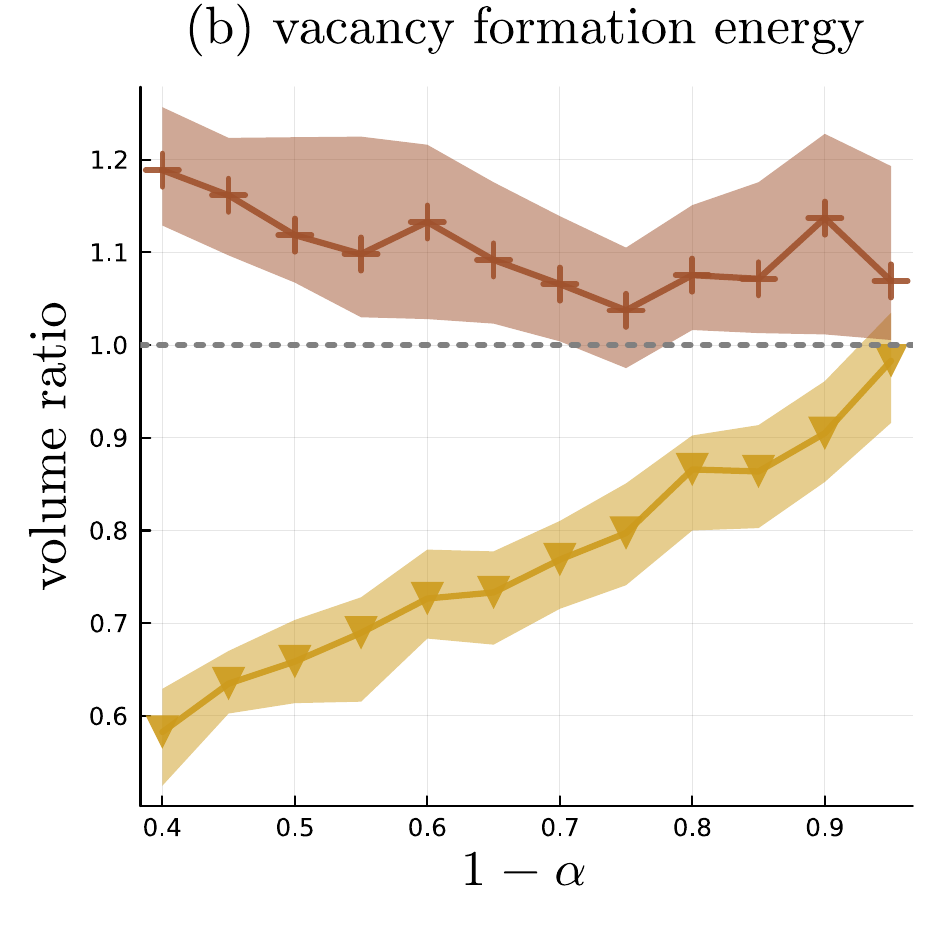} &
        \includegraphics[trim={0.55cm 0 0.55cm 0},width=0.31\linewidth]{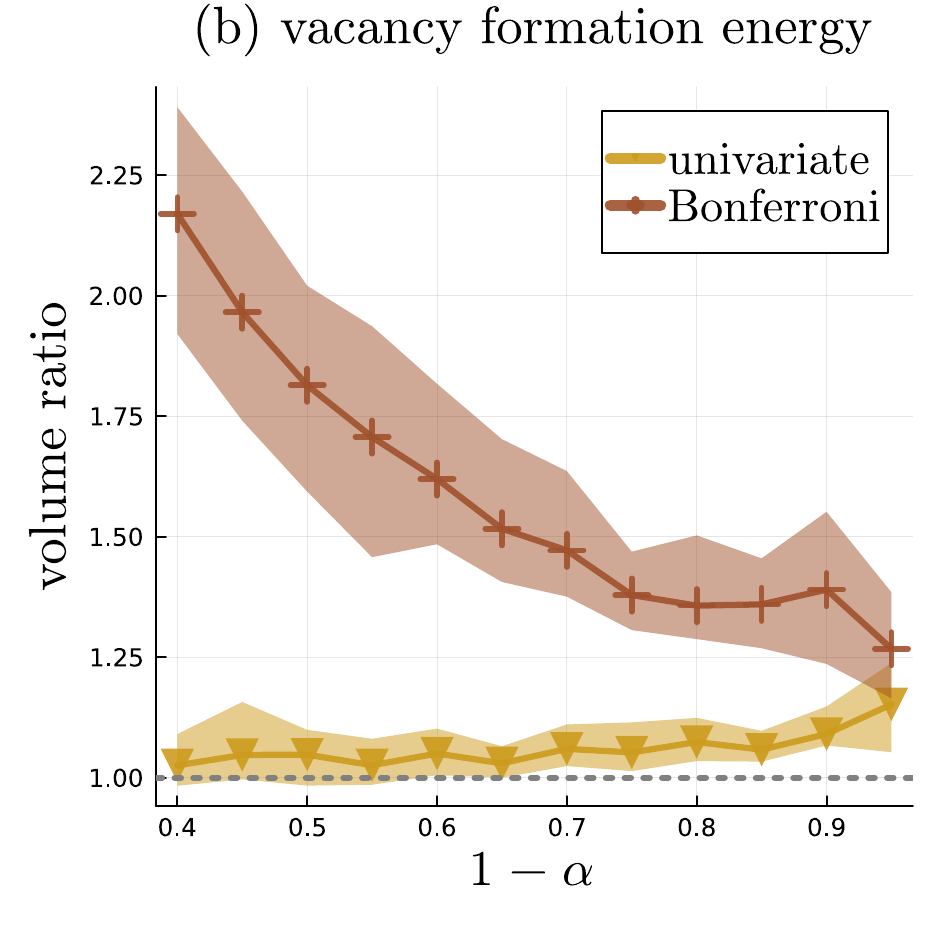} \\[6pt]
    \end{tabular}
\caption{\label{fig:vacancy_formation_n150} \textbf{Propagation to vacancy formation energy uncertainty}: Replication of results from Figure~\ref{fig:vacancy_formation} where the multitask dataset is trained with only $150$ configurations to inform the primary task, rather than $350$. (a): coverage of propagated conformal sets produced by univariate (gold triangles), Bonferroni-corrected univariate (red +s), multivariate (purple xs), and projection corrected multivariate (turquoise stars) approaches at tolerance $\alpha$. The $y=x$ line is dotted in gray. (b): Gold triangles mark the ratio between univariate and multivariate prediction interval length. Red $+s$ compare Bonferroni-corrected univariate to the  projection corrected multivariate method. A dotted horizontal line marks a ratio of $1$. (c): the comparison in volume ratio is repeated for univariate methods compared against the multivariate approach with propagation-corrected scores. Ribbons shade the region between the $0.25$ and $0.75$ quantiles. Results are obtained from $25$ randomly drawn training and calibration sets for each of $55$ silicon configurations. }
\end{figure}

Table~\ref{tab:vfe} reports the error distribution of the vacancy formation energy predictions used for our main text results in Section~\ref{ss:vacancy}. For Figure~\ref{fig:vacancy_formation} and Table~\ref{tab:vfe}, we trained our surrogate using $350$ bulk configurations. In Figure~\ref{fig:vacancy_formation_n150}, we demonstrate that the conformal prediction methods show similar qualitative behavior for a training set of $150$ training configurations. 

\begin{table}[h]
    \caption{\textbf{Propagation to vacancy formation energy}: Relative error of vacancy formation energy predictions against DFT reference. The $0.25$, $0.5$, $0.75$ quantiles are taken across $100$ random draws of the training and calibration sets from silicon diamond configurations with $\{2,16,54,128\}$ atoms. Calibration is performed in the space of the bulk configuration and vacancy energies and propagated to their scaled difference.}
    \centering
    \begin{tabular}{ccccc}
     & & quantiles & & \\
    $0.25$ & & $0.5$ & & $0.75$ \\ 
    \hline
      $0.0369428$  & & $0.0796825$ & & $0.140459$ \\
    \end{tabular}
    \label{tab:vfe}
\end{table}

\newpage

\end{document}